\documentclass[a4paper,11pt]{article}
\pdfoutput=1 % if your are submitting a pdflatex (i.e. if you have
\usepackage{jcappub} % for details on the use of the package, please
\usepackage[T1]{fontenc} % if needed
\usepackage{subcaption}

\title{\boldmath Lattice formulation of axion inflation. Application to preheating}

\def\lsim{\mathrel{\rlap{\lower3pt\hbox{\hskip0pt$\sim$}}
		\raise1pt\hbox{$<$}}}         %less than or approx. symbol
\def\gsim{\mathrel{\rlap{\lower4pt\hbox{\hskip1pt$\sim$}}
		\raise1pt\hbox{$>$}}}         %greater than or approx. symbol

\def\be{\begin{equation}} 
\def\ee{\end{equation}}
\def\bea{\begin{eqnarray}}
\def\eea{\end{eqnarray}}

\def \FFdual {F_{\mu\nu}\tilde{F}^{\mu\nu}}
\def \Bfour {B^{(4)}}
\def \Etwo {E^{(2)}}
\def \vertbar {{\big |}}
\def \dt {\text{dt}}

\newcommand{\mn}{{\mu\nu}}

\usepackage{color}
\definecolor{darkgreen}{RGB}{0, 100, 0}
\author{Jos\'e Roberto Canivete Cuissa,\,}
\author{Daniel G.~Figueroa}

\affiliation{Institute of Physics, Laboratory of Particle Physics and Cosmology (LPPC), \'Ecole Polytechnique F\'ed\'erale de Lausanne (EPFL), CH-1015 Lausanne, Switzerland.}

\emailAdd{jose.canivetecuissa@alumni.epfl.ch}
\emailAdd{daniel.figueroa@cern.ch}

\abstract{We present a lattice formulation of an interaction ${\phi\over\Lambda} F\tilde F$ between an axion and some $U(1)$ gauge sector with the following properties: it reproduces the continuum theory up to $\mathcal{O}(dx_\mu^2)$ corrections, it preserves exact gauge invariance and shift symmetry on the lattice, and it is suitable for self-consistent expansion of the Universe. The lattice equations of motion can no longer be solved by explicit methods, but we propose an implicit method to overcome this difficulty, which preserves the relevant system constraints down to arbitrary (tunable) precision. As a first application we study, in a comoving grid in $(3+1)$ dimensions, the last efolds of axion-inflation with quadratic potential and the preheating stage following afterwards. We fully account for the inhomogeneity and non-linearity of the system, including the gauge field contribution to the expansion rate of the Universe and its backreaction into the axion dynamics. We characterize in detail, as a function of the coupling, the energy transfer from the axion to the gauge field. Two coupling regimes are identified, sub- and super-critical, depending on whether the final energy fraction stored in the gauge field is below or above $\sim 50\%$ of the total energy. The Universe is very efficiently reheated for super-critical couplings, rapidly entering in a radiation dominated stage.  Our results on preheating confirm previously published results.}

\begin{document}
	
	%\begin{flushright}
	%UMN-TH/3720-18; 
	%\end{flushright}
	\maketitle
	\flushbottom
	
\section{Introduction}
	
Compelling evidence~\cite{Akrami:2018odb} supports the inflationary paradigm as the leading explanation for the initial conditions of the early Universe. The exact scenario responsible for inflation remains however unidentified. Inflation is often assumed to be driven by a singlet scalar field, the $inflaton$, with potential and initial conditions appropriately chosen to sustain a long enough period of accelerated expansion. From a model building point of view, inflationary constructions suffer typically from being very sensitive to unknown ultraviolet (UV) physics,  what can be viewed as a challenge to the robustness of the predictions. A simple way to avoid this problem relies on the introduction of some symmetry that naturally forbids certain interactions between the inflationary sector and other physics. In this respect, a promising candidate as an inflaton is a pseudo-scalar field or axion, that enjoys a shift-symmetry. If the inflationary sector is invariant under a constant shift of the field amplitude, the form of possible interactions of the inflaton with other species is severely restricted, ``protecting'' the inflationary dynamics from unknown UV physics. 
	
Axions-like fields are actually ubiquitous in particle physics, as they represent a possible solution to the strong CP problem~\cite{Peccei:1977hh}, potential dark matter candidates~\cite{Hui:2016ltb, Marsh:2015xka}, or interesting inflaton incarnations~\cite{Freese:1990rb,Sorbo:2011rz}. Several implementations of axion-driven inflation scenarios have been proposed, from Natural inflation~\cite{Freese:1990rb, Adams:1992bn} to $N$-flation~\cite{Dimopoulos:2005ac, Easther:2005zr, Bachlechner:2014hsa} and stringy oriented axion monodromy models ~\cite{McAllister:2008hb, Silverstein:2008sg}. There has been in fact a significant interest in the construction of viable axion inflation models within the framework of string theory~\cite{Baumann:2014nda,McAllister:2014mpa, Kaloper:2011jz, Marchesano:2014mla, Blumenhagen:2014gta, Hebecker:2014eua, Cai:2014vua,Pajer:2013fsa}. In this paper we will not be concerned, however, with the exact implementation of axion inflation. For simplicity we will consider a quadratic potential, corresponding to the simplest type of chaotic inflation~\cite{Linde:1983gd}. Even though these scenarios are under pressure by current CMB observations~\cite{Akrami:2018odb}, the addition of a small non-minimal gravitational coupling can easily reconcile them with the data~\cite{Tsujikawa:2013ila}. We consider a  quadratic potential simply as a working 'arena' for scenarios with a monomial shaped potential with a minimum after inflation. This choice will actually enable us to compare our results with others in the literature, where the same choice was made. Furthermore, as it will become clear later, a major aim of this paper is to introduce a robust lattice formulation of the interaction of a shift-symmetric field with a gauge field, in the context of an expanding Universe, and for this matter the choice of potential is simply irrelevant.
	
For any inflationary scenario, a reheating stage must follow after inflation, converting all the energy available into the particle species constituting all the matter and radiation in the Universe. Eventually, when the created particles dominate the total energy budget of the Universe and thermalize, the Universe enters into the thermal era characteristic of the ``hot Big Bang''  paradigm. If the inflaton has sufficiently large couplings to other species, the initial stages after inflation are typically characterized by non-perturbative particle production phenomena, typically known as {\it preheating}. Preheating mechanisms have been investigated in detail in the past: from parametric resonance~\cite{Kofman:1997yn,Figueroa:2016wxr}, spinodal instabilities~\cite{Felder:2000hj,Felder:2001kt} and instant-like effects~\cite{Felder:1998vq,Felder:1999pv} of scalar field species, to parametric excitation of fermions~\cite{Greene:1998nh,Greene:2000ew,Giudice:1999fb} or non-perturbative excitation of gauge fields~\cite{Dufaux:2010cf,Tranberg:2017lrx}. For reviews on preheating see~\cite{Allahverdi:2010xz,Amin:2014eta}. In this paper we will actually study the post-inflationary preheating dynamics in axion inflation. In general, once we have chosen the inflaton potential, the details of preheating depend directly on the choice of interactions between the inflaton and the fields coupled to the inflaton\footnote{Alternatively, the Universe can be reheated even in the absence of direct couplings of the inflaton to other field species, like in the context of gravitational reheating~\cite{Ford:1986sy} and variants~\cite{Figueroa:2016dsc,Dimopoulos:2018wfg} A recent criticism of standard gravitational reheating~\cite{Figueroa:2018twl}, reinforces however the idea that the Universe can be successfully reheated most naturally when the inflaton is coupled to other species.}. In the case of axion inflation, the shift symmetry severely constrains the form of the inflaton coupling to other species. 
	
In particular, the lowest dimensional operator compatible with a shift symmetry is the interaction term between the pseudo-scalar field and the \textit{Pontryagin} or {\it topological number} density $\mathcal{K} \equiv \FFdual$ of some gauge sector\footnote{One can also consider a derivative coupling of the axion inflaton to the axial current of the fermions~\cite{Kusenko:2014uta,Adshead:2015jza,Adshead:2015kza,Adshead:2018oaa,Domcke:2018eki}, what leads to very interesting phenomenology, including e.g.~the successful realization of baryogenesis.}. We will focus on an Abelian gauge group, exploring the particle production of the gauge field due to the inflaton oscillations after inflation, when the inflaton and the gauge field interact via $\phi\,\FFdual$. The phenomenology due to the presence of this coupling in axion inflation models is very rich, and it has been extensively studied in the literature. Namely, during slow-roll inflation, an exponential production of helical gauge bosons is expected~\cite{Garretson:1992vt, Prokopec:2001nc, Anber:2009ua,Barnaby:2010vf,Adshead:2013qp,Cheng:2015oqa}. The exited gauge fields can then lead to interesting phenomenology such as the production of non-Gaussian density perturbations~\cite{Barnaby:2011vw, Barnaby:2011qe,Pajer:2013fsa}, primordial black holes~\cite{Linde:2012bt, Pajer:2013fsa, Bugaev:2013fya, Cheng:2015oqa}, magnetic fields~\cite{Garretson:1992vt, Adshead:2016iae}, (chiral) gravitational waves~\cite{Sorbo:2011rz, Barnaby:2011qe, Cook:2013xea,Adshead:2013qp}, and a very efficient preheating mechanism~\cite{Adshead:2015pva,Adshead:2017xll}. These effects are very sensitive to the inflaton velocity, hence becoming larger towards the end of inflation and in particular during the preheating stage after inflation, as the inflaton velocity oscillates, rapidly changing strength and sign. Let us finally note that a later decay of helical gauge fields after inflation could be used to generate the baryon asymmetry of the Universe~\cite{Giovannini:1997eg,Anber:2015yca,Fujita:2016igl,Kamada:2016eeb,Jimenez:2017cdr}.
	
Some works have already studied (p)reheating after axion-driven inflation scenarios, considering peturbative decays~\cite{Blumenhagen:2014gta}, linearized analyses of an Abelian gauge field coupled to an oscillating scalar field~\cite{McDonough:2016xvu,ArmendarizPicon:2007iv, Braden:2010wd}, and a full non-linear analysis~\cite{Adshead:2015pva}. Ref.~\cite{Adshead:2015pva} is actually of particular relevance to us, as it has studied in full detail the non-linear dynamics of preheating after axion inflation with a coupling $\phi\,F_\mn\tilde F^\mn$. The aim of our present work is actually twofold: $a)$ to introduce a refined lattice formulation of axion inflation setup's, particularly suited for describing an interaction of the form $\phi\,F_\mn\tilde F^\mn$, and $b)$ to consider, as a first application of this technique, a review of the preheating stage after axion inflation studied in~\cite{Adshead:2015pva}, so that we can compare numerical results. 

In particular, we propose the use of the lattice formulation of an interaction term $\phi\,F_\mn\tilde F^\mn$ based on the formulation introduced in~\cite{Figueroa:2017qmv}, but adapted to include self-consistent expansion of the Universe, so that both (pseudo-)scalar and gauge fields contribute to the expansion rate. Our lattice technique reproduces the continuum limit of the theory to quadratic order in the lattice spacing, and more importantly it obeys the following properties on the lattice: $i)$ the system is exactly gauge invariant, and $ii)$ shift symmetry of the axion is exact. Property $i)$ implies that physical constraints such as the Gauss law or the Bianchi identities, are exactly verified on the lattice (up to machine precision). Property $ii)$ implies that the lattice formulation has naturally embedded a construction of the topological number density that admits a total (lattice) derivative representation $\mathcal{K} \equiv F_\mn\tilde F^\mn =  \Delta_\mu^+ K^\mu$, which reproduces the continuum expression $\mathcal{K} = \partial_\mu K^\mu \propto \vec E \cdot \vec B$ up to $\mathcal{O}(dx_\mu^2)$ corrections. Without this property the interaction $\phi\,F_\mn\tilde F^\mn$ cannot be interpreted as a derivative coupling, and hence it would not really be shift symmetric. As it is precisely shift symmetry that justifies the functional form of the interaction in first place, it seems therefore relevant to preserve exactly such symmetry at the lattice level. This leads however to an extra technical complication compared to standard formulations of Abelian gauge lattice theories: as we can no longer solve the lattice equations of motion by an explicit method, but rather by means of implicit methods that require numerically iterative procedures at each time step. 

Summarizing, the lattice representation we use preserves all the relevant physical properties of the system, i.e.~gauge invariance, shift symmetry and the topological nature of $F_\mn\tilde F^\mn$. Hence it represents the best approximation (to quadratic order in the lattice spacing) to the theory in the continuum. Equipped with our lattice formulation, we will study the last efolds of inflation and the preheating stage in axion-inflation, studying in detail the energy transfer from the inflaton to the Abelian gauge field as a function of the coupling strength. 
	
This paper is organized as follows. In Sect.~\ref{sec:theoretical_background} we briefly review the dynamics during inflation, introducing the standard equations of motion and considering both analytical and numerical solutions in the continuum. In Sect.~\ref{sec:lattice_approach} we introduce a non-compact lattice formulation of an Abelian gauge theory with topological term $\FFdual$, suitable for self-consistent expansion of the Universe. We explain in detail the virtues of this formulation and how to iterate the lattice equations of motion. In Sect.~\ref{sec:LatticeApplication} we study in the lattice the last efolds of inflation and the preheating stage following afterwards. In Sect.~\ref{subsec:Inflation} we discuss our strategy for initializing the fields on the lattice, determine what is the appropriate initial cutoff for the gauge field fluctuations, and study the consistency of the dynamics during the last efolds of inflation for simulations initiated at different times. In Sect.~\ref{subsec:Preheating} we study preheating after inflation, investigating the transfer of energy from the axion into the Abelian gauge field. We analyse in detail the evolution of the energy densities and the power spectra of the fields. We quantify the efficiency of preheating as a function of the coupling strength. In Sect.~\ref{subsec:Constraints} we demonstrate our numerical ability to preserve the relevant dynamical constraints on the lattice. In Sect.~\ref{sec:Discussion} we summarize our results and discuss future prospects and applications of our lattice formalism. In the appendix we present interesting aspects of our lattice implementation, and relevant tests passed by our simulations.
	
From now on {\small$m_{pl} = {1\over\sqrt{8\pi G}} \simeq 2.44\cdot10^{18}$ GeV} is the reduced Planck mass. The expanding Universe is described by a Friedmann-Lema\^itre-Robertson-Walker (FLRW) background metric $ds^2 = -dt^2 + a^2(t)d\vec x d\vec x$,  where {\small$a(t)$} is the scale factor, and {\small$t$} represents the cosmic time. Latin letters run over spatial dimensions, and greek letters over space-time dimensions, as usual. We assume summation over repeated indices in the continuum, but not in the lattice.
	
	\section{Axion inflation with $\phi\,\FFdual$ coupling}\label{sec:theoretical_background}
	
	Here we describe the continuum theory formulation of the scalar-gauge system with coupling $\phi F\tilde F$. We first introduce the basic field definitions and conventions in Sect.~\ref{subsec:DefinitionsConventions}, and then discuss the field dynamics during axion inflation in Sect.~\ref{subsec:ContinuumDynamics}, reviewing analytical results from the literature, both in the absence and presence of backreaction of the gauge field into the dynamics. In Sect.~\ref{subsec:BackreactionLess} we describe numerical solutions in the absence of backreaction, that will serve as the basis for our approach to initialize the gauge field on the lattice. 
	
	\subsection{Basic framework}
	\label{subsec:DefinitionsConventions}
	
	Let us begin by considering the action in a FLRW background for an Abelian gauge field coupled to a pseudo-scalar field $\phi$, which will be referred to as an \textit{axion} from now on. Given our choice of signature $(-,+,+,+)$, we have
	\begin{eqnarray}
	S &=& \int d^4x \sqrt{-g} \left(\frac{1}{2}m_{pl}^2R -\frac{1}{2}\partial_{\mu}\phi\partial^{\mu}\phi - \frac{1}{2}m^2\phi^2 - \frac{1}{4} F_{\mu\nu}F^{\mu\nu}  + \frac{\phi}{4\Lambda}\FFdual \right)\text{,   }~~ \label{eq:action_C_curved} 
	\end{eqnarray}
	where $\Lambda$ is a mass scale controlling the scalar-gauge coupling strength. The quadratic potential supports slow-roll inflation at super-Planckian field values, while the mass of the axion is fixed to $m = 1.5\cdot10^{13}$ GeV, in order to explain the amplitude of the CMB temperature anisotropies. The field strength of the gauge field, and its dual counterpart are defined as usual
	\be
	F_{\mu\nu} \equiv \partial_{\mu}A_{\nu} - \partial_{\nu}A_{\nu}~~ \text{,  } ~~ \tilde{F}_{\mu\nu} \equiv \frac{1}{2}\epsilon_{\mu\nu\rho\sigma}F^{\rho\sigma}\text{  ,  }  \label{eq:fmunu}
	\ee
	however notice that as we are in FLRW, $\epsilon_{\mu\nu\rho\sigma}$ is the completely anti-symmetric tensor in curved space-time
	\be
	\epsilon^{0123} \equiv \frac{1}{\sqrt{-g}} = {1\over a^3(t)}\text{ .}
	\ee
	
	Action~(\ref{eq:action_C_curved}) is invariant under gauge transformations $A_{\mu} \longrightarrow A_{\mu} + \partial_{\mu}\alpha(x)$, where $\alpha(x) \in \mathbb{R}$ is an arbitrary function, whilst the axion enjoys a shift symmetry [explicitly broken though by the mass term of Eq.~(\ref{eq:action_C_curved})] as $\phi \rightarrow \phi + c$, with $c \in \mathbb{R}$ an arbitrary real constant. Defining the axion conjugate momenta, and electric and magnetic fields, as usual by
	\bea 
	\pi_{\phi} \equiv \dot{\phi}\,, ~~~~  E_i \equiv \dot{A}_i - \partial_iA_0\,, ~~~~ B_i \equiv \epsilon_{ijk}\partial_jA_k\,, \label{eq:electricandmagneticfields}
	\eea
	%With these definitions we obtain $-\frac{1}{4}F_{\mu\nu}F^{\mu\nu} = \frac{1}{2a^4}\left(a^2\vec{E}^2 - \vec{B}^2\right)$, and $\frac{\phi}{4\Lambda}\FFdual = \frac{\phi}{\Lambda a^3}\vec{E}\cdot\vec{B}$. Since $\sqrt{-g} = a^3$, the above Lagrangian 
	the action can be re-written in a vectorial form like
	\begin{eqnarray}
	S_{m} &=& \int d^4x \left( \frac{1}{2}a^3\pi_{\phi}^2 - \frac{1}{2}a(\vec{\nabla}\phi)^2 - \frac{1}{2}a^3m^2\phi^2 + \frac{1}{2}a (\vec{E}^2 - a^{-2}\vec{B}^2) + \frac{\phi}{\Lambda}\vec{E}\cdot\vec{B}  \right)\text{.  }~~~~~ \label{eq:actionVEC_C_curved}
	\end{eqnarray}
	From here it is worth noticing that the interaction term between the axion and the gauge field is not coupled to the space-time background. 
	
	Varying the action Eq.~(\ref{eq:action_C_curved}) with respect to the axion and gauge field, leads to the respective equations of motion (EOM)
	\begin{eqnarray}
	\partial_0\partial_0 \phi  &=&  -3H\partial_0 \phi + \frac{1}{a^2}\sum_i\partial_i\partial_i \phi - m^2\phi + \frac{1}{4\Lambda}\FFdual \text{  ,  } \label{eq:axionEOM_C_curved} \\
	%%%
	\partial_{\mu}\left(\sqrt{-g}F^{\mu\nu} \right) &=& \frac{1}{\Lambda} \partial_{\mu}\left(\sqrt{-g}\phi\tilde{F}^{\mu\nu} \right)\,, \label{eq:gaugeEOM_C_curved} 
	\end{eqnarray}
	where $H = \dot{a}/a$ is the Hubble parameter. The vectorial counterparts are
	\begin{eqnarray}
	& &\dot{\pi}_{\phi} =  -3H\pi_{\phi} + \frac{1}{a^2}\vec{\nabla}^2 \phi - m^2\phi + \frac{1}{a^3 \Lambda}\vec{E}\cdot\vec{B}\text{,} \label{eq:axionEOM_vec_C_curved} \\
	& &\dot{\vec{E}} = -H\vec{E} - \frac{1}{a^2}\vec{\nabla} \times \vec{B}  - \frac{1}{a\Lambda}\pi_{\phi}\vec{B} + \frac{1}{a\Lambda} \vec{\nabla}\phi \times \vec{E} - \frac{\phi}{a\Lambda}\underbrace{\left(\dot{\vec{B}} - \vec{\nabla}\times\vec{E}\right)}_\textrm{=0}\text{,  }~~ \label{eq:gaugeEOM_vec_C_curved} \\
	& &\vec{\nabla}\cdot\vec{E} =  - \frac{1}{a\Lambda}\vec{\nabla}\phi\cdot\vec{B} - \frac{\phi}{a\Lambda}\underbrace{\vec{\nabla}\cdot\vec{B}}_\textrm{=0} \text{  ,  } \label{eq:gausslaw_vec_C_curved} 
	\end{eqnarray}
	where Eq.~(\ref{eq:gausslaw_vec_C_curved}) is not dynamical and represents the Gauss constraint.  Note that in  Eqs.~(\ref{eq:gaugeEOM_vec_C_curved})-(\ref{eq:gausslaw_vec_C_curved}) we have kept terms that in reality are identically zero due to the Bianchi identities
	\begin{eqnarray}
	\partial_{\mu}(\sqrt{-g}\tilde{F}^{\mu\nu}) = 0 ~~\Longleftrightarrow~~ \begin{cases}
	\dot{\vec{B}} - \vec{\nabla}\times\vec{E} = 0 \vspace*{0.25cm} \\
	\vec{\nabla}\cdot\vec{B} = 0 \\
	\end{cases} \text{  .  } \label{eq:bianchiId}
	\end{eqnarray}
	We keep these null terms just to remind the reader that, in the lattice, obtaining vanishing Bianchi identities is not guaranteed by default, as it depends on the lattice representation of the terms involving the gauge field in the Lagrangian. The lattice formulation of $\phi\FFdual$ that we will describe in Sect.~\ref{sec:lattice_approach}, is based precisely on the ability to obtain vanishing Bianchi identities, and a Pontryagin density of topological nature (i.e.~admitting a total derivative representation).
	
	Together with Eqs.~(\ref{eq:axionEOM_vec_C_curved})-(\ref{eq:gausslaw_vec_C_curved}), we also need to take into account Friedmann equations governing the expansion of the Universe,
	\begin{eqnarray}
	\frac{\ddot{a}}{a} &=& \frac{-1}{6m_{pl}^2}(3p + \rho) \text{  ,  } \label{eq:friedmann2_C}\\
	\left(\dot a\over a\right)^2 &=& \frac{1}{3m_{pl}^2}\rho \text{  ,  } \label{eq:friedmann1_C}
	\end{eqnarray}
	where the first equation describes the dynamics of the scale factor, whereas the second represents a constraint to which we will refer to as the Hubble constraint. Here $\rho$ and $p$ are the background energy and pressure densities, defined from the stress-energy tensor of a perfect fluid  $T_{\mu\nu} \equiv (\rho + p)u_{\mu}u_{\nu} + pg_{\mu\nu}$, like 
	\begin{eqnarray}
	\rho = T_{00}\,,~~~~ p = \frac{1}{3a^2}\sum_iT_{ii}\,~. \label{eq:rhoandp_def}
	\end{eqnarray}
	In our problem at hand, the explicit form of these quantities can be obtained from the energy-momentum tensor in curved space-time
	\begin{eqnarray}
	T_{\mu\nu} 	&=& \frac{-2}{\sqrt{-g}}\frac{\delta( \sqrt{-g}\mathcal{L}_{m})}{\delta g^{\mu\nu}}\\
	&=& g_{\mu\nu}\left(-\frac{1}{2}g^{\alpha\beta}\partial_\alpha\phi\partial_\beta\phi - \frac{1}{2}m^2\phi^2 - \frac{1}{4}g^{\alpha\rho}g^{\beta\sigma}F_{\alpha\beta}F_{\rho\sigma}\right) + \partial_{\mu}\phi\partial_{\nu}\phi + g^{\alpha\beta}F_{\mu\alpha}F_{\nu\beta}\text{,  }\nonumber \label{eq:stress-energytensor_C} 
	\end{eqnarray}
	where $\mathcal{L}_m$ is the Lagrangian in Eq.~(\ref{eq:action_C_curved}). The energy and pressure densities then read
	\begin{eqnarray}
	\rho &\equiv& \frac{1}{2}\pi_{\phi}^2 + \frac{1}{2a^2}\sum_i (\partial_i\phi)^2 + \frac{1}{2}m^2\phi^2 + \frac{1}{2}\left( \sum_i \frac{E_i^2}{a^2} + \sum_i \frac{B_i^2}{a^4} \right) \text{  ,  }\label{eq:energy_density_C} \\
	p &\equiv& \frac{1}{3a^2}\sum_j T_{jj} = \frac{1}{2}\pi_{\phi}^2 - \frac{1}{6a^2}\sum_i (\partial_i\phi)^2 - \frac{1}{2}m^2\phi^2 + \frac{1}{6}\left( \sum_i \frac{E_i^2}{a^2} + \sum_i \frac{B_i^2}{a^4} \right)\text{  .  } \label{eq:pressure_density_C} 
	\end{eqnarray}
	
\subsection{Axion inflation dynamics}
\label{subsec:ContinuumDynamics}
	
The potential $V(\phi) = {1\over2}m^2\phi^2$ naturally sustains a \textit{slow-roll} regime of inflation for large field amplitudes. Introducing the slow-roll parameter
	\be
	\epsilon_H = -\frac{\dot{H}}{H^2}\,, \label{eq:slow_roll_parameter}
	\ee
	we expect $\epsilon_H \ll 1$ for most of the inflationary period, when the energy is dominated by the potential energy of the inflaton. As inflation goes on and the inflaton rolls down its potential, the slow-roll parameter grows (as the Hubble rate $H$ decreases), and eventually it reaches $\epsilon_H = 1$, what signals the end of inflation. 
	
Deep inside inflation, Eqs.~(\ref{eq:axionEOM_vec_C_curved})-(\ref{eq:gaugeEOM_vec_C_curved}) can be simplified by neglecting the gradients of the axion,  $| \nabla^2\phi| \ll m^2\phi$ and $|\nabla \phi \times  \vec E| \ll |\pi_\phi\vec B|$, as well as the backreaction of the gauge field into the axion dynamics, $|\vec E\cdot \vec B| \ll a^3m^2\Lambda\phi$. Moreover, the contribution of the gauge field to the energy and pressure densities appearing in Eqs.~(\ref{eq:friedmann2_C})-(\ref{eq:friedmann1_C}) should also be negligible, $a^2{\vec E}^2 + {\vec B}^2 \ll a^4m^2\phi^2$. The expansion of the Universe in such an approximation is dictated by the dynamics of an homogeneous axion, 
\begin{eqnarray}\label{eq:EOMinflatonHomogeneous}
	\ddot\phi + 3H\dot\phi + m^2\phi = 0\,,~~~~ {\ddot a\over a} = {1\over 3m_{pl}^2}\left(m^2\phi^2-\dot\phi^2 \right)~.
\end{eqnarray}
At the onset of inflation the initial value of the inflaton is super-Planckian $\phi_i \gg m_{pl}$ and from the slow-roll conditions we can infer that $\dot\phi_i \simeq -V_{,\phi}/3H_i$, $H_i \simeq (\phi_i/\sqrt{6})({m/m_{pl}})$. 
	
The inflaton slowly rolls its potential reducing its amplitude, until it becomes eventually of order $\phi \sim m_{pl}$, just when $\epsilon_H \gtrsim 1$. More precisely, from the condition $\epsilon_H(\phi_{\rm end}) = 1$ we obtain $\phi_{\rm end} = 1.039~ m_{pl}$. The number of efolds before the end of inflation is simply obtained as
\be\label{eq:NvsPhi}
	N \equiv -\ln\frac{a_{end}}{a} = -{1\over \sqrt{6}m_{pl}}\int_t^{t_{\rm end}}dt'\sqrt{\dot\phi^2(t') + m^2\phi^2(t')}\sim -\left(\frac{\phi}{2m_{pl}}\right)^2\,,  
\ee
where we use the convention that during inflation $N<0$, and in the last equality we have used properties of slow-roll, so it should only be taken as indicative\footnote{We have actually checked that the expression in Eq.~(\ref{eq:NvsPhi}) as a function of the inflaton amplitude, predicts the number efolds before the end of inflaton up the last efold, with only $\sim 3 \%$  error.}. In order to ensure that the inflationary period lasts $\sim 50-60$ efolds, the value of the inflaton at the moment when CMB fluctuations exited the Hubble radius must have been of the order $\phi_{\rm CMB} \sim 14-15~m_{pl}$. 
	
Deep inside inflation, say $\mathcal{O}(10)$ efolds before the end, the homogeneous axion domination is very well justified. However, because of the axionic-coupling $\phi\FFdual$, the gauge field will be excited during inflation. Thus, even if we start with a gauge field amplitude given by quantum fluctuations, as we approach the end of inflation, the gauge field will be largely amplified, and it might -- depending on the scalar-gauge coupling -- impact  on the inflationary dynamics. Analytically, it has been studied that the backreaction of the gauge field into the axions can potentially delay the end of inflation in a significant way, see e.g.~\cite{Barnaby:2011qe}. 

Let us briefly describe the status of the art on the analytical understanding of the gauge field dynamics and its backreaction. We can start by considering the gauge field evolution imposing homogeneity of the axion field. Switching to conformal time  $\tau \equiv \int {dt'\over a(t')}$, Eq.~(\ref{eq:gaugeEOM_C_curved}) becomes 
\begin{eqnarray}
	\nu = 0 &\rightarrow& \partial_{j}\partial_{j}A_0 = \partial_{\tau}\partial_{j}A_j\text{  ,  } \label{eq:gausslaw_initcond}\\
	\nu = i &\rightarrow& \partial_{\tau}^2A_i - \partial_{j}\partial_{j}A_i + \frac{1}{\Lambda}\epsilon_{ijk}\partial_{\tau}\phi\partial_{j}A_k = 0\text{  .  } \label{eq:gauge_EOM_initcond}
\end{eqnarray}
In order to solve these equations, it is useful to work in Fourier space, where we decompose each mode in a basis of helicity states
\be\label{eq:ModeDecomposition}
	\vec{A}(\tau, \mathbf{x}) = \int\frac{d^3k}{(2\pi)^3}\vec{A}(\tau, \mathbf{k}) e^{i \mathbf{k}\cdot\mathbf{x}}         = \sum_{\lambda = \pm}\int \frac{d^3k}{(2\pi)^3}A^{\lambda}(\tau,k)\vec{\varepsilon}^{\hspace{0.05cm}\lambda}( \hat{\mathbf{k}})e^{i \mathbf{k}\cdot\mathbf{x}}\text{  ,  }
\ee
where the helicity vectors $\varepsilon_i^{\pm}(\hat{\mathbf{k}})$ are defined in such a way that they satisfy the following properties 
\begin{eqnarray}
	%\begin{array}{rl}
	\hspace{-0.25cm}
	k_i \varepsilon_i^{\pm}(\hat{\mathbf{k}}) = 0\,,~~
	\epsilon_{ijk}k_j\varepsilon_k^{\pm}(\hat{\mathbf{k}}) = \mp ik\varepsilon_i^{\pm}(\hat{\mathbf{k}})\,,~~
	\varepsilon_i^{\pm}(\hat{\mathbf{k}})^{\ast} = \varepsilon_i^{\pm}(-\hat{\mathbf{k}})\,,~~
	\varepsilon_i^{\lambda}(\hat{\mathbf{k}})\varepsilon_i^{\lambda'}(-\hat{\mathbf{k}}) = \delta_{\lambda\lambda'}\,.
	%\end{array}
\end{eqnarray}
Using this mode decomposition and the properties of the helicity vectors just exposed, the differential equations for the two polarizations of the gauge field read
\be
	\left(\partial_{\tau}^2 + k^2 \mp \frac{k \dot{\phi}}{\Lambda H \tau}\right) A^{\pm}(\tau,k) = 0 \text{  ,  } \label{eq:gaugemodes_EOM}
	\ee
where we have used the slow-roll condition $\tau \approx -1 / aH$. Because of the different sign in Eq.~(\ref{eq:gaugemodes_EOM}), if $\dot\phi < 0$ ($\dot\phi > 0$) the polarization $A^+$ ($A^-$) will be exponentially amplified, while $A^-$ ($A^+$) will not, and hence can be neglected. A solution for the functional form of the excited polarization (say $A^+$ if we choose $\dot\phi < 0$), normalized to match the form of quantum fluctuations in the ultraviolet limit, is given by~\cite{Anber:2009ua}
\be
	A^+(\tau, k) = \frac{e^{\frac{\pi}{2}\xi}}{\sqrt{2 k}}W_{ -i\xi, \frac{1}{2}}(2 i k\tau) \simeq \frac{1}{\sqrt{2k}}\left(\frac{k}{2|\xi| aH}\right)^{1/4}e^{\pi|\xi| - 2\sqrt{2|\xi| k/aH}} \text{  ,  } \label{eq:approx_A+_mode}
	\ee
where the second expression is valid for $k|\tau| \ll 2|\xi|$, with the parameter $\xi$ is defined as
\be
	\xi \equiv \frac{\dot{\phi}}{2H\Lambda}\text{  .  } \label{eq:xi_def}
\ee
The expression in Eq.~(\ref{eq:approx_A+_mode}) is expected to approximate well the real solution to the gauge field mode function in the limit where $\xi = \text{const}$, which is well justified deep inside inflation, but less and less accurate towards the end of it. The second expression in Eq.~(\ref{eq:approx_A+_mode}) exhibits clearly the fact that the gauge field is mostly amplified around the Hubble scale at each time. 
	
Using Eq.~(\ref{eq:approx_A+_mode}) one can compute the mean electromagnetic energy density and the topological term, as a function of $\xi$~\cite{Anber:2009ua},
\begin{eqnarray}
	\label{eq:approximated_EM_density}
	\frac{1}{2a^4} \langle a^{2}E^2 + B^2 \rangle \simeq 1.4\cdot10^{-4}\frac{H^4}{|\xi|^3}e^{2\pi|\xi|}\,,\\
	\label{eq:approximated_EB}
	{1\over a^3}\langle \vec{E}\cdot\vec{B} \rangle \simeq -2.4\cdot10^{-4}\frac{H^4}{|\xi|^4}e^{2\pi|\xi|}\,.
\end{eqnarray}
In the Hartree approximation, the dynamics are described by the equations
\begin{eqnarray}\label{eq:expansionU_axion}
	\ddot{\phi} &=& -3H\dot{\phi} - m^2\phi + \frac{1}{a^3\Lambda}\langle \vec{E}\cdot\vec{B} \rangle \text{  ,  } \\
	\label{eq:expansionU_friedmann}
	{\ddot a\over a} &=& \frac{1}{3m_{pl}^2}\left(\frac{1}{2}m^2\phi^2 - \dot{\phi}^2 - \frac{1}{2a^4} \langle a^2E^2 + B^2 \rangle\right) \text{  , } 
\end{eqnarray}
so by plugging Eqs.~(\ref{eq:approximated_EM_density}), (\ref{eq:approximated_EB}) in Eqs.~(\ref{eq:expansionU_axion}), (\ref{eq:expansionU_friedmann}), one can quantify the backreaction of the gauge field into both the axion and expanding Universe dynamics within such approximation. One can estimate in that way the moment during inflation (for a given coupling $\Lambda^{-1}$) when the terms $\langle \vec{E}\cdot\vec{B} \rangle$ and $\langle a^2E^2 + B^2 \rangle$ become relevant. 

Before we move further into other considerations, let us discuss the allowed range of values for the scale $\Lambda$. In particular, to respect CMB limits on the amount of primordial non-Gaussianity, the function $\xi$ defined by Eq.~(\ref{eq:xi_def}) should respect  $\xi \lesssim \xi_{\rm CMB} \simeq 2.5$ at CMB scales~\cite{Ade:2015lrj,Ade:2015ava}, which translates (given our choice of potential ${1\over2}m^2\phi^2$) into $1/\Lambda \lesssim 35\,m_{pl}^{-1}$. However, to ensure that primordial black holes are not over-produced by large density fluctuations, a more stringent constraint actually emerges, requiring $\xi \lesssim \xi_{CMB} \simeq 1.5-1.7$ \cite{Linde:2012bt,Bugaev:2013fya}. This implies that $1/\Lambda \lesssim 22-25~m_{pl}^{-1}$. In practice, as the primary focus of the application of our lattice formalism in this paper will be preheating, we will not need to reach so large couplings: as we will show in Sect.~\ref{subsec:Preheating}, the preheating efficiency does not improve further as soon as we reach $1/\Lambda \gtrsim 10~m_{pl}^{-1}$. Hence, a range like $1/\Lambda \leq 15~m_{pl}^{-1}$, which from now on we adopt as a fiducial range, suffices for our purposes.

\subsection{Backreaction-less (numerical) solution}\label{subsec:BackreactionLess}
	
The non-linearity of the system makes actually the dynamics very complicated, and probably inaccurate, to be dealt with an analytical approach. Hence the importance of our lattice study (Sect.~\ref{subsec:Inflation}, \ref{subsec:Preheating}), which will capture in full detail the non-linearities of the system. However, to be able to address the problem from the lattice point of view, we need to have control over the field configuration that will be introduced in the lattice as an initial condition. 

As the solution Eq.~(\ref{eq:approx_A+_mode}) may not be so precise towards the end of inflation, we will use a different approach in order to introduce an initial condition of the fields in the lattice. In particular, back to cosmic time, but still keeping the axion homogeneous, Eq.~(\ref{eq:gaugemodes_EOM}) becomes
\begin{eqnarray}\label{eq:AplusMathematica}
\ddot{A}_\pm(t,k) + \left({\dot a\over a}\right){\dot A}_{\pm}(t,k) +\left[{k^2\over a^2}\pm \left(\frac{k\dot{\phi}}{a\Lambda}\right)\right] A_{\pm}(t,k) = 0\,,
\end{eqnarray}
where we have used again the helicity mode decomposition from Eq.~(\ref{eq:ModeDecomposition}). Once again, given our convention $\dot\phi < 0$, only $A_+$ becomes tachyonic, so we will only worry about this polarization. Note that since we are working in cosmic time, no approximations have been done apart from assuming the homogeneity of the axionic field. Furthermore, since this equation is linear in $k$, we can solve it numerically in a 1-dimensional grid of discretized comoving momenta $\lbrace k_1, k_2, ..., k_n\rbrace$, initializing each mode $k_i$ at a time when its wavelength is well inside the Hubble radius. Introducing a conformal transformation of the gauge field, $A_+ \equiv a^{-1/2}{\tilde A}_+$, we can get rid off the friction term in Eq.~(\ref{eq:AplusMathematica}). We then obtain that modes which are deep inside the Hubble radius $k \gg aH$ during inflation, evolve according to $\ddot{\tilde A}_+ + \omega_k^2 \tilde A_+ \simeq 0$, with $\omega_k^2 \simeq {k^2\over a^2} \gg H^2, {\ddot a\over a}$. Choosing the Bunch-Davies vacuum, ${\tilde A}_+(t,k \gg aH) \simeq  {1\over \sqrt{2\omega_k}}e^{-i\omega_k t}$, we can use the following initial condition,
\begin{eqnarray}\label{eq:InitialCondition_A}
	%\begin{array}{rcl}
	&&\hspace*{-11mm}A_+(k_i,t_i) = {\tilde A_+(k_i,t_i)\over \sqrt{a(t_i)}} \simeq {e^{-i\omega_{i}t_i}\over \sqrt{2a(t_i)\omega_{i}}} \equiv {1\over \sqrt{2k_i}}\left[\cos(\omega_{i}t_i)-i\sin(\omega_{i}t_i)\right]\,, \\%& ~~~& {\rm Im}\lbrace A_+(k_i,t_i)\rbrace = 0\\
	%{\rm Re}\lbrace \dot A_+(k_i,t_i)\rbrace = 0\,, & ~~~&  
	\label{eq:InitialCondition_dA}
	&&\hspace*{-11mm}\dot A_+(k_i,t_i) = {d\over dt}\left\lbrace{\tilde A_+(k_i,t_i)\over \sqrt{a(t_i)}}\right\rbrace\simeq -i{\omega_{i}\over \sqrt{2k_i}}e^{-i\omega_{i}t_i} \equiv -{1\over a(t_i)}\sqrt{k_i\over 2}\left[\sin(\omega_{i}t_i)+i\cos(\omega_{i}t_i)\right],
	% \end{array}
\end{eqnarray}
where $\omega_i \equiv k_i/a$. Furthermore, the initial time for each mode is obtained from the condition
\begin{eqnarray}
	k_i \equiv x_i a(t_i)H(t_i)\,,~~~ x_i \gg 1\,,
\end{eqnarray}
with $x_i$ some large positive value determining the initial ``penetration'' of the scale $1/k$ inside the Hubble radius. In practice, we choose $x_i = 1000$ for all the modes, making sure in this way that we initialize the gauge field sufficiently deep inside the Hubble radius.

\begin{figure} % "[t!]" placement specifier just for this example
	\begin{subfigure}{0.49\textwidth}
		\includegraphics[width=\linewidth]{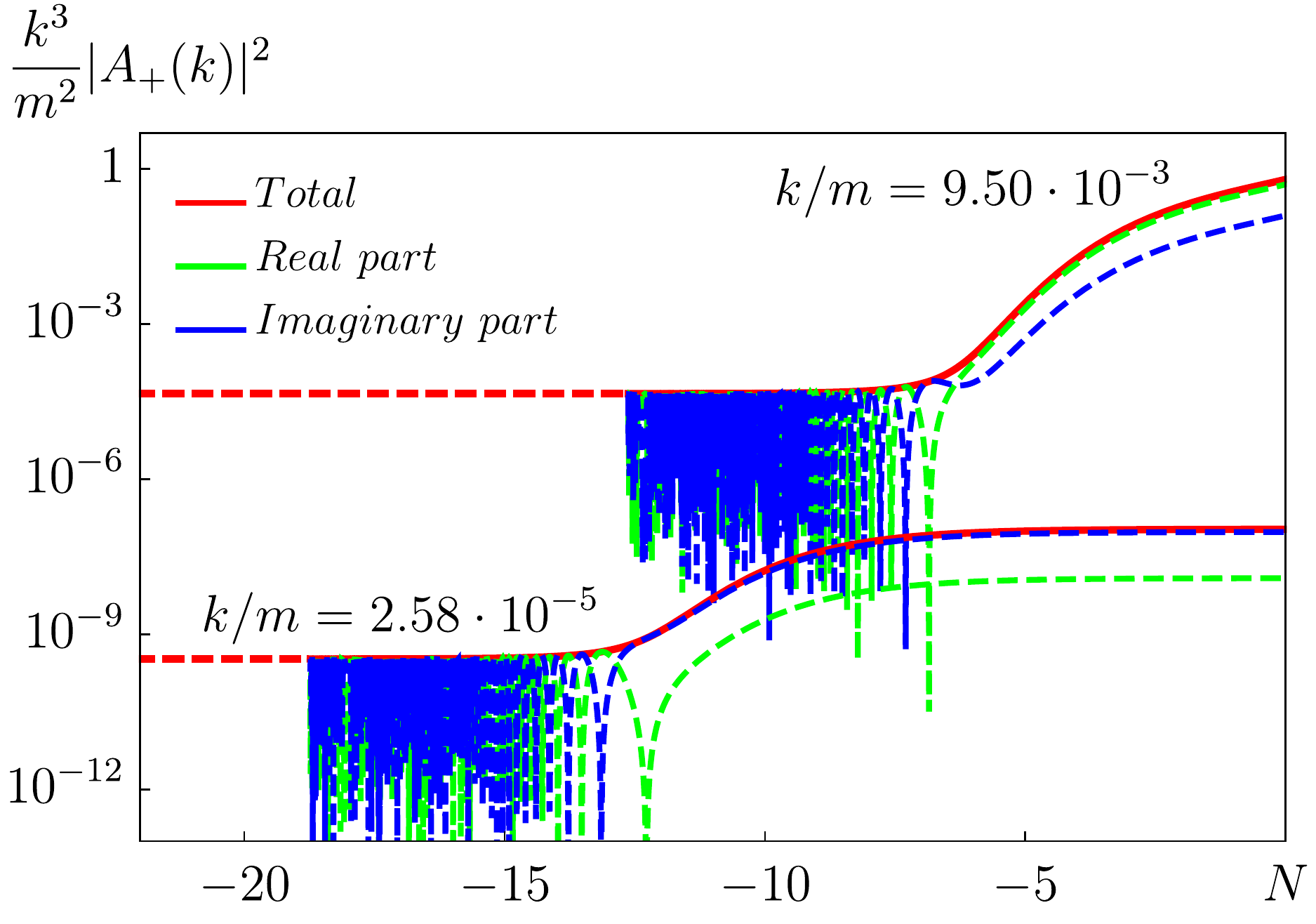}
		\caption{} \label{fig:ModeEvolution}
	\end{subfigure}\hspace*{\fill}
	\begin{subfigure}{0.49\textwidth}
		\includegraphics[width=\linewidth]{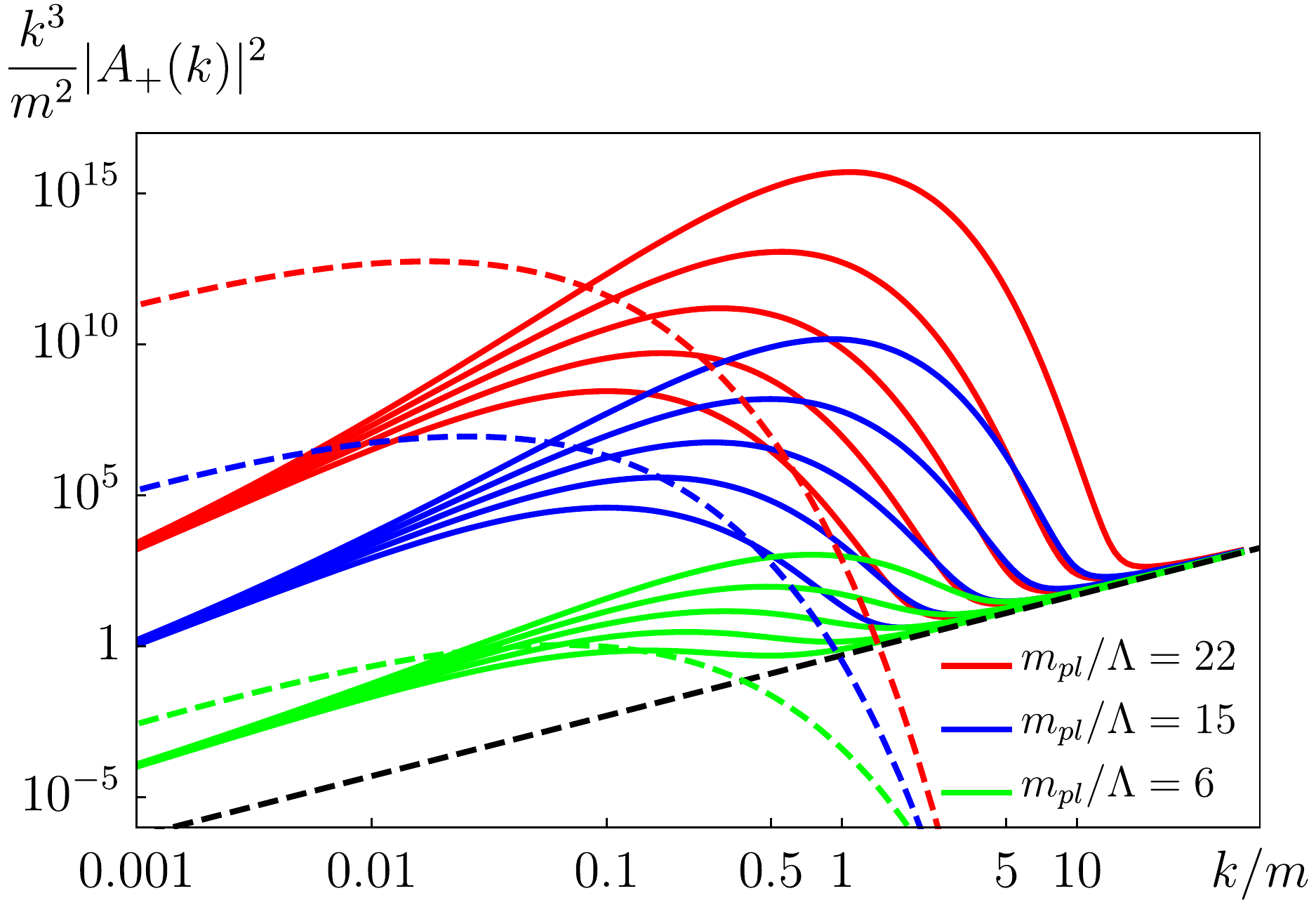}
		\caption{} \label{fig:InitialSpectrasubplot}
	\end{subfigure}
	\caption{Panel (a): The evolution through inflation of the power spectrum amplitude of the backreaction-less solution for two modes, $k/m = 2.58\cdot 10^{-5}$ (lower curve) and $k/m = 9.50\cdot 10^{-3}$ (upper curve). Panel (b): The gauge field power spectrum for different couplings at $N = -2.0, -1.5, -1.0, -0.5,~ 0.0$ efolds before the end of inflation, built from the backreaction-less solution. The dashed lines represent the analytical solution Eq.(\ref{eq:approx_A+_mode}) at $-N_i = 2$ efolds before inflation ends. }  \label{fig:InitialSpectra}
\end{figure}

In Fig.~\ref{fig:ModeEvolution} we show the numerical evolution in time of the gauge field power spectrum ${k^3\over m^2}|A_+(k,t)|^2$ for two modes, $k/m = 2.58\cdot 10^{-5}$ (lower curve) and $k/m = 9.50\cdot 10^{-3}$ (upper curve). We decompose each curve in real and imaginary contributions so one can see the mode oscillations deep inside the Hubble scale. As the physical wavelength approaches the Hubble radius, the field mode amplitude becomes excited, growing during few efolds. As it becomes more and more super-Hubble, its amplitude asymptotically freezes to a constant value, as expected from an interaction $\propto k\dot\phi$. In Fig.~\ref{fig:InitialSpectrasubplot} we illustrate the evolution in time of gauge field power spectra for different values of the coupling.  As expected, the larger the axionic coupling $\Lambda^{-1}$, the more excited the gauge field is. Let us remark that these spectra, built from the mode evolution displayed in the panel (a), are obtained by numerically evolving Eq.~(\ref{eq:AplusMathematica}) with initial conditions~(\ref{eq:InitialCondition_A}), (\ref{eq:InitialCondition_dA}), considering that the inflationary background dynamics are entirely due to the homogeneous axion, according to Eqs.~(\ref{eq:EOMinflatonHomogeneous}). In other words, numerical solutions shown in Fig.~\ref{fig:InitialSpectra} ignore the possible backreaction of the gauge field itself in the inflationary dynamics. We will refer from now on to this approximation as the ``backreaction-less'' solution, which differs more and more from the truly physical solution the closer we get to the end of inflation, or the larger the strength of the coupling $\Lambda^{-1}$. This numerical backreaction-less solution is however very useful for two purposes: $a)$ it will serve as the initial condition on the lattice before backreaction becomes relevant, and $b)$ it shows clearly that the analytical solution fails noticeably towards the end of inflation (see Fig.~\ref{fig:InitialSpectrasubplot}), even if the gauge field backreaction is still negligible (see dicussion in Sect.~\ref{subsec:InitialCondition}).

%%%%%%%%%%%%%%%%%%%%%%%%%%%%%%%%%%%%%%%%%%%%%%%%%%%%%%%%%%%%%%%%%%%%%%
	
	\section{Lattice formulation of axion inflation with $\phi\,\FFdual$ coupling}
	\label{sec:lattice_approach}
	
	In this section we present a lattice formulation of an axionic-coupling $\phi F\tilde F$ for a $U(1)$ gauge group based on~\cite{Figueroa:2017qmv}, but adapted to self-consistent expansion of the Universe. Armed with this formulation, we will derive the equivalent lattice equations to simulate numerically the field dynamics described by Eqs.~(\ref{eq:axionEOM_vec_C_curved})-(\ref{eq:gaugeEOM_vec_C_curved}), together with Eq.~(\ref{eq:friedmann2_C}) describing the expansion of the Universe sourced by (pseudo-)scalar and gauge fields as in Eqs.~(\ref{eq:energy_density_C})-(\ref{eq:pressure_density_C}), while preserving the constraint equations on the dynamics Eq.~(\ref{eq:gausslaw_vec_C_curved}), and on the expansion rate Eq.~(\ref{eq:friedmann1_C}). As the resulting lattice equations of motion are more complicated than usual and require implicit solution methods, we also present our methodology to solve them.
	
	\subsection{Lattice action in flat space-time}
	
	We will start by considering the interaction $\phi\FFdual$ in flat space-time, so we can set up notation and introduce the proper problems arising in the lattice formulation of such term, postponing for later other problems due to the inclusion of the expansion of the Universe in the lattice. We label each lattice point as $n = (t, \vec{n}) = (t, n_1, n_2, n_3)$, the lattice spacing and time step as $\Delta x$ and $\Delta t$, and indicate a displacement by a lattice unit in direction $\hat{\mu}$ (spatial or temporal) as $n + \hat{\mu}$, or simply $+\hat{\mu}$.  To avoid confusion, we do not consider summation over repeated indices in lattice variables. We use a non-compact formulation of lattice gauge theory, where gauge fields live on the links between two neighbouring sites, $A_{\mu} \equiv A_{\mu}(n + \frac{1}{2}\hat{\mu})$.  We define the lattice versions of electric and magnetic fields as%, and axion conjugate momenta, as
	\begin{eqnarray}
	E_i \equiv (\Delta_0^+A_i - \Delta_i^+A_0)\vertbar_{l = n + \frac{\hat{i}}{2} + \frac{\hat{0}}{2}}\,, ~~~~B_i \equiv \sum_{j,k}\epsilon_{ijk}\Delta_j^+A_k\vertbar_{l = n + \frac{\hat{i}}{2} + \frac{\hat{j}}{2}}\,,
	%~~~~\pi_{\phi} = \Delta_0^-\phi\vertbar_{l = n + \frac{\hat{i}}{2} + \frac{\hat{0}}{2}}\,,
	\end{eqnarray}
	and, for later convenience, the more elaborated versions
	\begin{eqnarray}
	\Etwo_i \equiv \frac{1}{2}(E_i + E_{i, -i})\vertbar_{l = n + \frac{\hat{0}}{2}}\,,~~~~ %\text{  ,  } \label{eq:E2_definition} \\
	\Bfour_i \equiv \frac{1}{4}(B_i + B_{i, -j} +  B_{i, -k} + B_{i, -j -k})\vertbar_{l = n} \text{  ,  } \label{eq:B4_definition} 
	\end{eqnarray}    
	where in the right hand side we specify the natural site where each field naturally lives.
	
	In flat space-time  the action of our problem [c.f.~Eq.~(\ref{eq:actionVEC_C_curved}) with $a = 1$] reads
	\begin{eqnarray}
	S_{m} &=& \int d^4x \left( \frac{1}{2}\pi_{\phi}^2 - \frac{1}{2}(\vec{\nabla}\phi)^2 - \frac{1}{2}m^2\phi^2 + \frac{1}{2}(\vec{E}^2 - \vec{B}^2) + \frac{\phi}{\Lambda}\vec{E}\cdot\vec{B}  \right)\text{. } \label{eq:actionVEC_C_flat}
	\end{eqnarray}
	Following~\cite{Figueroa:2017qmv}, an appropriate lattice action mimicking Eq.~(\ref{eq:actionVEC_C_flat}) is 
	\begin{eqnarray}
	S_L %&=& S_A + S_{ED} + S_{A-C} \nonumber \\
	&=& \Delta t \Delta x^3 \sum_{t, \vec{n}} {\Big \{ } ~~\frac{1}{2}\left( \Delta_0^- \phi\right)^2 - \frac{1}{2}\sum_i\left( \Delta_i^+ \phi\right)^2 - \frac{1}{2}m^2 \phi^2 + \frac{1}{2}\sum_i \left(\Delta_0^+A_i - \Delta_i^+A_0\right)^2 \nonumber \\
	& & \hspace{2.1cm}  - \frac{1}{4}\sum_{i,j} \left(\Delta_i^+ A_j - \Delta_j^+ A_i\right) + \frac{\phi}{\Lambda}\sum_i \frac{1}{2}E_i^{(2)}\left(B_i^{(4)} + B_{i,+\hat{0}}^{(4)}\right) {\Big \} }\text{  ,  } \label{eq:action_flat_L}	
	\end{eqnarray}
	where $\Delta_{\mu}^{\pm}\varphi \equiv \frac{\pm 1}{dx}\left(\varphi_{\pm \mu} -  \varphi\right)$ are standard forward and backward lattice derivatives. We refer the interested reader to~\cite{Figueroa:2017qmv} for details on the derivation of this action. Here we simply repeat its virtues. In particular, Eq.~(\ref{eq:action_flat_L}) ensures that:
	\begin{enumerate}
		\item Gauge invariance is manifest with the $U(1)$ lattice  transformation $A_{\mu} \rightarrow A_{\mu}+ \Delta_{\mu}^+ \alpha$, where $\alpha(n)$ an arbitrary real function. 
		
		\item All terms reproduce the continuum limit to order $\mathcal{O}(dx^2)$. Variation of the action will therefore lead to lattice equations of motion that reproduce their continuum counterparts also to order $\mathcal{O}(dx^2)$, see below Eqs.~(\ref{eq:axion_EOM_flat_L})-(\ref{eq:GL_EOM_flat_L}).
		
		\item Variation with respect to the gauge field leads to $\sum_{j,k}\epsilon_{ijk}(\Delta^+_j + \Delta^-_j)(E_k^{(4)}+E_{k,+k}^{(4)}) = (\Delta_o^+ +\Delta_o^-)(B_i^{(4)}+B_{i,+i}^{(4)})$ and $\sum_i\Delta_i^-(B_i^{(4)}+B_{i,+i}^{(4)}) = 0$, i.e.~vanishing Bianchi identities on the lattice. 
		
		\item The topological term $ \mathcal{K_L} \equiv (\FFdual)_L \equiv {4}\sum_i {1\over2}E_i^{(2)}(B_i^{(4)}+B_{i,+0}^{(4)})$ admits a discrete total derivative representation $\mathcal{K}_L = \sum_{\mu}\Delta_\mu^+ K_L^\mu$, with $K_L^0 = - K_0^L \equiv {2}\sum_i A_{i}^{(2)}B_{i}^{(4)}$,  $K^i_L = K_i^L \equiv -\sum_{j,k}\epsilon_{ijk}\left(E_{j}^{(2)}A_{k,-i}^{(2)}+E_{j,-i}^{(2)}A_{k}^{(2)}\right)$.
	\end{enumerate} 
	
	The equations of motion describing the real time dynamics on a lattice follow from variation of action Eq.~(\ref{eq:action_flat_L}). Fixing the gauge to $A_0 = 0$, we obtain~\cite{Figueroa:2017qmv}
	\begin{eqnarray}
	\Delta_0^+ \pi_{\phi} &=& \sum_i \Delta_i^- \Delta_i^+ \phi_{+\frac{\hat{0}}{2}} -  m^2 \phi_{+\frac{\hat{0}}{2}} + \frac{1}{\Lambda}\sum_i \frac{1}{2}E^{(2)}_{i,+\frac{\hat{0}}{2}}\left(B_i^{(4)} + B_{i,+\hat{0}}^{(4)}\right) \label{eq:axion_EOM_flat_L} \\
	\Delta_0^- E_{i,+\frac{\hat{0}}{2}} &=& -\sum_{j,k}\epsilon_{ijk}\Delta_j^-B_k - \frac{1}{2\Lambda}\left(\pi_{\phi}\Bfour_i + \pi_{\phi,+i}\Bfour_{i,+i}\right) \nonumber \\
	& & \hspace{-1.7cm} + \frac{1}{8\Lambda}(2 + dx\Delta_i^+)\sum_{\pm}\sum_{j,k} \epsilon_{ijk} \left\lbrace [ (\Delta_j^{\pm}\phi)\Etwo_{k,\pm j} ]_{+\frac{\hat{0}}{2}} + [(\Delta_j^{\pm} \phi)\Etwo_{k,\pm j}]_{-\frac{\hat{0}}{2}} \right\rbrace \label{eq:gauge_EOM_flat_L} \\ 
	\sum_i \Delta_i^-E_{i,+\frac{\hat{0}}{2}} &=& - \frac{1}{4\Lambda}\sum_{\pm}\sum_i \Delta_i^{\pm} \phi_{+\frac{\hat{0}}{2}}\left(\Bfour_i + \Bfour_{i,+\hat{0}}\right)_{\pm i} \text{  ,  } \label{eq:GL_EOM_flat_L} 
	\end{eqnarray}
	where the electric field $E_{i,+\frac{\hat{0}}{2}} \equiv \Delta_0^+ A_i$ lives at semi-integer time steps,  whereas the axion conjugate field $\pi_{\phi} \equiv \Delta_0^- \phi$ lives at integer times, as the axion field itself $\phi_{+\frac{\hat{0}}{2}}$ lives at semi-integer times, due to its pseudo-scalar nature. Eq.~(\ref{eq:GL_EOM_flat_L}) represents the Gauss constraint of the system. It can be shown that the set of Eqs.~(\ref{eq:axion_EOM_flat_L})-(\ref{eq:GL_EOM_flat_L}) reproduces correctly the continuum equations of motion to order $\mathcal{O}(dx^2)$. 
	
	\subsubsection{Implicit (iterative) solution for the electric fields} \label{subsec:Explicit_approximation_and_iterative_method}
	
	Standard lattice equations of a gauge system, e.g.~an Abelian-Higgs model, typically admit an explicit scheme for their solution: field amplitudes at a given time can be obtained from the corresponding values at the previous time step,~i.e
	\begin{eqnarray}
	\phi_{+\frac{\hat{0}}{2}} = \phi_{-\frac{\hat{0}}{2}} + \dt \pi_{\phi}\,,~~~~
	A_{i,+\hat{0}} &=& A_i + \dt E_{i,+\frac{\hat{0}}{2}}  \text{  ,  }
	\end{eqnarray} 
	and similarly the evolution of conjugate momenta can be obtained from previous ones as
	\begin{eqnarray}
	\pi_{\phi,+\hat{0}} = \pi_{\phi} + \dt \mathcal{Q}_{\frac{\hat{0}}{2}} \,,~~~~
	E_{i,+\frac{\hat{0}}{2}} = E_{i,-\frac{\hat{0}}{2}} + \dt \mathcal{P}_i \text{  ,  } 
	\end{eqnarray} 
	with $\mathcal{Q}_{+{\hat{0}}/{2}}$ and $\mathcal{P}_i$ characteristic kernels of the system of equations. This simple leap-frog scheme assumes that $\mathcal{Q}_{+{\hat{0}}/{2}}$ and $\mathcal{P}_i$ are functions only of the field amplitudes. The set of Eqs.~(\ref{eq:axion_EOM_flat_L})-(\ref{eq:GL_EOM_flat_L}) does not admit however such explicit scheme of evolution (unless the axion is forced to remain homogeneous). Inspecting Eq.~(\ref{eq:gauge_EOM_flat_L}), we observe that on the right hand side there are terms containing $E_{k,\pm j + {\hat{0}}/{2}}$. This means that to solve for the electric field $E_{i, +{\hat{0}}/{2}}$ on the left hand side, one needs to know  the neighbouring electric fields $E_{k, \pm j +{\hat{0}}/{2} }$ at the same evolved time. This makes it impossible to solve explicitly for electric fields. %The problem obviously does not arise in the case of a homogeneous axion. Even though in general we do not expect the gradients of the axion to be very large (say the axion gradient energy is typically subdominant versus the kinetic or potential energy terms), since the problematic term contains spatial derivatives of the axion field.
	
	A potential solution to this problem is to obtain an explicit scheme by using an approximated version of the lattice equations of motion. Making the following approximation on the right hand side of Eq.~(\ref{eq:gauge_EOM_flat_L}),
	\be
	\left\lbrace [ (\Delta_j^{\pm}\phi)\Etwo_{k,\pm j} ]_{+\frac{\hat{0}}{2}} + [(\Delta_j^{\pm} \phi)\Etwo_{k,\pm j}]_{-\frac{\hat{0}}{2}} \right\rbrace \approx 2[(\Delta_j^{\pm} \phi)\Etwo_{k,\pm j}]_{-\frac{\hat{0}}{2}}\text{  ,  } \label{eq:explicit_approx_flat_L}
	\ee 
	leads to an explicit, but approximated, equation of motion for the electric fields
\begin{eqnarray}
&& \hspace*{1cm}	E_{i,+\frac{\hat{0}}{2}}{\big |}_1 =  E_{i,-\frac{\hat{0}}{2}} + \dt \mathcal{C}_{i} + \frac{\dt}{8\Lambda}(2 + dx\Delta_i^+)\sum_{\pm}\sum_{j,k}\epsilon_{ijk}[(\Delta_j^{\pm} \phi)\Etwo_{k,\pm j}]_{-\frac{\hat{0}}{2}} \label{eq:E_1_EOM_flat_L}\,,\\
	%\end{eqnarray}
	%where
	%\begin{eqnarray}\vspace*{-0.2cm}
&& \hspace*{-0.5cm} {\rm where}\nonumber\\	
&&	\mathcal{C}_{i} \equiv -\sum_{j,k}\epsilon_{ijk}\Delta_j^-B_k  - \frac{\left(\pi_{\phi}\Bfour_i + \pi_{\phi,+i}\Bfour_{i,+i}\right)}{2\Lambda} + \frac{(2 + dx\Delta_i^+)}{8\Lambda}\sum_{\pm}\sum_{j,k} \epsilon_{ijk}[(\Delta_j^{\pm} \phi)\Etwo_{k,\pm j}]_{-\frac{\hat{0}}{2}}\,.\nonumber\\
	\end{eqnarray}
	The error caused in Eq.~(\ref{eq:E_1_EOM_flat_L}) is of order $\mathcal{O}(\dt)$. Therefore, the set of coupled equations Eqs.~(\ref{eq:axion_EOM_flat_L}), (\ref{eq:E_1_EOM_flat_L}) will only reproduce the continuum limit of the dynamics up to order $\mathcal{O}(dx_\mu)$, instead of $\mathcal{O}(dx_\mu^2)$. Furthermore, as the left and right hand sides of Eq.~(\ref{eq:GL_EOM_flat_L}) are not well balanced anymore, Eq.~(\ref{eq:E_1_EOM_flat_L}) leads into a violation of the Gauss law (which accumulates in time), and hence the whole scheme may exhibit inaccuracies in the numerical solution, which are out of control. Solving the dynamics with Eq.~(\ref{eq:E_1_EOM_flat_L}) instead of Eq.~(\ref{eq:gauge_EOM_flat_L}), does not represent therefore an appealing solution.
	
	In order to preserve an order $\mathcal{O}(dx^2)$ in the accuracy of the equations of motion, and to verify the Gauss law to machine precision, \cite{Figueroa:2017qmv}  proposed an implicit method for solving the electric fields iteratively. In practice the solution $E_{i, +\frac{\hat{0}}{2}}{\big |}_1$ of Eq.~(\ref{eq:E_1_EOM_flat_L}) is simply viewed as an intermediate value of the updated electric fields at each lattice site. Then, by inserting this solution into the original Eq.~(\ref{eq:gauge_EOM_flat_L}), we obtain a new and more precise value of it, which will be called $E_{i,+\frac{\hat{0}}{2}}{\big |}_2$. Then, in a following iteration $E_{i,+\frac{\hat{0}}{2}}{\big |}_2$ acts as an intermediate solution to compute $E_{i,+\frac{\hat{0}}{2}}{\big |}_3$, and so on. The iteration can be done $n$ times, obtaining a solution 
	\begin{eqnarray}
	E_{i, +\frac{\hat{0}}{2}}{\big |}_n = E_{i,-\frac{\hat{0}}{2}} + \dt \mathcal{C}_{i}  +  \frac{\dt}{8\Lambda}(2 + dx\Delta_i^+)\sum_{\pm}\sum_{j,k} \epsilon_{ijk}  (\Delta_j^{\pm} \phi)_{\frac{\hat{0}}{2}}\Etwo_{k,\pm \hat j+\frac{\hat{0}}{2}}{\Big |}_{n-1} \,,\label{eq:E_n_EOM_flat_L}
	%&=& E_{i,-\frac{\hat{0}}{2}} + \dt \left( -\sum_{j,k}\epsilon_{ijk}\Delta_j^-B_k - \frac{1}{2\Lambda}\left(\pi_{\phi}\Bfour_i + \pi_{\phi,+i}\Bfour_{i,+i}\right) \right) \nonumber \\
	%& & \hspace{-2.5cm}+ \dt \left(\frac{1}{8\Lambda}(2 + dx\Delta_i^+)\sum_{\pm}\sum_{j,k} \epsilon_{ijk} \left\lbrace [(\Delta_j^{\pm} \phi)\Etwo_{k,\pm j}]_{-\frac{\hat{0}}{2}}  + [(\Delta_j^{\pm} \phi)\Etwo_{k,\pm j}{\big |}_{n-1}]_{+\frac{\hat{0}}{2}} \right\rbrace \right) \text{ ,  }\label{eq:E_n_EOM_flat_L} 
	\end{eqnarray}
	where $E_{i,+\frac{\hat{0}}{2}}{\big |}_0 \equiv E_{i,-\frac{\hat{0}}{2}}$. In the limit $n \rightarrow \infty$ this solution approaches the exact solution $E_{i, +\frac{\hat{0}}{2}}$ to Eq.~(\ref{eq:gauge_EOM_flat_L}). To reproduce the continuum theory up to order $\mathcal{O}(dx^2)$, it should be enough to iterate two times. In practice, to verify the Gauss constraint Eq.~(\ref{eq:GL_EOM_flat_L}) to machine precision, we need a solution as close as possible to the exact solution to Eq.~(\ref{eq:gauge_EOM_flat_L}), and hence it is not clear a priori how many iterations are required for such purpose. Approximating the solution of the electric fields by $E_{i, +\frac{\hat{0}}{2}} \approx E_{i, +\frac{\hat{0}}{2}}{\big |}_n$ will lead to a degree of violation of the Gauss constraint, the smaller the larger the number of iterations $n$. We can therefore exploit this fact and determine (by trial an error in a given simulation) the number of iterations needed to verify the Gauss law to certain degree of violation. In practice, we observe that $n = 10$ iterations is typically enough to verify the Gauss law with machine precision, as presented in Sect.~\ref{sec:constraints}.
	% We present our tests on this in Sect.~\ref{sec:InitialCondition}, showing that preserving the Gauss law to a certain degree of accuracy is crucial for the real time evolution, since its conservation is essential to ensure the physical stability of the system.
	
	\subsection{Adding expansion of the Universe in the lattice}
	
	Having laid the foundations of an interaction $\phi \FFdual$ on the lattice, let us implement now the expansion of the Universe. We build a lattice action that reproduces Eq.~(\ref{eq:action_C_curved}) in the continuum using the same axionic coupling as in Eq.~(\ref{eq:actionVEC_C_flat}), 
	\begin{eqnarray}
	S_L &=& \Delta t \Delta x^3 \sum_{t, \vec{n}} {\bigg \{ } ~~\frac{1}{2}a^3\left( \Delta_0^- \phi_{+\frac{\hat{0}}{2}}\right)^2 - \frac{1}{2}a_{+\frac{\hat{0}}{2}}\left( \Delta_i^+ \phi_{+\frac{\hat{0}}{2}}\right)^2 - \frac{1}{2}a_{+\frac{\hat{0}}{2}}^3m^2 \phi_{+\frac{\hat{0}}{2}}^2 \\
	& & \hspace*{0.5cm}+ ~\frac{1}{2}a_{+\frac{\hat{0}}{2}}\sum_i \left(\Delta_0^+A_i - \Delta_i^+A_0\right)^2 - \frac{1}{4a}\sum_{i,j} \left(\Delta_i^+ A_j - \Delta_j^+ A_i\right) \nonumber\\ 
	& & \hspace*{0.5cm} + ~\frac{\phi}{\Lambda}\sum_i \frac{1}{2}E_i^{(2)}\left(B_i^{(4)} + B_{i,+\hat{0}}^{(4)}\right) {\bigg \} }\,,\nonumber \label{eq:action_curved_L} 
	\end{eqnarray}
	where we choose that the scale factor $a_{+\frac{\hat{0}}{2}}$ lives at semi-integer times, so that at integeter times we write
	\be
	a \equiv \frac{1}{2}\left(a_{-\frac{\hat{0}}{2}}  + a_{+\frac{\hat{0}}{2}} \right) \text{  .  } \label{eq:a_integer_times}
	\ee 
	From here, we can derive the discrete equations of motion in an expanding background. Fixing the gauge to  $A_0 = 0$, we obtain
	\begin{eqnarray}
	\Delta_0^+\left(a^3\pi_{\phi}\right) &=& a_{+\frac{\hat{0}}{2}}\sum_i \Delta_i^- \Delta_i^+ \phi_{+\frac{\hat{0}}{2}} -  a^3_{+\frac{\hat{0}}{2}}m^2 \phi_{+\frac{\hat{0}}{2}}   \nonumber\\
	& & + \frac{1}{\Lambda}\sum_i \frac{1}{2}E^{(2)}_{i,+\frac{\hat{0}}{2}}\left(B_i^{(4)} + B_{i,+\hat{0}}^{(4)}\right)\,, \label{eq:axion_EOM_curved_L} \\
	%%%%%%%%%%%
	\Delta_0^-\left(a_{+\frac{\hat{0}}{2}} E_{i,+\frac{\hat{0}}{2}}\right) &=& -\frac{1}{a}\sum_{j,k}\epsilon_{ijk}\Delta_j^-B_k - \frac{1}{2\Lambda}\left(\pi_{\phi}\Bfour_i + \pi_{\phi,+i}\Bfour_{i,+i}\right) \nonumber \\
	& & \hspace{-2cm} + \frac{1}{8\Lambda}(2 + dx\Delta_i^+)\sum_{\pm}\sum_{j,k}\left\lbrace \epsilon_{ijk}[ (\Delta_j^{\pm}\phi)\Etwo_{k,\pm j} ]_{+\frac{\hat{0}}{2}} + [(\Delta_j^{\pm} \phi)\Etwo_{k,\pm j}]_{-\frac{\hat{0}}{2}} \right\rbrace\,,  \label{eq:gauge_EOM_curved_L}\\ 
	%%%%%%%%%%%
	a_{+\frac{\hat{0}}{2}}\sum_i \Delta_i^-E_{i,+\frac{\hat{0}}{2}} &=& - \frac{1}{4\Lambda}\sum_{\pm}\sum_i \left(\Delta_i^{\pm} \phi_{+\frac{\hat{0}}{2}}\right)\left(\Bfour_i + \Bfour_{i,+\hat{0}}\right)_{\pm i}\text{  ,  } \label{eq:GL_EOM_curved_L} 
	\end{eqnarray} 
	where Eq.~(\ref{eq:GL_EOM_curved_L}) represents the Gauss law, and the conjugate momenta are 
	\begin{eqnarray}
	\pi_{\phi} \equiv \Delta_0^- \phi_{+\frac{\hat{0}}{2}}\,,~~~~~ %\label{eq:pi_EOM_curved_L} \\
	%%%%%%%%%%%
	E_{i,+\frac{\hat{0}}{2}} \equiv \Delta_0^+ A_i\,. \label{eq:E_EOM_curved_L}
	\end{eqnarray}
	By expanding each term  in Eqs.~(\ref{eq:axion_EOM_curved_L})-(\ref{eq:GL_EOM_curved_L}) around their natural site, it can be shown that this set of equations reproduce to order $\mathcal{O}(dx_\mu^2)$ the equations describing the continuum dynamics, Eqs.~(\ref{eq:axionEOM_vec_C_curved})-(\ref{eq:gausslaw_vec_C_curved}).
	
	Let us consider now the expansion of the Universe. In light of the continuum equations Eqs.~(\ref{eq:friedmann2_C}),~(\ref{eq:friedmann1_C}), and given our choice of a scalar factor living at semi-integer times, lattice counterpart of Friedmann equations can then be written as
	\begin{eqnarray}
	\left({\Delta_0^+a_{-\hat 0/2}}\right)^2 &=& \frac{a^2}{3m_{pl}^2}\rho_{L}%\left(\bar{H}^{kin} + \frac{1}{a^2}\bar{H}^{grad} + \bar{H}^{pot} + \frac{1}{a^2}\bar{H}^{E} + \frac{1}{a^4}\bar{H}^{B}\right) \text{  ,  } 
	\,,\label{eq:Hubble_constraint_L}\\
	{\Delta_0^-\Delta_0^+a_{+\hat 0/2}} &=& - \frac{a_{+\hat 0/2}}{6m_{pl}^2}%\left(4\bar{H}^{kin} - 2\bar{H}^{pot} + 2(\frac{1}{a^2}\bar{H}^{E} + \frac{1}{a^4}\bar{H}^B)\right) \text{  ,  } 
	(\rho_{L}+3p_{L})_{+\hat 0/2}~\,,
	\label{eq:friedmann_L}
	\end{eqnarray}
	where Eq.~(\ref{eq:Hubble_constraint_L}) is the lattice version of the Hubble constraint.
	From Eqs.~(\ref{eq:energy_density_C}), (\ref{eq:pressure_density_C}) we can identify the mean value (volume average) of the energy and pressure densities in the lattice as
	\begin{eqnarray}
	&&\rho_{L} =  \bar{H}^{kin}
	+ \frac{1}{a^2}{1\over2}(\bar{H}^{grad}_{-\hat 0/2}+\bar{H}^{grad}_{+\hat 0/2}) + {1\over2}(\bar{H}^{pot}_{-\hat 0/2}+\bar{H}^{pot}_{+\hat 0/2}) + \frac{1}{a^2}{1\over 2}(\bar{H}^{E}_{-\hat 0/2} + \bar{H}^{E}_{+\hat 0/2}) + \frac{1}{a^4}\bar{H}^{B}\text{  ,  } \label{eq:energy_density_L} \nonumber\\
	&&(\rho_{L}+3p_{L})_{+\hat0/2} = %4\times{1\over2}
	2(\bar{H}^{kin}+\bar{H}^{kin}_{+\hat 0})  - 2\bar{H}^{pot}_{+\hat 0/2} + %2\times
	\frac{2}{a_{+{\hat{0}}/{2}}^{2}}\bar{H}^{E} + %2\times
	\frac{1}{a_{+{\hat{0}}/{2}}^{4}}%{1\over2}
	(\bar{H}^{B}+\bar{H}^{B}_{+\hat 0})\,,\label{eq:pressure_density_L} 
	\end{eqnarray}
	with
	\begin{eqnarray}
	&&\bar{H}^{kin} = \frac{1}{N^3}\sum_{\vec{n}}\frac{\pi_{\phi}^2}{2}\,,~~~~ \bar{H}^{grad} = \frac{1}{N^3}\sum_{\vec{n}}\frac{1}{2}\sum_i(\Delta_i^+\phi_{+\frac{\hat{0}}{2}})^2\,,~~~~
	\bar{H}^{pot} = \frac{1}{N^3}\sum_{\vec{n}}\frac{1}{2}m^2\phi_{+\frac{\hat{0}}{2}}^2\, \nonumber \\
	&&\hspace*{1cm}\bar{H}^{E} = \frac{1}{N^3}\sum_{\vec{n}}\sum_i \frac{1}{2}E_{i,+\frac{\hat{0}}{2}}^2\,,~~~~~~ 
	\bar{H}^{B} = \frac{1}{N^3}\sum_{\vec{n}}\sum_{i,j}\frac{1}{4}(\Delta_i^+A_j - \Delta_j^+A_i)^2\,,\label{eq:Hterms}
	\end{eqnarray}
	so that $\bar{H}^{kin}, \bar{H}^{grad}, \bar{H}^{pot}$ are the terms associated with the axion's kinetic, gradient and potential energy densities, and $\bar{H}^{E}, \bar{H}^{B}$ are those associated with the gauge field electric and magnetic energy densities. Note that $\bar{H}^{grad}$, $\bar{H}^{pot}$ and $\bar{H}^{E}$ naturally live at semi-integer times, whereas $\bar{H}^{kin}$ and $\bar{H}^B$ live at integer times. Thus, we choose the semi-sums above so that $lhs$ and $rhs$ in each of the Eqs.~(\ref{eq:Hubble_constraint_L}), (\ref{eq:friedmann_L}) are consistently living at the same time (given our choice of scale factor living at semi-integer times), and reproduce the continuum Eqs.~(\ref{eq:friedmann2_C}), (\ref{eq:friedmann1_C}) to order $\mathcal{O}(dx_\mu^2)$.
	
	Let us emphasize that it is only Eq.~(\ref{eq:friedmann_L}) that controls the evolution of the scalar factor; Eq.~(\ref{eq:Hubble_constraint_L}) instead, is not a dynamical equation, but rather a constraint that should be conserved by the real time evolution of the field. Its violation indicates the degree of violation of ``energy conservation'' in an expanding Universe.  %To be more precise, as much as the lattice Gauss constraint reflects how well gauge invariance is preserved, the Hubble constraint Eq.~(\ref{eq:Hubble_constraint_L}) will serve as an indicator of how well energy is 'conserved' in an expanding Universe.
	
	\subsubsection{Implicit (iterative) solution for the scale factor} \label{subsec:iterative_method_for_the_scale_factor}
	
	Eqs.~(\ref{eq:friedmann_L}), (\ref{eq:pressure_density_L}) and (\ref{eq:Hterms}) show that there is no explicit scheme for evolution of the scale factor. To see this, let us label $b$ the lattice time derivative of the scale factor $a_{+0/2}$,~i.e.
	\be
	\Delta_0^+ a_{+\frac{\hat{0}}{2}} = b_{+\hat{0}}\text{  .  } \label{eq:b_EOM_L}
	\ee
	By inspecting Eq.~(\ref{eq:axion_EOM_curved_L}) we observe that to evolve the axion's conjugate momentum
	\be
	\pi_{\phi,+\hat{0}} = \frac{1}{a_{+\hat{0}}^3}\left( a^3\pi_{\phi} + \dt \mathcal{\tilde{Q}}_{+\frac{\hat{0}}{2}}\right) \text{  ,  }
	\ee
	we need $a_{+\hat{0}} = \frac{1}{2} ( a_{+\frac{\hat{0}}{2}} + a_{+\frac{3}{2}\hat{0}} )$,  which contains $a_{+\frac{3}{2}\hat{0}}$, i.e.~an unknown quantity which has not been computed yet at that moment. We could find $a_{+\frac{3}{2}\hat{0}}$ by using Eq.~(\ref{eq:b_EOM_L}) once $b_{+\hat{0}}$ is known, but to calculate $b_{+\hat{0}}$ we need $\pi_{\phi,+\hat{0}}$. Therefore, there is no consistent scheme of evolution. 
	
	It is worth stressing that if we had chosen a scale factor living instead at integer times, an analogous problem would still appear but related with the evolution of $E_{i, + \frac{3}{2}\hat{0}}$, instead of  $\pi_{\phi, +\hat{0}}$. This can be traced back to the fact that in our system, the pseudo-scalar field is forced to live at semi-integer times due to the nature of the axionic-coupling interaction, see~\cite{Figueroa:2017qmv} for further discussion on this. %If this was not the case, i.e. we let the scale factor live at integer times, this problem would not arise and an explicit scheme becomes possible.
	
	In order to solve for the scale factor, we will then use an implicit method similar to the one used in Sect.~\ref{subsec:Explicit_approximation_and_iterative_method} to solve for the electric field. In particular, we first make the following approximation,
	\be
	a_{+\hat{0}} \equiv \frac{1}{2}\left(a_{+\frac{\hat{0}}{2}} + a_{+\frac{3}{2}\hat{0}}\right) \approx a_{+\frac{\hat{0}}{2}} \text{  ,  }\label{eq:explicit_approx_a_L}
	\ee
	from where we compute a first approximated value of the evolved conjugate momenta as 
	\be
	\pi_{\phi,+\hat{0}}{\big |}_1 = \frac{1}{a_{+\frac{\hat{0}}{2}}^3}\left( a^3\pi_{\phi} + \dt\tilde{\mathcal{Q}}_{+\frac{\hat{0}}{2}} \right)\text{  .  } \label{eq:pi_EOM_explicit_approx_L}
	\ee
	An approximate value of the parameter $b_{+\hat{0}}\vertbar_1$ then follows as
	\begin{eqnarray}
	&& b_{+\hat{0}}{\big |}_1 = b -c_{+{\hat{0}}/{2}}\,\dt - \dt\frac{a_{+{\hat{0}}/{2}}}{3m_{pl}^2}\left[\bar{H}^{kin} + \bar{H}^{kin}_{+\hat{0}}{\big |}_1\right]\,, \label{eq:b_EOM_explicit_approx_L}\\
	{\rm where}&&\hspace*{0.5cm} c_{+{\hat{0}}/{2}} \equiv \frac{a_{+{\hat{0}}/{2}}}{3m_{pl}^2}{\Bigg [}{\bar{H}_{+{\hat{0}}/{2}}^{E}\over a^{\,2}_{+{\hat{0}}/{2}}} + {\left({\bar{H}_{+\hat{0}}^B+\bar{H}^B} \right)\over 2a^{\,4}_{+{\hat{0}}/{2}}} - \bar{H}^{pot}_{+{\hat{0}}/{2}} {\Bigg ]}\,,
	\end{eqnarray}
	%where we average values of the kinetic and magnetic energy density components since the Eq.~(\ref{eq:friedmann_L}) naturally lives at semi-integer times. 
	from where we can then also compute a first approximation of the evolved scale factor
	\be
	a_{+\frac{3}{2}\hat{0}}{\big |}_1 = a_{+\frac{\hat{0}}{2}} + \dt b_{+\hat{0}}{\big |}_1 \text{  .  } \label{eq:a_EOM_explicti_approx_L}
	\ee
	An iterative scheme can then be put forward as
	\begin{eqnarray}
	\pi_{\phi,+\hat{0}}{\big |}_n &=& \frac{a^3_{+\hat{0}}{\big |}_{n-2}}{a^3_{+\hat{0}}{\big |}_{n-1}}\pi_{\phi,+\hat{0}}{\big |}_{n-1} \text{  ,  }\nonumber \\
	b_{+\hat{0}}{\big |}_n &=& b -c_{+{\hat{0}}/{2}}\,\dt  - \dt\frac{a_{+{\hat{0}}/{2}}}{3m_{pl}^2}\left[\bar{H}^{kin} +\bar{H}^{kin}_{+\hat{0}}{\big |}_{n}\right] \text{  ,  }\nonumber \\
	a_{+\frac{3}{2}\hat{0}}{\big |}_n &=& a_{+\frac{\hat{0}}{2}} + \text{dt}b_{+\hat{0}}{\big |}_n \text{  .  } \label{eq:a_iterative_scheme_L}
	\end{eqnarray}
	For a sufficiently large number of iterations $n$, we expect to reproduce the continuum evolution up to order $\mathcal{O}(dx^2)$, ensuring a sufficiently good accuracy in the conservation of the Hubble constraint Eq.~(\ref{eq:Hubble_constraint_L}).
	
\subsection{Lattice power spectrum of fields}

Let us discuss here briefly the lattice formulation of the power spectrum of a generic field $f$, with ensemble average $\langle f^2 \rangle$ in the continuum defined as 
\be
		\langle f^2 \rangle = \int d\log k~\mathcal{P}_f(k)~~, ~~~ \langle f_k f_{k^{\prime}} \rangle = (2\pi)^3 {2\pi^2\over k^3}\mathcal{P}_f(k) \delta (\mathbf{k}-\mathbf{k^{\prime}})~. \label{eq:continuumPS}
\ee 
In the lattice the ensemble average is substituted with a volume average, 
\be
		\langle f^2 \rangle_V = \frac{dx^3}{V}\sum_{n} f^2(n)~\,.
\ee
Using the following convention for the discrete Fourier transform, 
\be
		f(n) \equiv {1\over N^3}\sum_{\tilde{n}}e^{-i{2\pi \over N} \tilde{n}n}\tilde{f}(\tilde{n})~\,, ~~~ \tilde{f}(\tilde{n}) \equiv \sum_n e^{+i{2\pi \over N}\tilde{f}(\tilde{n})}f(n)~\,,
\ee
one obtains 
\be
		\langle f^2 \rangle_V = \frac{1}{2\pi^2}\sum_{|\tilde{n}|}\Delta\log k(\tilde{n}) ~k^3(\tilde{n})\left(\frac{dx}{N}\right)^3 \big\langle \big|\tilde{f}(\tilde{n})\big|^2\big\rangle_{R(\tilde{n})}~\,,
\ee
where $\langle ( ... ) \rangle \equiv \frac{1}{4\pi|\tilde{n}|^2}\sum_{\tilde{n}^{\prime}\in R(\tilde{n})}( ... )$ is an angular average over the spherical shell of radius $\tilde{n}^{\prime}\in [|\tilde{n}|,|\tilde{n}+ \Delta\tilde{n}|]$, with $\Delta \tilde{n}$ a given radial binning, and we used the following lattice definitions: $\Delta \log k(\tilde{n}) \equiv \frac{k_{IR}}{k(\tilde{n})}\,,~ \mathbf{k}(\tilde{n}) \equiv k_{IR}\tilde{n}$ and $ k_{IR} \equiv \frac{2\pi}{L}$.
By identification with Eq.~(\ref{eq:continuumPS}), we conclude that
	\be
		 \mathcal{P}_f(k) \equiv \frac{k^3(\tilde n)}{2\pi^2}\left(\frac{dx}{N}\right)^3 \big\langle \big|\tilde{f}(\tilde{n})\big|^2\big\rangle_{R(\tilde{n})}\,.
	\ee

The lattice spectra of the axion, gauge field amplitude and its associated electric and magnetic fields, are given by
\begin{eqnarray}
\mathcal{P}_\phi(k(\tilde n)) &=& \frac{k^3(\tilde n)}{2\pi^2}\left(\frac{dx}{N}\right)^3 \big\langle \big|\tilde{\phi}(\tilde{n})\big|^2\big\rangle_{R(\tilde{n})}~\,,  \\
\mathcal{P}_A(k(\tilde n)) &=& \frac{k^3(\tilde n)}{2\pi^2}\left(\frac{dx}{N}\right)^3 \sum_i\big\langle \big|\tilde{A}_i(\tilde{n})\big|^2\big\rangle_{R(\tilde{n})}~\,,  \\
\mathcal{P}_E(k(\tilde n)) &=& \frac{k^3(\tilde n)}{2\pi^2}\left(\frac{dx}{N}\right)^3 \sum_i\big\langle \big|\dot{\tilde{A}}_i(\tilde{n})\big|^2\big\rangle_{R(\tilde{n})}~\,, \\
\mathcal{P}_B(k(\tilde n)) &=& \frac{k^5(\tilde n)}{2\pi^2}\left(\frac{dx}{N}\right)^3 \sum_i \big\langle \big|\tilde{A}_i(\tilde{n})\big|^2\big\rangle_{R(\tilde{n})}~.
\end{eqnarray}

Given our choice of quadratic potential $V(\phi) = {1\over2}m^2\phi^2$, we note that we measure all observables in units of $m$, as this is a natural choice in our problem, given that $a)$ the mass of the axion is the only dimensionful scale in the problem while the gauge field backreaction remains negligible, and $b)$ it will determine the frequency of oscillations of the axion field around the minimum of its potential during the preheating stage. Hence we measure e.g.~the gauge field amplitude or its energy in the corresponding powers of $m$, $\Large[\Large\langle{\vec A}^{\,2} \Large\rangle \Large] = m^2$, $\Large[\Large\langle{\vec E}^2 \Large\rangle  \Large] =\Large[\Large\langle{\vec B}^2 \Large\rangle \Large] = m^4$. Correspondingly their logarithmic spectra have identical dimensions, $\Large[\mathcal{P}_A\Large] = m^2$, $\Large[\mathcal{P}_E\Large] = \Large[\mathcal{P}_B\Large] = m^4$, etc.

\section{Dynamics on the lattice}
\label{sec:LatticeApplication}

In this section we study the evolution of the fields during the last efolds of inflation (Sect.~\ref{subsec:Inflation}), and during the preheating stage that follows afterwards (Sect.~\ref{subsec:Preheating}).

\subsection{Part I. Inflation (last efolds)}
\label{subsec:Inflation}	

Here we concentrate first in the study of the last efolds of inflation. As long as excitation of the gauge field remains ``small'' -- we will quantify what this means in a precise manner --, %, e.g.~the energy density in the gauge field is much subdominant compared to the energy density of the axion 
we can ignore the backreaction of the gauge field in the inflationary dynamics and safely use the initial conditions described in Sect.~\ref{subsec:BackreactionLess}. Whereas this should represent a safe procedure deep inside inflation, it is expected to become less and less accurate the closer we are to the end of inflation, as the excitation of the gauge field is larger and the axion may also develop some degree of inhomogeneity. However, due to the natural limitations of using a finite volume, we cannot start our simulation too many efolds before the end of inflation, say $- N \gg 1$. In fact, the initial modes captured in the lattice would be red-shifted too fast out of the box and we would miss the successive relevant ones within our limited dynamical range. It is clear that we can only aim to start our simulations some $- N \sim O(1)$ number of efolds before the end of inflation. 

\subsubsection{Initial conditions}
\label{subsec:InitialCondition}

\begin{figure} % "[t!]" placement specifier just for this example
	\begin{subfigure}{0.49\textwidth}
		\includegraphics[width=\linewidth]{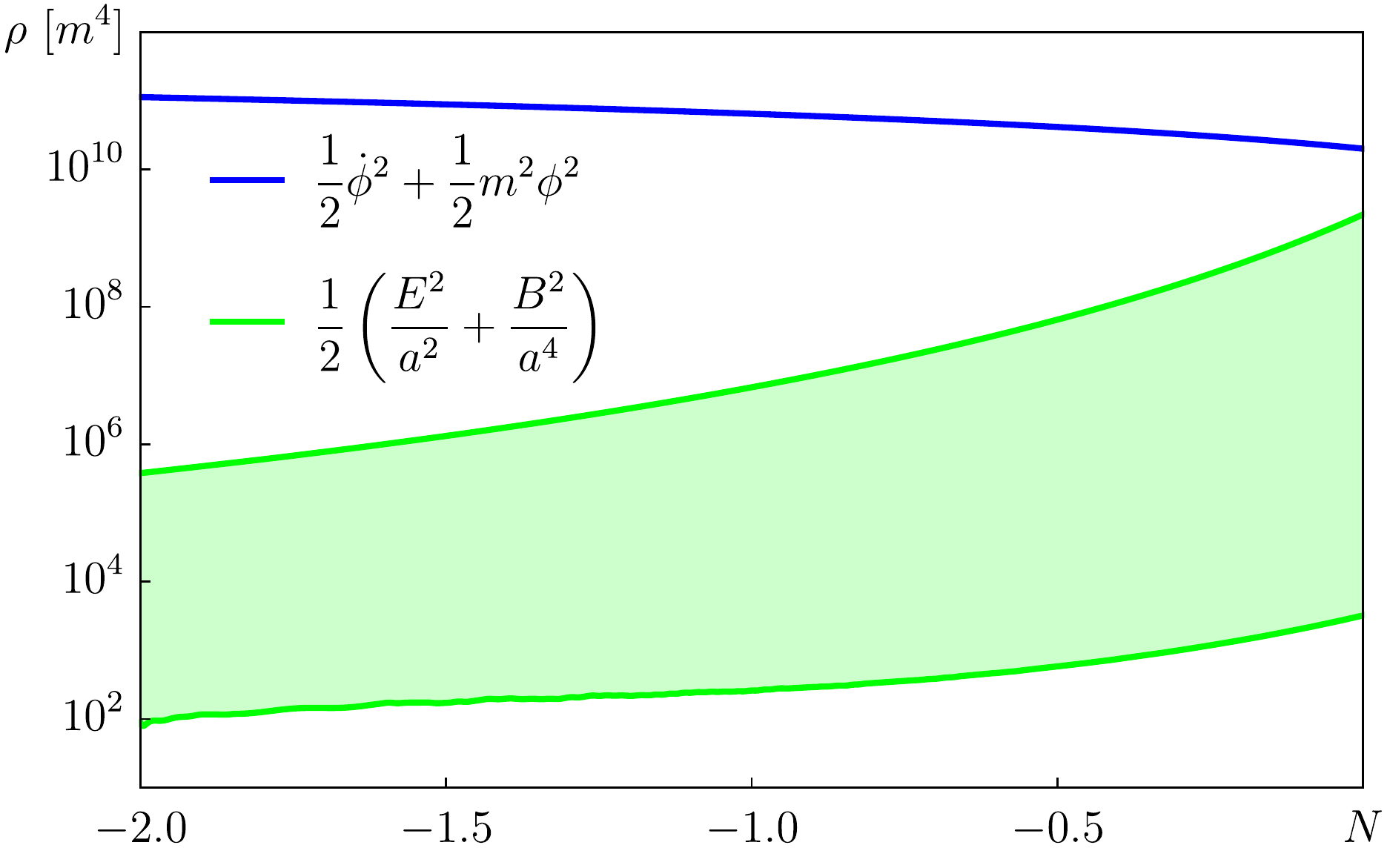}
		\caption{} \label{fig:AnalysisIC_Energy}
	\end{subfigure}\hspace*{\fill}
	\begin{subfigure}{0.49\textwidth}
		\includegraphics[width=\linewidth]{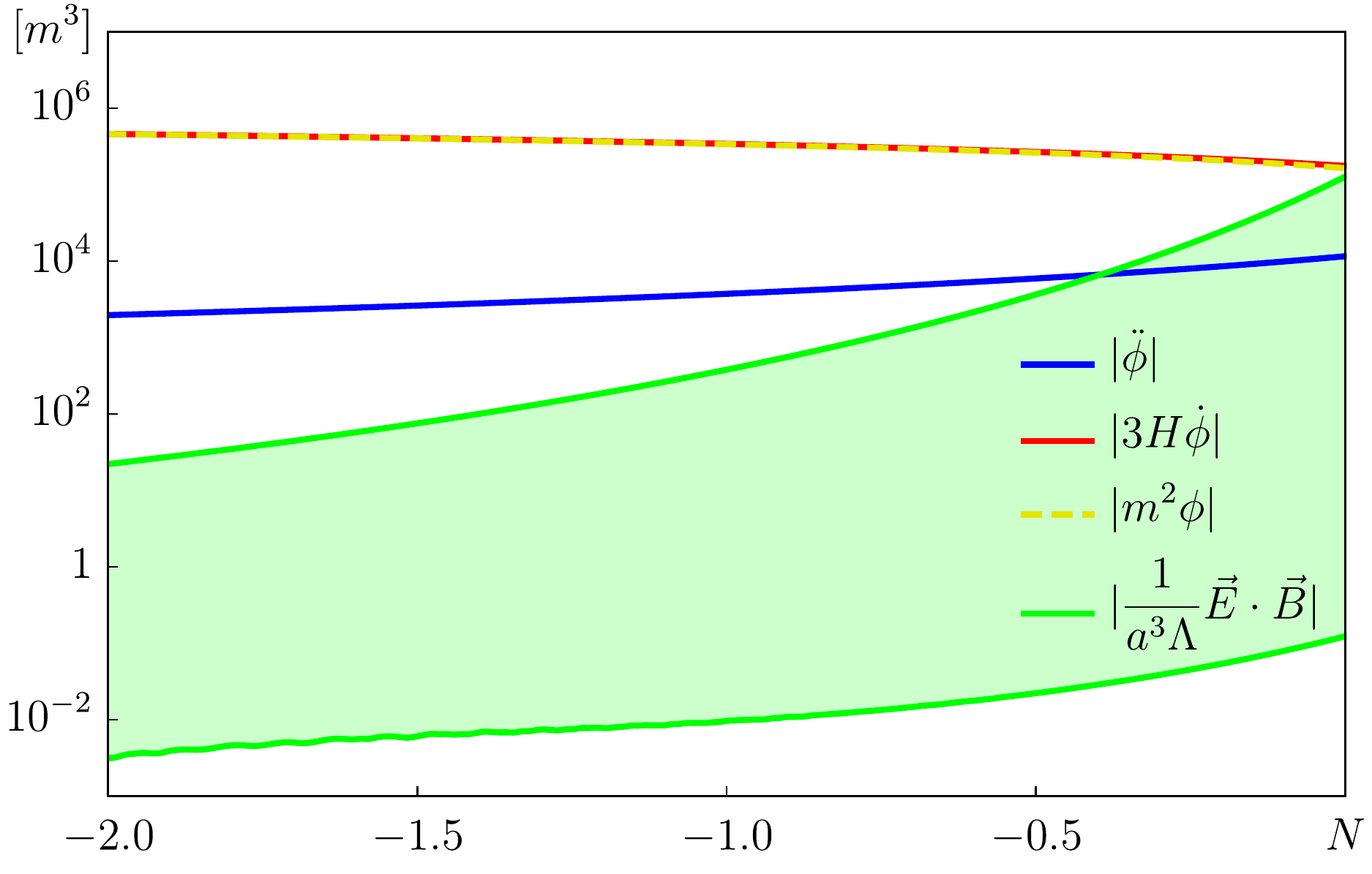}
		\caption{} \label{fig:AnalysisIC_EOM}
	\end{subfigure}
	\caption{Panel (a): The evolution of the axionic and electromagnetic energy densities for $1/\Lambda$ varying between $6~m_{pl}^{-1}$ (lower bound) and $15~m_{pl}^{-1}$ (upper bound). Panel (b): The magnitude of the different terms appearing in Eq.~(\ref{eq:axionEOM_vec_C_curved}). We computed the evolution of the  $\frac{1}{a^3\Lambda}|\vec{E}\cdot\vec{B}|$ term for $1/\Lambda$ which varies between $6~m_{pl}^{-1}$ (lower bound) and $15~m_{pl}^{-1}$ (upper bound) }  \label{fig:EnergyDensity_Nefolds}
\end{figure}

To determine, for a given coupling, when the backreaction-less solution described in Sect.~\ref{subsec:BackreactionLess} becomes a bad approximation to the real dynamics, we need to determine the moment when the contribution of the gauge field to the expansion of the Universe, as well as its backreaction on the axion dynamics, become non-negligible. In  Fig.~\ref{fig:AnalysisIC_Energy} we show the evolution of the electromagnetic energy density built from the gauge field backreaction-less numerical solution, together with the energy density components of the axion field, during the last efolds of inflation. In Fig.~\ref{fig:AnalysisIC_EOM}, we plot the different terms involving the gauge field and the homogeneous axion in Eq.~(\ref{eq:axionEOM_vec_C_curved}). For the range of couplings exhibited, $6 \leq m_{pl}/\Lambda \leq 15$, we see that the axionic energy components always dominate until the end of inflation. The ratio of the electromagnetic to the axion energy density is always very suppressed, for instance, at $N=-1$ the ratio ranges between $\sim \mathcal{O}(10^{-8})-\mathcal{O}(10^{-4})$. Even at the end of inflation the suppression still persists as $\sim \mathcal{O}(10^{-4})$ for the lowest coupling and $\sim \mathcal{O}(10^{-1})$ for $1/\Lambda = 15~m_{pl}^{-1}$. At the same time, we see that the terms $m^2\phi$ and $3H\dot{\phi}$ in the axion field equation of motion dominate over $\ddot{\phi}$ (slow-roll condition) all the way until $\epsilon_H = 1$, whereas the electromagnetic backreaction term $\frac{1}{a^3\Lambda}\vec{E}\cdot\vec{B}$ starts to be significant for the largest couplings only at $-N \lesssim 0.5$ efolds. We see therefore that choosing as initial time for our lattice simulations any moment between $N \sim -2$ and $N \sim -1$ efolds before the end of inflation guarantees that both the energy density and backreaction of the gauge field have a negligible contribution (orders or magnitude suppressed) into the inflationary dynamics for all the relevant coupling range we considered.

Gauge field backreaction effects will be automatically accounted for, as small as they might be, once we plug in the backreaction-less solution as an initial condition in our lattice simulations. In fact, our code will solve the full fields dynamics in a $3+1$ space-time according to the discretized version of the equations of motion, naturally accounting for the gauge field contribution to the inflationary dynamics. Furthermore, let us remind that the above discussion about the weight of gauge field effects on the inflationary dynamics, based on the backreaction-less solution, also assumes that the gradient contributions from the axion field are negligible. That is a very reasonable assumption as the axion is only expected to develop inhomogeneities whenever the backreaction of the gauge field becomes noticeable. Our code will account for possible effects due to the development of axion gradients, allowing us to quantify in a self-consistent manner the validity of neglecting axion inhomogeneities in first place. 

Finally, we note that we have chosen to initialize the electric field to zero, $E_i(t_i) = 0$. This, together with the fact that we introduce the axion initially as homogeneous, $\phi(t_i,\vec x) = \phi_0(N_i)$, guarantees that the lattice Gauss constraint given by Eq.~(\ref{eq:GL_EOM_curved_L}) is automatically satisfied by the initial conditions. Once the Gauss constraint is verified initially, it will remain conserved through all the running time of the simulations. Depending on the number of iterations used in the implicit method to solve for the electric field (recall Sect.~\ref{subsec:Explicit_approximation_and_iterative_method}), we will preserve the Gauss law with worse or better accuracy, even down to computer precision if desired. In Sect.~\ref{subsec:Trajectories} we will elaborate further on the consistency of taking initial vanishing electric field. In Sect.~\ref{sec:constraints} we will elaborate on our numerical ability to preserve the Gauss law Eq.~(\ref{eq:GL_EOM_curved_L}) and the Friedmann constraint Eq.~(\ref{eq:Hubble_constraint_L}) [c.f.~Eqs.~(\ref{eq:gausslaw_vec_C_curved}), (\ref{eq:friedmann1_C}) for the continuum counterparts.]

\subsubsection{Overlapping trajectories and fluctuations' cutoff}
\label{subsec:Trajectories}

In the previous section we have concluded that introducing the backreaction-less solution as an initial condition $-N \sim 1-2$ efolds before the end of inflation seems consistent with the idea that all gauge field contributions to $i)$ the axion dynamics, $ii)$ to the expansion of the Universe, and $iii)$ to the induction of axion gradients, are negligible. Of course, the deeper inside inflation, the less excited the gauge field is, and hence the better such consideration becomes. From that point of view, the larger the number of efolds before the end of inflation to introduce the initial condition in the code, the better we should capture the real dynamics. However, in practice, starting at an arbitrary number of efolds before the end of inflation, say at $N < -2$, does not allow to capture well the whole dynamical range of mode excitation. Modes are red-shifted away out of our simulation box very fast during inflation, so to capture the whole dynamical range all the way from the beginning of the simulation until the end of preheating, is only possible if we start sufficiently close to the end of inflation. Actually, given that we are limited to lattice sizes of $\leq 256^3$ sites, starting our simulation at $-N = 2$ is already barely acceptable, as we will show in a moment. 

To quantify the appropriated choice of the initial number of efolds $-N$ before the end of inflation to introduce the initial condition in our code, we can do the following self-consistency exercise: for each given coupling we introduce initial conditions based on the backreaction-less solution at two different moments. We shall identify these moments $t_2$ and $t_1$, which are naturally associated to $-N_2$ and $-N_1$ efolds before the end of inflation. Let us suppose that $t_2 < t_1$ (equivalently $-N_2 > -N_1$). Then we evolve the lattice equations of motion for each initial condition and we label each dynamical solution for the fields as $\lbrace \phi^{(2)}(t,\bar x), A_\mu^{(2)}(t,\bar x)\rbrace$ and $\lbrace \phi^{(1)}(t,\bar x), A_\mu^{(1)}(t,\bar x)\rbrace$. We shall obtain each solution until the condition $\epsilon_H(t_{\rm end}) = 1$ is reached numerically in each case. If the backreaction is truly negligible at both initial times, essentially we should obtain numerically that $\lbrace \phi^{(2)}(t,\bar x), A_\mu^{(2)}(t,\bar x)\rbrace = \lbrace \phi^{(1)}(t,\bar x), A_\mu^{(1)}(t,\bar x)\rbrace$ for all the overlapping time range $t_1 \leq t \leq t_{end}$, where $t_{end}$ signals the end of inflation. %Furthermore, given our initial normalization of the scale factor, $a_2(0) \neq a_1(0)$, the numerical value of the scale factor should be however the same at the end of each simulation, $a_2(t_{\rm end}^{(2)}) = a_1(t_{\rm end}^{(1)})$. 
We will refer to this situation as correct ``overlapping dynamical histories'' between two initial conditions started at $t_2$ and $t_1$. If on the contrary, between $t_2$ and $t_1$ the backreaction of the gauge field becomes noticeable, the solution initialized at $t_2$ will not coincide later at $t_1$, with the one just initialized in that moment, i.e. $\lbrace \phi^{(2)}(t_1,\bar x), A_\mu^{(2)}(t_1,\bar x)\rbrace \neq \lbrace \phi^{(1)}(t_1,\bar x), A_\mu^{(1)}(t_1,\bar x)\rbrace$. In this case, we then have to initialize the fields at yet another earlier time $t_3 < t_2$, and check again whether the dynamical field trajectories overlap or not for $t_2 \leq t \leq t_{end}$. Nevertheless, let us remember that we cannot push $t_3$ too deep inside inflation because of the finite size of our lattice box. Hence, we expect the trajectories to fail overlapping also when the initialization of fields happens too many efolds before the end of inflation.  

\begin{figure} % "[t!]" placement specifier just for this example
	\begin{subfigure}{0.49\textwidth}
		\includegraphics[width=\linewidth]{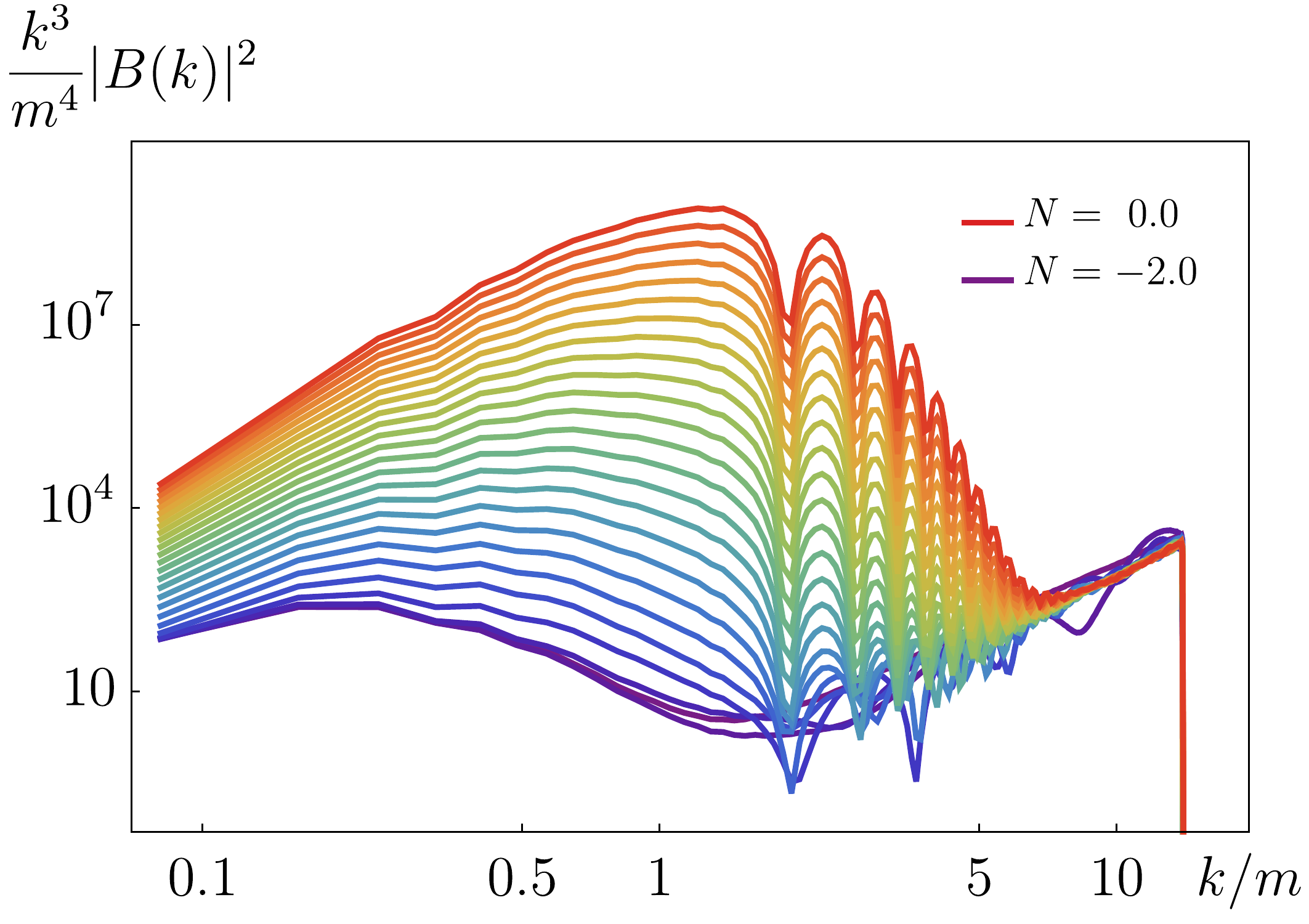}
		\caption{} \label{Inflation_MagSpectra_Evolution_15}
	\end{subfigure}
	\hspace*{\fill}
	\begin{subfigure}{0.49\textwidth}
		\includegraphics[width=\linewidth]{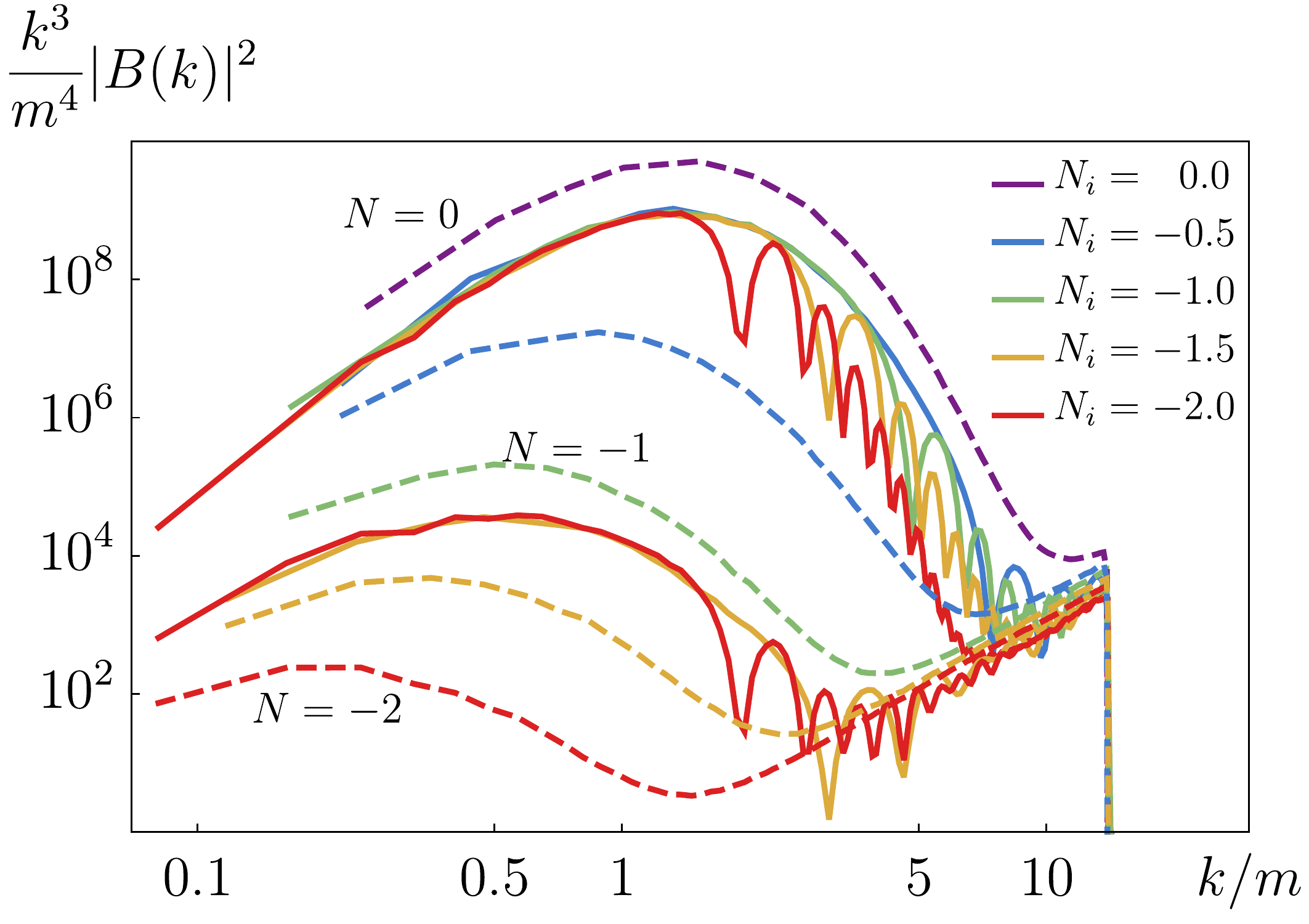}
		\caption{} \label{fig:CatchUp_B2PS_15_Inflation}
	\end{subfigure}
	
	\caption{Panel (a): The evolution during the last 2 efolds of inflation of the magnetic power spectra for the coupling $1/\Lambda = 15~m_{pl}^{-1}$ and initialization at $N_i = -2.0$ efolds. Panel (b): The ``overlap'' of magnetic power spectra at $N = -2, -1, 0$ for different initialization moments $N_i = -2.0,-1.5,-1.0,-0.5,~0.0$ efolds. Dashed lines represent the initial power spectra.}  \label{fig:SpectraInInflation}
\end{figure}

In order to perform the ``overlapping trajectories'' test, we need first to determine what is the typical range of excitation of the gauge field fluctuations during inflation. Based on the backreaction-less solution, we can already anticipate that within the allowed range of coupling values $m_{pl}/ \Lambda \lesssim 22$, we only expect to excite the gauge field up to modes $k/m \lesssim 20$  up to the end of inflation, see Fig.~\ref{fig:InitialSpectra}. However, for our application to preheating, see Sect.~\ref{subsec:Inflation}, we will only be interested in the coupling range $ 6 \leq m_{pl}^{-1}/\Lambda \leq 15$, and hence only modes up to $k/m \lesssim 10$ will be excited towards the end of inflation. During preheating, however, higher modes will be excited, and thus we need to guarantee that the natural UV cutoff $k_{UV} = \sqrt{3}{\pi\over dx}$ of the lattice, is large enough to encompass the physically excited modes.

Let us emphasize here that whenever we say we start a simulation at a given number $-N_i$ of efolds before the end of inflation, this refers to the would be efolds in a Universe dominated only by the axion. In reality, for the largest a couplings, the contribution of the gauge field to the expansion rate towards the end of inflation may become noticeable, and hence the number of efolds till we reach the condition $\epsilon_H = 1$, will actually be larger than $-N_i$. Having this in mind, in Fig.~\ref{Inflation_MagSpectra_Evolution_15} we show the evolution of the magnetic energy density power spectra for $m_{pl}/\Lambda = 15$ from a lattice simulation started at $N = -2$ efolds before the end of inflation (in purple) up to the end of inflation (in red). It can clearly be seen that the spectrum shifts towards the ultraviolet and grows in amplitude, as expected. In Fig.~\ref{fig:CatchUp_B2PS_15_Inflation} we illustrate the overlap of magnetic power spectra at $N = -1, 0$, superimposing the evolution of simulations initialized at $N_i = -2.0, -1.5, -1.0, -0.5, ~0.0$, and plotting also also spectra given by initial conditions in dashed lines. From this figure we can notice how the dynamical range varies according to the initialization moment $N_i$. In fact, the deeper inside inflation we initialize the fields, the more red-shifted the lattice cutoffs $k_{IR}$ and $k_{UV}$ are. This is done to ensure that we are able to capture whole range of relevant modes at the initial condition and during the simulation. It is clear that, because of the ultraviolet shift of excitations, if we were to use a fixed dynamical range for all initializations, we would not capture either the initial spectra deep in inflation, or the excited modes towards the end of it and afterwards. In light of these considerations, we decided to choose as an IR lattice cutoff adapted to the initialization moment $N_i$ by fixing $k_{IR} = 0.5 ~a(N_i)H(N_i)$. This automatically leads to a lattice cutoff $k_{UV} = \sqrt{3}N k_{IR}/2$ which, for the coupling range considered, suffices to later encompass all the physically excited UV modes during preheating.

\begin{figure} % "[t!]" placement specifier just for this example
	\begin{subfigure}{0.49\textwidth}
		\includegraphics[width=\linewidth]{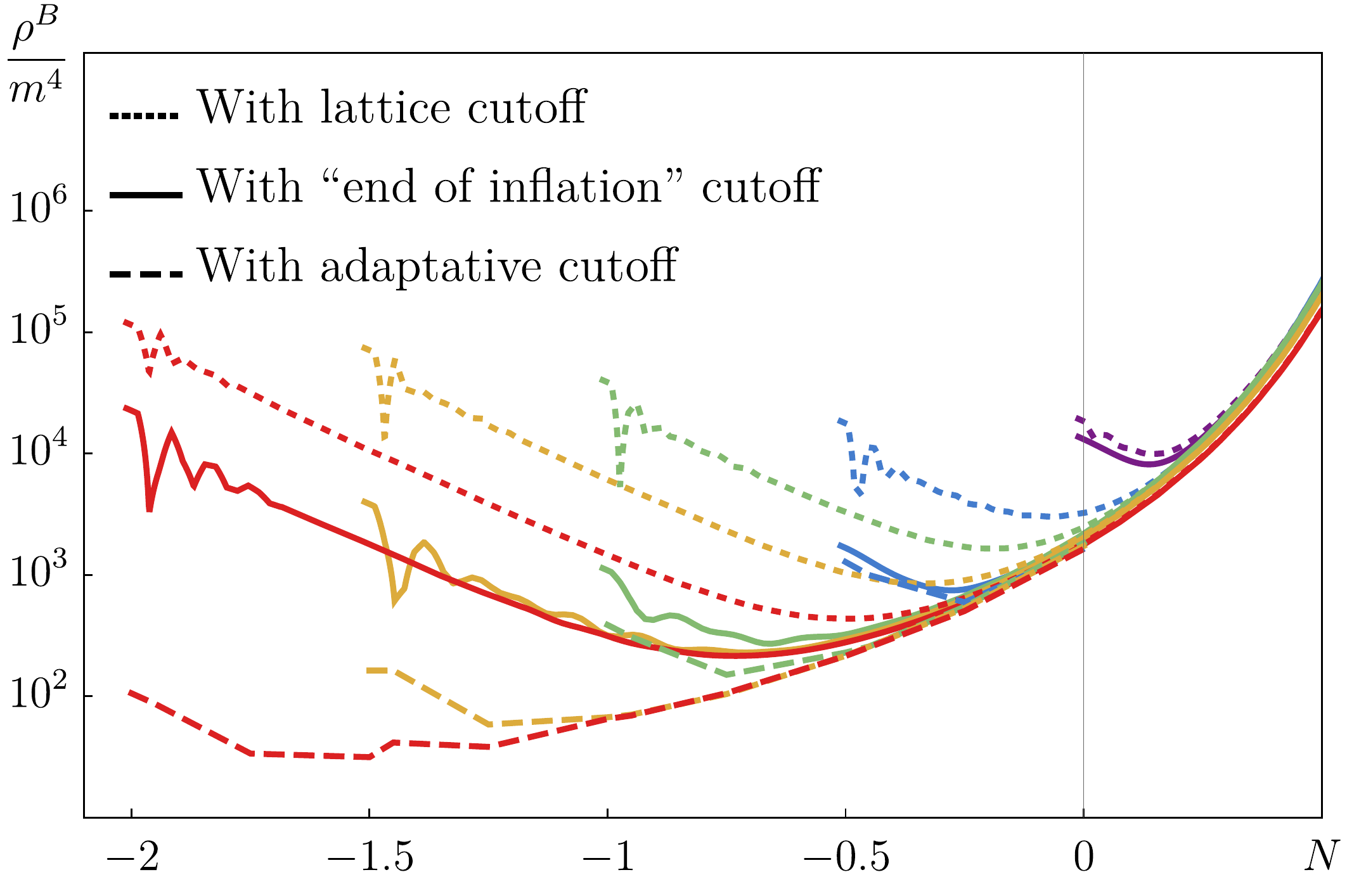}
		\caption{} \label{fig:CutoffAnalysisMAG_95}
	\end{subfigure}
	\hspace*{\fill}
	\begin{subfigure}{0.49\textwidth}
		\includegraphics[width=\linewidth]{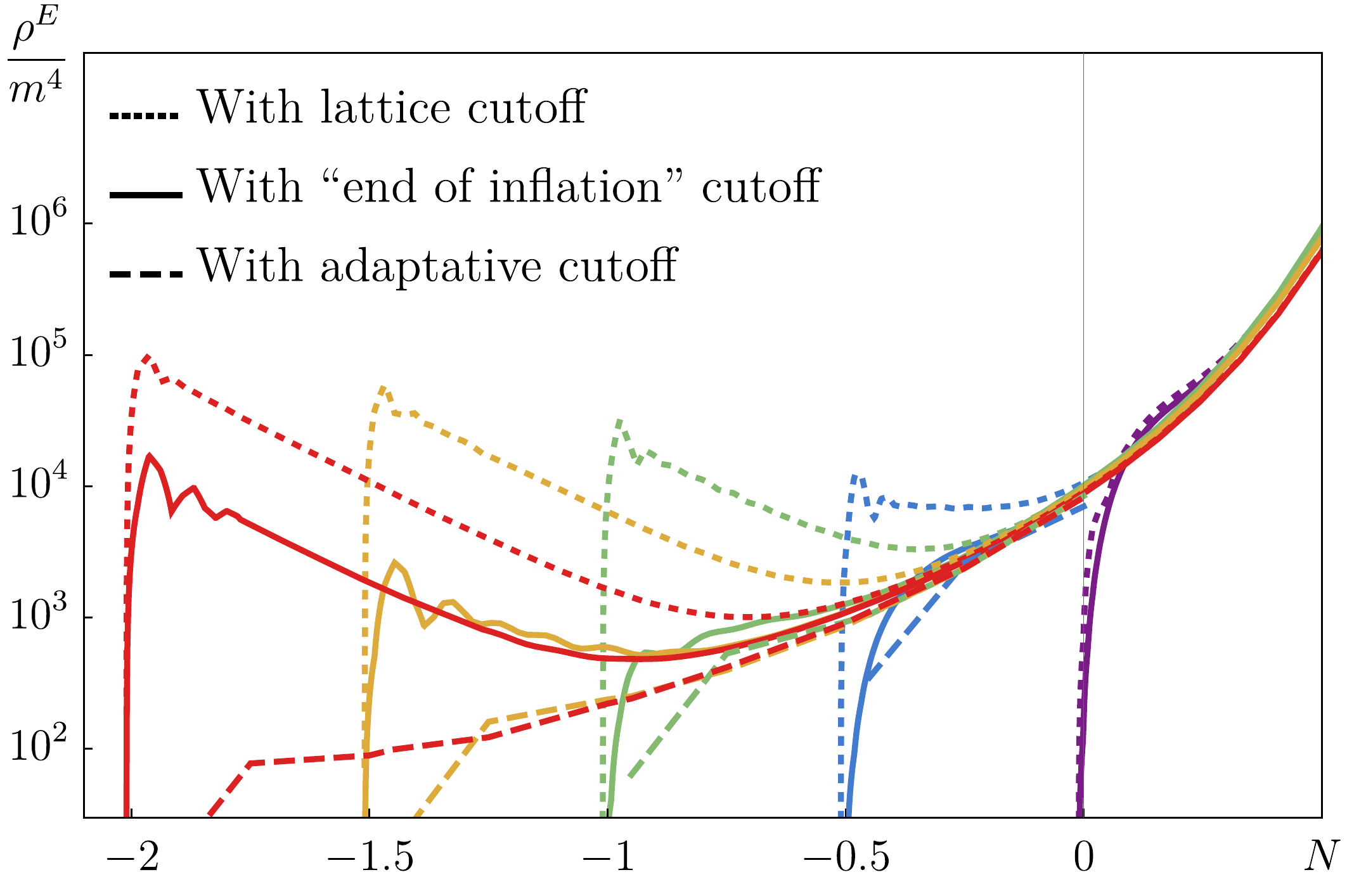}
		\caption{} \label{fig:CutoffAnalysisELE_95}
	\end{subfigure}
	
	\caption{Panel (a): The ``overlapping of magnetic energy density trajectories'' for the coupling $1/\Lambda = 9.5~m_{pl}^{-1}$ initialized at $N_i = -2.0 ~\text{(red)}, ...,~0.0 ~\text{(purple)}$ efolds. To compute the trajectories we make use of three different UV cutoffs: the lattice cutoff, the ``end of inflation'' cutoff and an adaptative cutoff.  }  \label{fig:OverlappingHistories}
\end{figure}

Another important aspect concerns the modes which are in vacuum at the moment of initialization and how to deal with them. Since we are interested in the classical simulation of the system, these modes should be eliminated (set to zero) above an initial cutoff $k \geq k_*$.  Whereas at $N = -2$ the gauge field spectrum shows fluctuations excited up to $k\lesssim  1\,m$, at the end of inflation the cutoff has grown to $k\lesssim  7-10\,m$. However, we cannot introduce as initial cutoff $k_*\Large|_{_{N_i}}$ the one inferred from the backreaction-less initial condition at the initial moment $N_i$, because there will be always harder modes $k \gtrsim k_*\Large|_{_{N_i}}$ in vacuum at that initial time, that will be successively excited later on as we approach the end of inflation. This is precisely the meaning of the ultraviolet displacement of the spectra observed in Fig.~\ref{fig:SpectraInInflation}. Therefore we need to introduce a sufficiently large cutoff,
otherwise modes in vacuum that are meant to be excited during the last efolds before the end of inflation will not be excited. On the other hand, the initial cutoff cannot be arbitrary large, since the energy density obtained from integrating over the logarithmic spectrum until such cutoff should be dominated by the excited part of the spectrum. The contribution to the energy density of vacuum modes should be sub-dominant because they are not physical in our classical simulation.  In practice, we will choose to introduce as initial cutoff the inferred one from the spectra at the end of inflation based on the excitation range of the backreaction-less solution (for each coupling $\Lambda^{-1}$). This way we guarantee to cover the dynamical range of gauge field excitation, which is the most important criteria for the correct evolution of the system. However this choice implies that deeper inside inflation there will be a larger and larger contribution to the energy densities coming from the ultraviolet vacuum tail. This effect becomes more clear when looking at the overlap of dynamical trajectories started at successive earlier times. In Fig.~\ref{fig:OverlappingHistories} we plot the evolution in time of the  magnetic (a) and electric (b) energy density components of the gauge field, for $1/\Lambda = 9.5~m_{pl}^{-1}$. There, it is actually manifest the failure of the overlapping of trajectories during inflation started at different initial times if the natural UV lattice cutoff is the one considered for the initial fluctuation spectrum (i.e.~if we do not introduce any initial cutoff). When instead a shorter UV cutoff is introduced according to the prescription explained before, we see that the trajectories overlap well, as they should. However, it is noticeable an unnatural decay in time from the onset of each simulation. This is a reflection of the fact that some vacuum modes are still contributing to the total energy, in particular those between the physical cutoff and our initial cutoff. This effect could be removed by means of an adaptative cutoff, that dynamically shifts towards the ultraviolet at the same speed as the real cutoff of the physical spectrum, hence capturing only the range of physically excited modes, leaving out the undesired vacuum ones. This approach is shown in dashed lines in both panels of Fig.~\ref{fig:OverlappingHistories} and there we see how the trajectories correctly grow during inflation and successively overlap. Nonetheless, we notice that after inflation ends, all trajectories overlap relatively well independently of whether we chose the prescribed initial cutoff or an adaptative one, as the UV vacuum tail contribution becomes irrelevant once the spectrum is dominated by the physically excited modes.

Note that in Fig.~\ref{fig:CutoffAnalysisELE_95}, the evolution in time of the electric energy density goes up abruptly initially, and then follows a smooth overlapping trajectory (when the right cutoff is chosen). This is because, as mentioned in Sect.~\ref{subsec:InitialCondition}, in order to have the Gauss law initially verified, we set the electric field to zero as initial condition, even though the amplitude of the gauge field is introduced according to the backreaction-less solution. The electric field is however rapidly generated from the equation of motion of the gauge field, as it is not possible to sustain the magnetic field given by the initial condition of the gauge field, without restoring the electric counterparts. Therefore for few time steps after the onset of the initial condition, the evolution of both the gauge and electric field amplitudes do not follow the real physical dynamics, since the electric field need first to catch up with its physical value.
 However, this occurs very fast and, since we expect the gauge field dynamics to be initially linear, very soon the spectrum of gauge field fluctuations starts shifting towards the UV. Thus the system will completely lose memory of this artefact after a very short period of adjustment. However, thanks to this artefact, we are able to preserve the Gauss law from the onset to the end of our simulation. Once the restoration is completed by the dynamics, the electric field energy density enters into a correct regime of overlapping dynamical trajectories initiated at different times, as clearly exhibited in Fig.~\ref{fig:CutoffAnalysisELE_95}. % Moreover this explains why in Fig.~\ref{fig:CatchUp_B2PS_15_Inflation} the 
% As the gauge field backreaction is negligible, for the coupling range used, at $-N = 2$ efolds before the end of inflation, we have also made sure that the initial evolution (for few time steps after electric field restoration) of both the electric and magnetic field spectra, that coincides with the equivalent spectra built from the backreaction-less solution.
In conclusion, the initially vanishing electric field trick has no impact in the dynamics, however allowing for a simple initialization of the fields preserving the Gauss law from the beginning. 

%In the left lower panel of Fig.~\ref{fig:OverlappingHistories}, we plot the evolution in time of the total electromagnetic energy density, exhibiting (again) how the trajectories of runs started at succesive earlier times, overlap very well on top of each other as they succesively catch up the previous ones. As in this case we are plotting the contribution from both the electric and magnetic energy density componets, the initial dynamical recreation of the electric field is barely visible. For completion, we also show in the right lower panel of Fig.~\ref{fig:OverlappingHistories}, the superposition of time trajectories of the backreaction term $\propto \vec E\cdot\vec B$. Even though this term is built up from both magnetic and electric field, as soon as the the artificial initial fast growth of the electric field reaches to and end, it exhibits also a clear pattern of overlapping trajectories in time. 

\subsubsection{Choice of initialization moment}

\begin{figure} % "[t!]" placement specifier just for this example
	\begin{subfigure}{0.49\textwidth}
		\includegraphics[width=\linewidth]{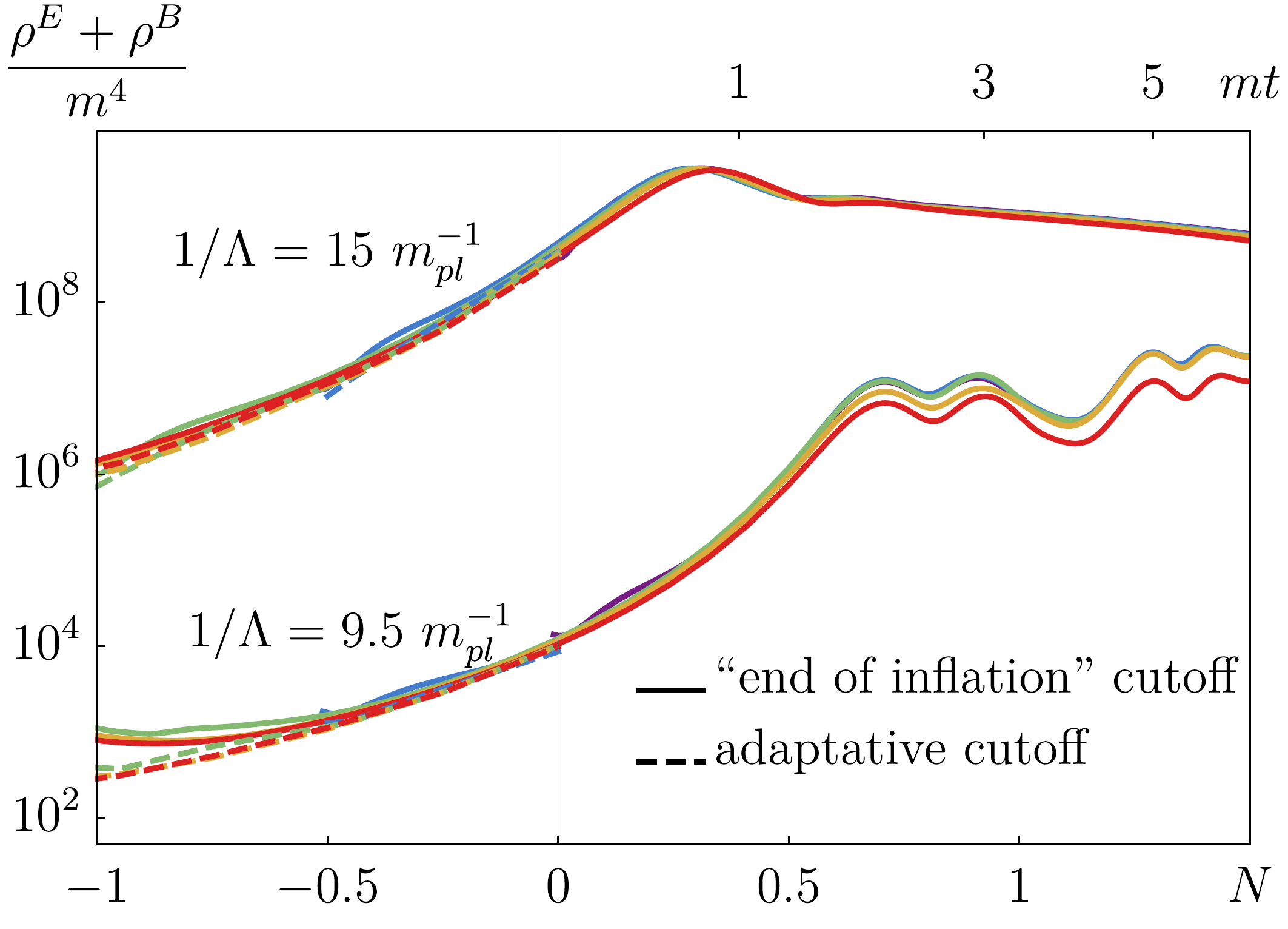}
		\caption{} \label{fig:CatchUp_Preheating_EMenergy}
	\end{subfigure}
	\hspace*{\fill}
	\begin{subfigure}{0.49\textwidth}
		\includegraphics[width=\linewidth]{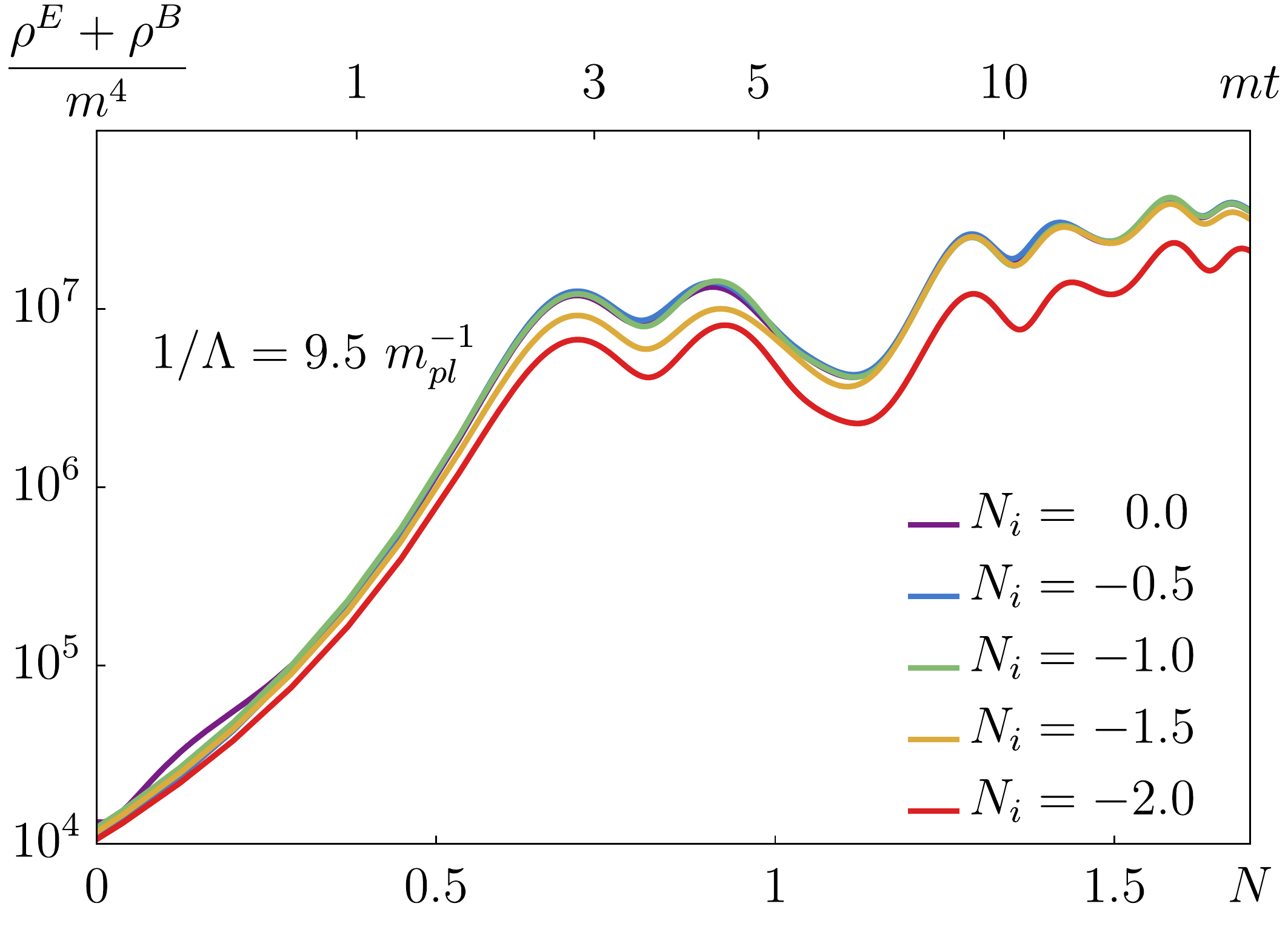}
		\caption{} \label{fig:CatchUp_Preheating_EMenergy_95}
	\end{subfigure}
	\medskip
	\begin{subfigure}{0.49\textwidth}
		\includegraphics[width=\linewidth]{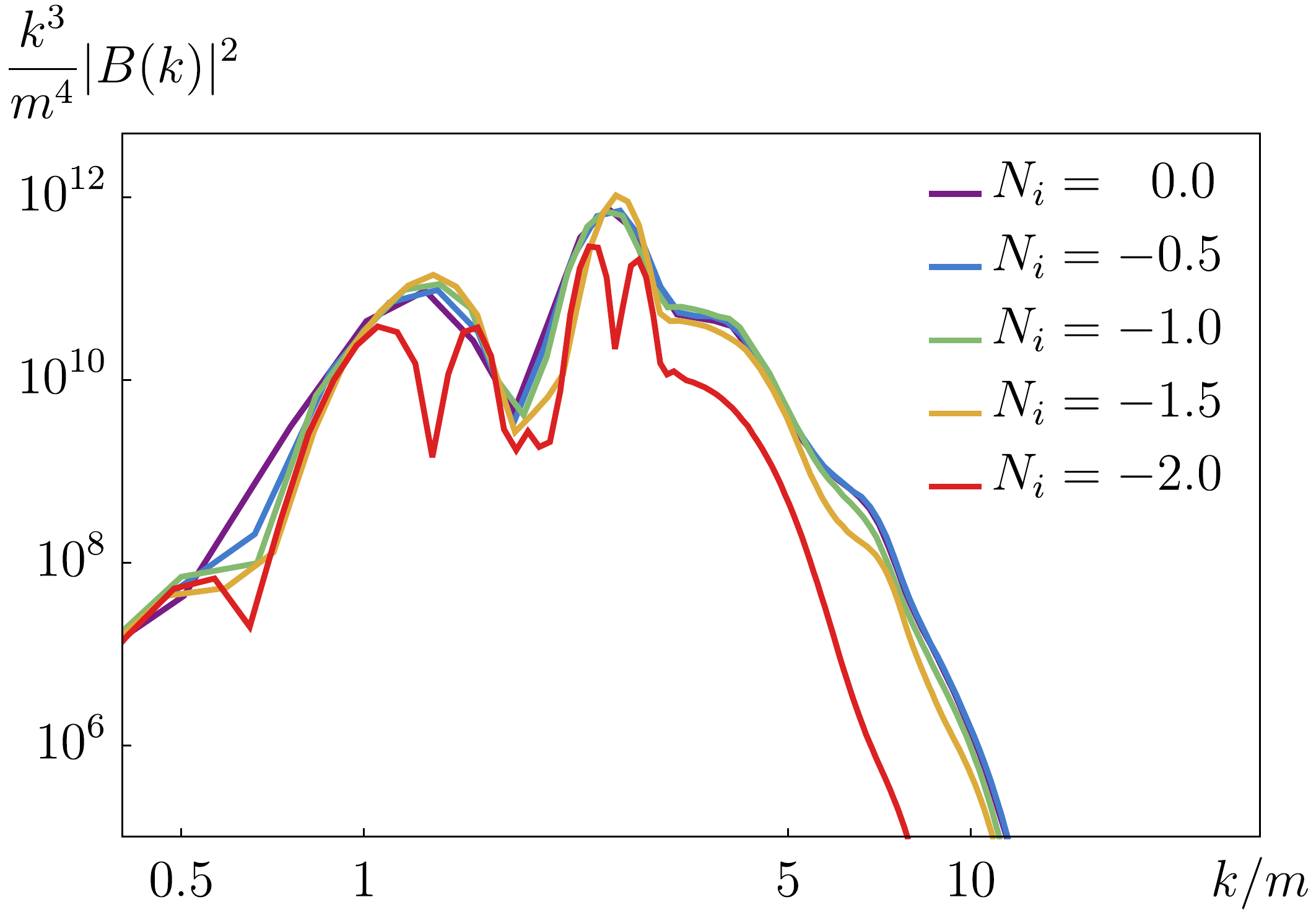}
		\caption{$1/\Lambda = 9.5~m_{pl}^{-1}$} \label{fig:CatchUp_Preheating_20mt_95}
	\end{subfigure}
	\hspace*{\fill}
	\begin{subfigure}{0.49\textwidth}
		\includegraphics[width=\linewidth]{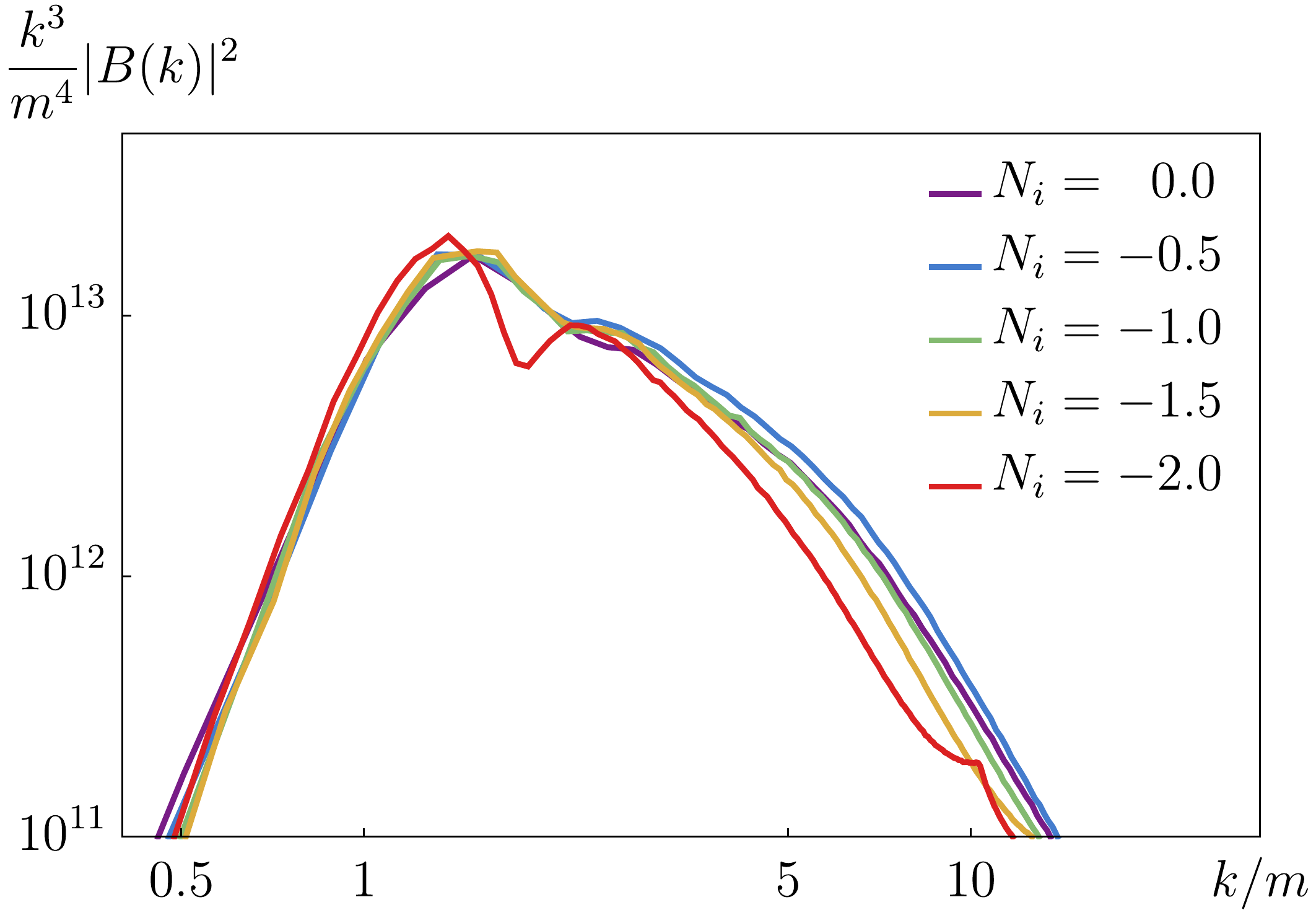}
		\caption{$1/\Lambda = 15~m_{pl}^{-1}$} \label{fig:CatchUp_Preheating_20mt_15}
	\end{subfigure}
	
	\caption{Panel (a): The ``overlapping of electromagnetic energy density trajectories'' for two couplings during the last efolds of inflation and first stages of preheating. Panel (b): The same overlapping during preheating for $1/\Lambda = 15~m_{pl}^{-1}$. Panels (c) and (d): The overlapping of magnetic power spectra for the same coupling at $mt = 1$ (c) and $mt = 20$ (d).}  \label{fig:CatchUp_Preheating}
\end{figure}

In Sect.\ref{subsec:Trajectories} we have shown that during inflation the trajectories of electromagnetic energy densities overlap consistently for initialization moments going from $N_i = -2.0$ to $N_i = 0.0$ efolds. However, from Fig.~\ref{fig:CatchUp_B2PS_15_Inflation} we notice a fall in power for the simulation that started at $N_i = -2.0$ efolds, particularly in the UV tail of the spectrum. This is a symptom of lack of dynamical range. Now we will show how well the trajectories overlap during preheating, and from there decide how much we can initialize the simulations inside inflation without losing the dynamical range of the system.

In Fig.~\ref{fig:CatchUp_Preheating_EMenergy} we can see evolution and overlap of electromagnetic energy density during the onset of preheating for the couplings $1/\Lambda = 9.5~m_{pl}^{-1}$ and $1/\Lambda = 15~m_{pl}^{-1}$. We notice that during the last stages of inflation the trajectories overlap very well, but eventually some discrepancies emerge, in particular for the simulations starting at $N_i = -2.0$ and $N_i = -1.5$ efolds. In Fig.~\ref{fig:CatchUp_Preheating_EMenergy_95} we have zoomed in the overlapping trajectories for the coupling $1/\Lambda = 9.5~m_{pl}^{-1}$, from where we can see more in detail the discrepancy between the $N_i = -1.0, -0.5, ~0.0$ efolds and the $N_i = -2.0, -1.5$ efolds initializations. We notice that the field dynamics starting at $N_i = -2.0$ exhibit a clear discrepancy after $mt \approx 1$, while the $N_i = -1.5$ case presents differences between $mt \approx 2$ and $\approx 5$, but eventually it catches up the other trajectories. From Figs.~\ref{fig:CatchUp_Preheating_20mt_95} and \ref{fig:CatchUp_Preheating_20mt_15} we can inspect the magnetic power spectra at $mt = 20$ for $1/\Lambda = 9.5~m_{pl}^{-1}$ and $1/\Lambda = 15~m_{pl}^{-1}$. As notice before, the $N_i = -2.0$ case manifests a decrease of power in the UV modes for both couplings. Moreover we can see the presence of extra wiggles which are not present or not very noticeable in the other spectra. From these two plots we can also convince ourselves that the $N_i = -1.5$ case is borderline since for IR modes we have a good overlap but the UV modes start to suffer from the same fall in power encountered for the $N_i = -2.0$ case.

The reason for this is very simple: when we initialize the fields as early as $N_i = -2.0$ efolds before the end of inflation, we need to make larger the volume of the lattice. As we actually take as our lattice IR cutoff as $k_{\rm IR} = 0.5aH$, the earlier we start a simulation the more we reduce our lattice IR cutoff to capture the initial spectrum of gauge modes at that time. This reduces automatically the lattice UV cutoff, which is always given by $k_{\rm UV} = \frac{\sqrt{3}N}{2}k_{\rm IR}$. Therefore, the problem with starting our simulation at $N_i = -2.0$ efolds is that we barely capture well the UV range that is really needed later on for describing the gauge field excitation during preheating. As a result, the simulations started at $N_i = -2.0$ efolds, may eventually differ from others started later, specially in the case of relatively large couplings $m_{pl}/\Lambda \gtrsim 9$, where the enhancement of gauge field is not negligible. Whereas $N_i = -2.0$ as a starting time is barely acceptable, we have seen that the dynamically evolved solutions starting at earlier times, say $N_i < -2.0$, exhibit clear differences which are only due to the lack of a good dynamical coverage. Therefore, cannot consider them.

Based on the discussion presented so far, we could take as a typical initializing time of our simulations $N_i = -1.5$ or $N_i = -1.0$ efolds before the end of inflation. For the largest couplings simulations starting at $N_i = -1.5$, the results are however borderline acceptable, as they still exhibit some lack of power in UV. Because of this, we have finally decided to take as our canonical starting time as $N_i = -1.0$. This choice guarantees for all the couplings that the initial backreaction is negligible, based on Fig.~\ref{fig:AnalysisIC_EOM}, and at the same time allows to use a sufficiently large dynamical range in our simulations, so that all the relevant excited modes from the start to the end of the simulation, are well captured. Once again, let us notice that $N_i = -1.0$ refers to a Universe dominated by an axion sector only; for the largest couplings however the physical number of efolds till the end of inflation can be larger, as the gauge field backreaction in the inflationary dynamics is not negligible.

\subsection{Part II. Preheating}
\label{subsec:Preheating}

In this section we investigate the preheating stage after inflation. In particular, we analyse the time evolution of the different energy density components and of the gauge field power spectra, and we quantify the efficiency of energy transfer from the axion to the gauge field, as a function of the coupling $1/\Lambda$. All simulations presented here start according to the prescription discussed in Sect.~\ref{subsec:InitialCondition}, using a grid of $N^3 = 256^3$ points, and initialized at our canonical choice of $N_i = -1.0$ efolds before end of inflation (as viewed from the axion-only dominated dynamics).

\subsubsection{Energy densities and power spectra}

\begin{figure} % "[t!]" placement specifier just for this example
	\begin{subfigure}{0.32\textwidth}
		\includegraphics[width=\linewidth]{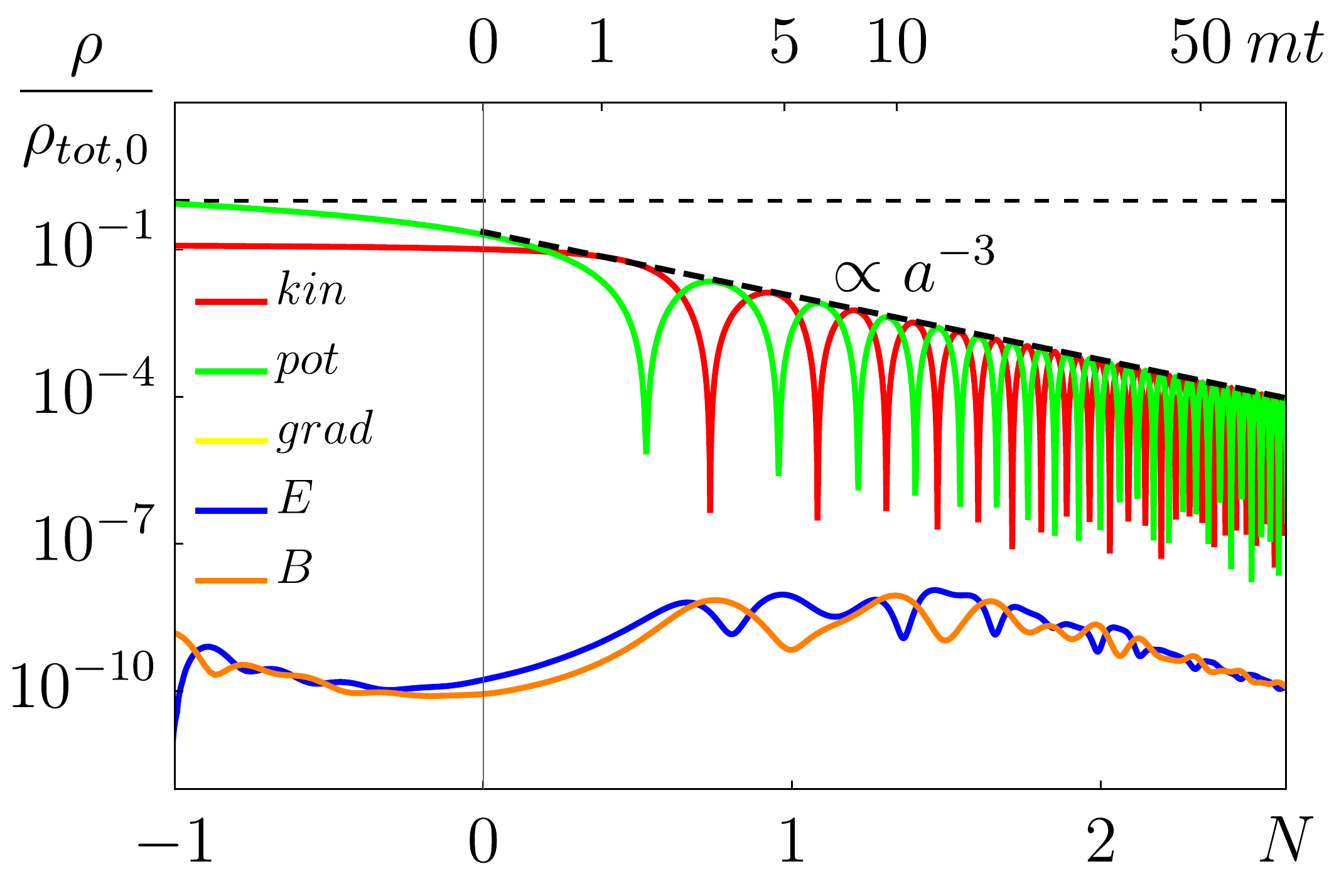}
		\caption{$1/\Lambda = 6~m_{pl}^{-1}$} \label{Total_energies_1_6}
	\end{subfigure}\hspace*{\fill}
	\begin{subfigure}{0.32\textwidth}
		\includegraphics[width=\linewidth]{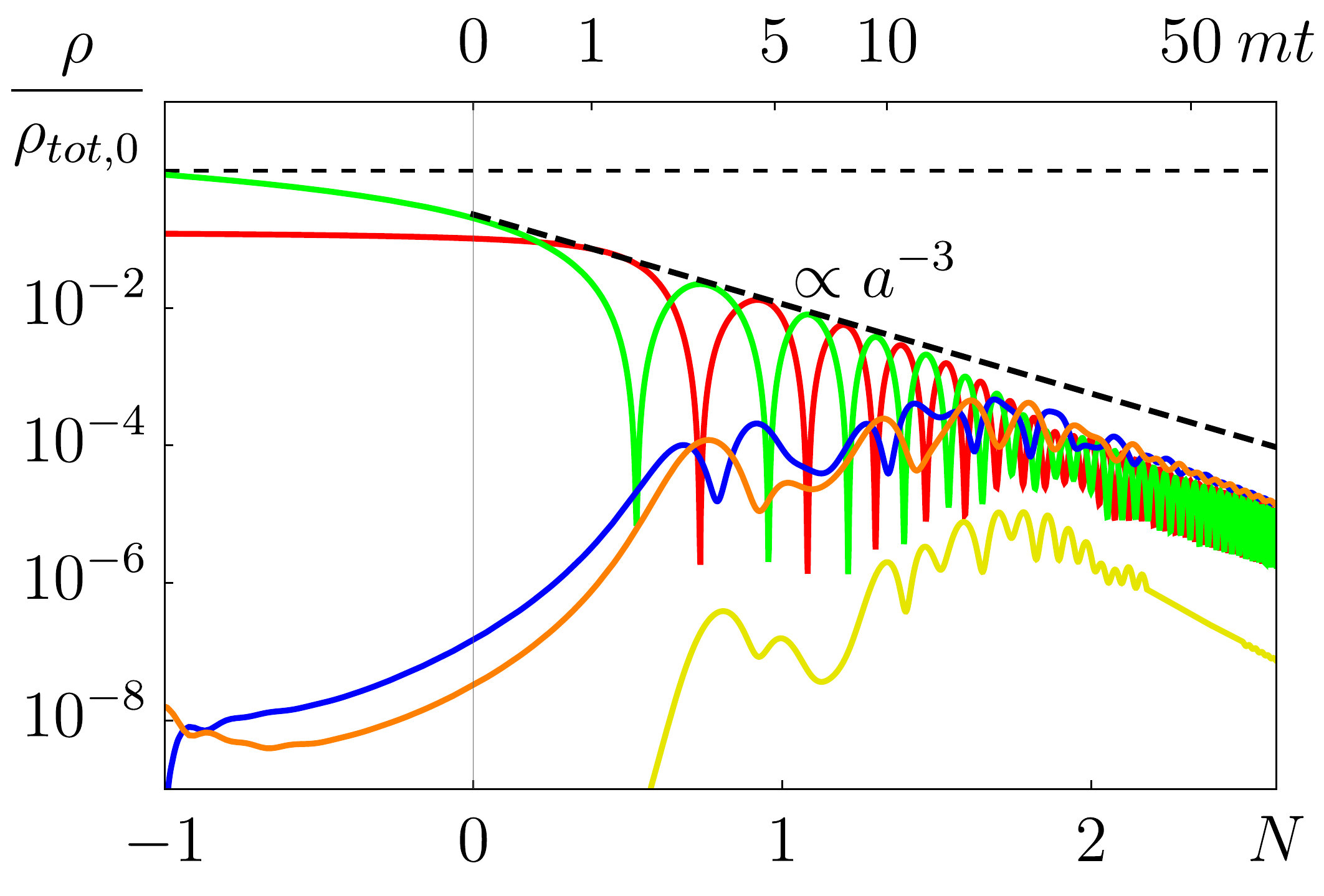}
		\caption{$1/\Lambda = 9.5~m_{pl}^{-1}$} \label{Total_energies_1_95}
	\end{subfigure}
	\hspace*{\fill}
	\begin{subfigure}{0.32\textwidth}
		\includegraphics[width=\linewidth]{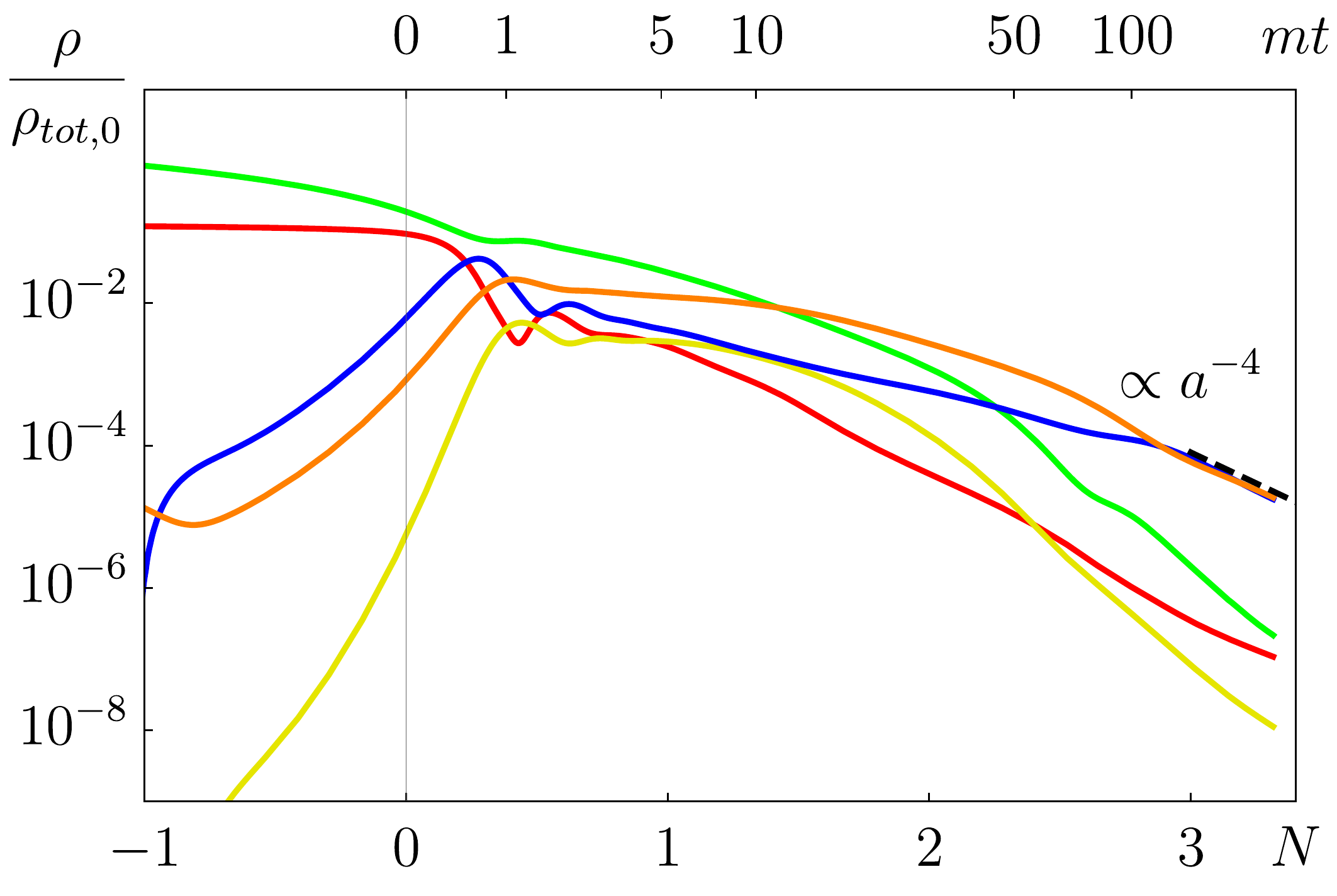}
		\caption{$1/\Lambda = 15~m_{pl}^{-1}$} \label{Total_energies_1_15}
	\end{subfigure}
	\hspace*{\fill}
	\caption{Complete evolution of the energy density components, normalized by the initial one, for three different representative couplings. The black and dashed lines represent the expected scalings of the dominant sector in an expanding Universe.}  \label{fig:total_energies}
\end{figure}

We start by studying the evolution in time of the volume averaged energy density components. We will consider three different coupling values, representative of three distinct regimes, characterized by the fraction of energy transferred from the axion into the gauge field. In Fig.~\ref{Total_energies_1_6} we see the evolution of the axion and gauge field energy density components for $1/\Lambda = 6~m_{pl}^{-1}$. This value is representative of the ``sub-critical'' coupling regime, for which the energy density of the gauge field remains always sub-dominant with respect to the axionic one through the whole evolution (during and after inflation). For sub-critical couplings the gradient energy density is tiny compared to the other components, so we do not display it in Fig.~\ref{Total_energies_1_6}. We observe that the kinetic and potential energy densities of the axion behave as those of an harmonic oscillator with frequency $m$ in an expanding Universe. This is expected as in this weak coupling regime the axionic sector dominates the energy budget of the Universe (and hence the expansion rate). In other words, the backreaction of the gauge field remains mostly negligible. As expected, the oscillation-averaged kinetic and potential energy densities correctly scale as $\rho~\propto~a^3$ (say the envelope of the oscillation maxima), as if the Universe was matter dominated with $a \propto~t^{2/3}$. Concerning the gauge field, we see that the evolution of electric and magnetic energy densities are characterized by wiggles around a common trajectory. These wiggles appear after the end of inflation during the axion oscillation regime, and decrease in amplitude towards late times, as the velocity of the axion also becomes gradually smaller due to the expansion of the Universe. After few axion oscillations (approximately $\sim$ 1-2 efolds after the end of inflation) the dilution of the gauge field due to the expansion of the Universe becomes stronger than the tachyonic instability from the axionic coupling, and hence on the overall the electromagnetic energy density starts decaying in time.

\begin{figure} % "[t!]" placement specifier just for this example
	\begin{subfigure}{0.32\textwidth}
		\includegraphics[width=\linewidth]{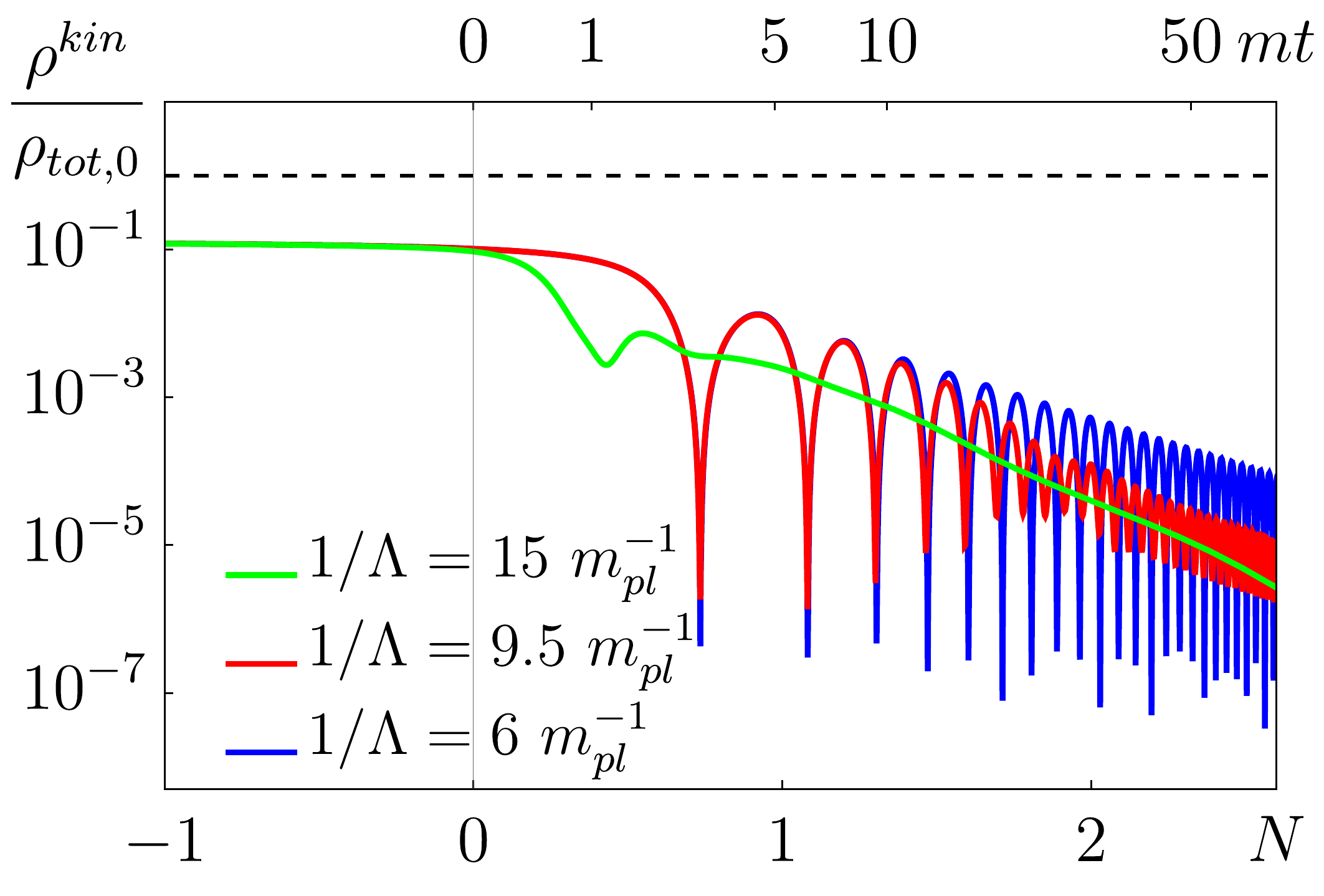}
		\caption{Kinetic energy} \label{fig:Dynamics_Kin_3couplings}
	\end{subfigure}\hspace*{\fill}
	\begin{subfigure}{0.32\textwidth}
		\includegraphics[width=\linewidth]{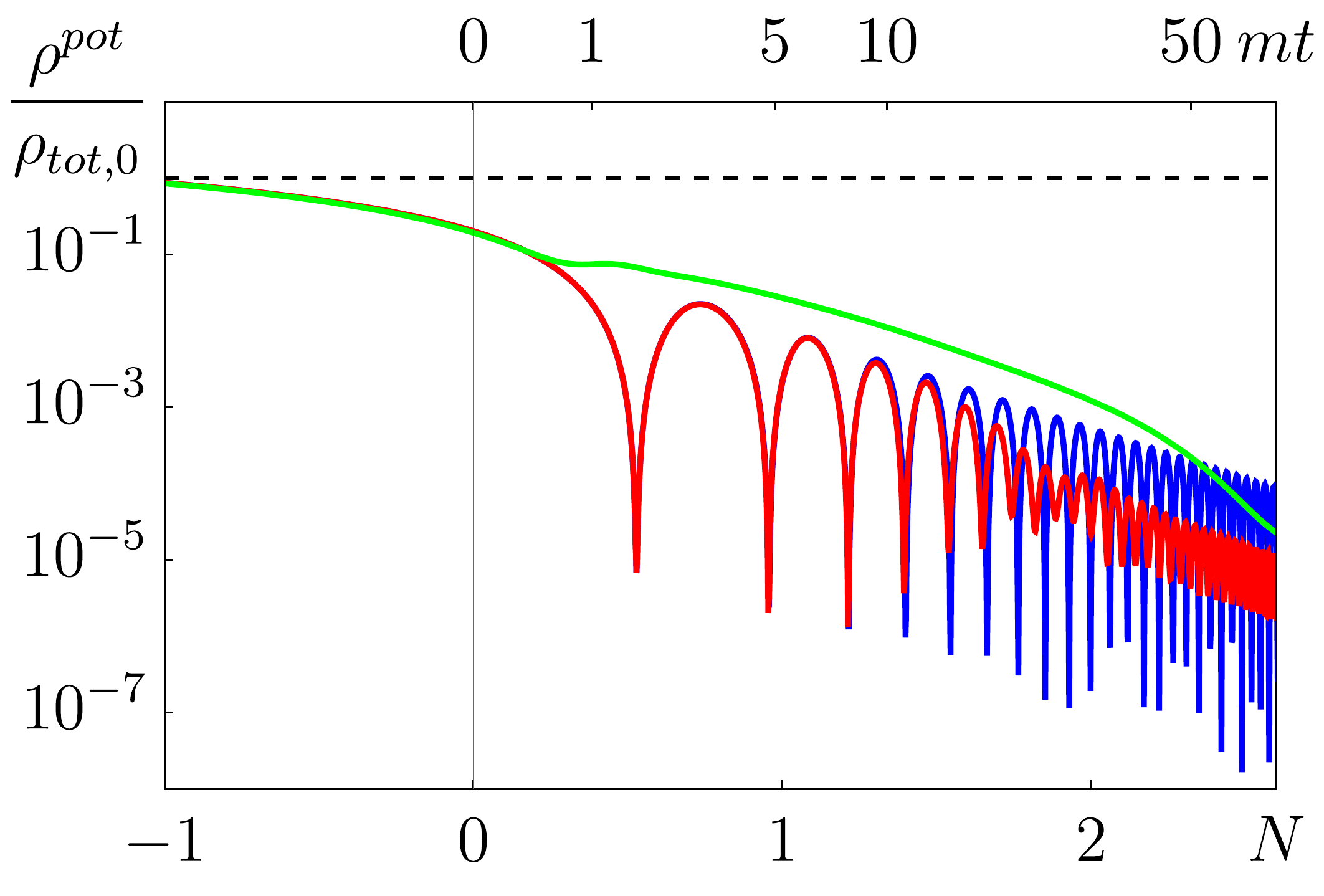}
		\caption{Potential energy} \label{fig:Dynamics_Pot_3couplings}
	\end{subfigure}\hspace*{\fill}
	\begin{subfigure}{0.32\textwidth}
		\includegraphics[width=\linewidth]{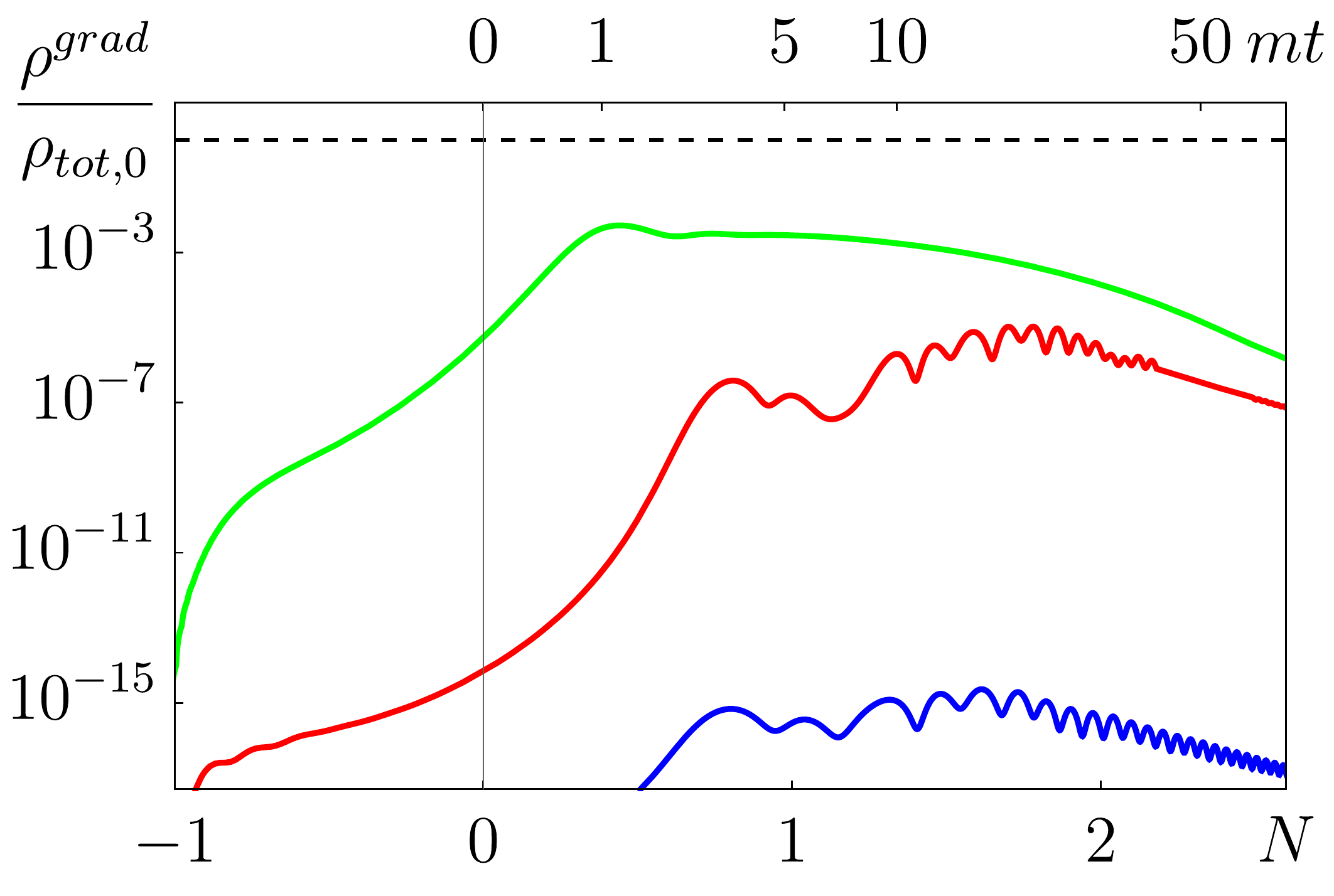}
		\caption{Gradient energy} \label{fig:Dynamics_El_3couplings}
	\end{subfigure}
	\medskip
	\begin{subfigure}{0.32\textwidth}
		\includegraphics[width=\linewidth]{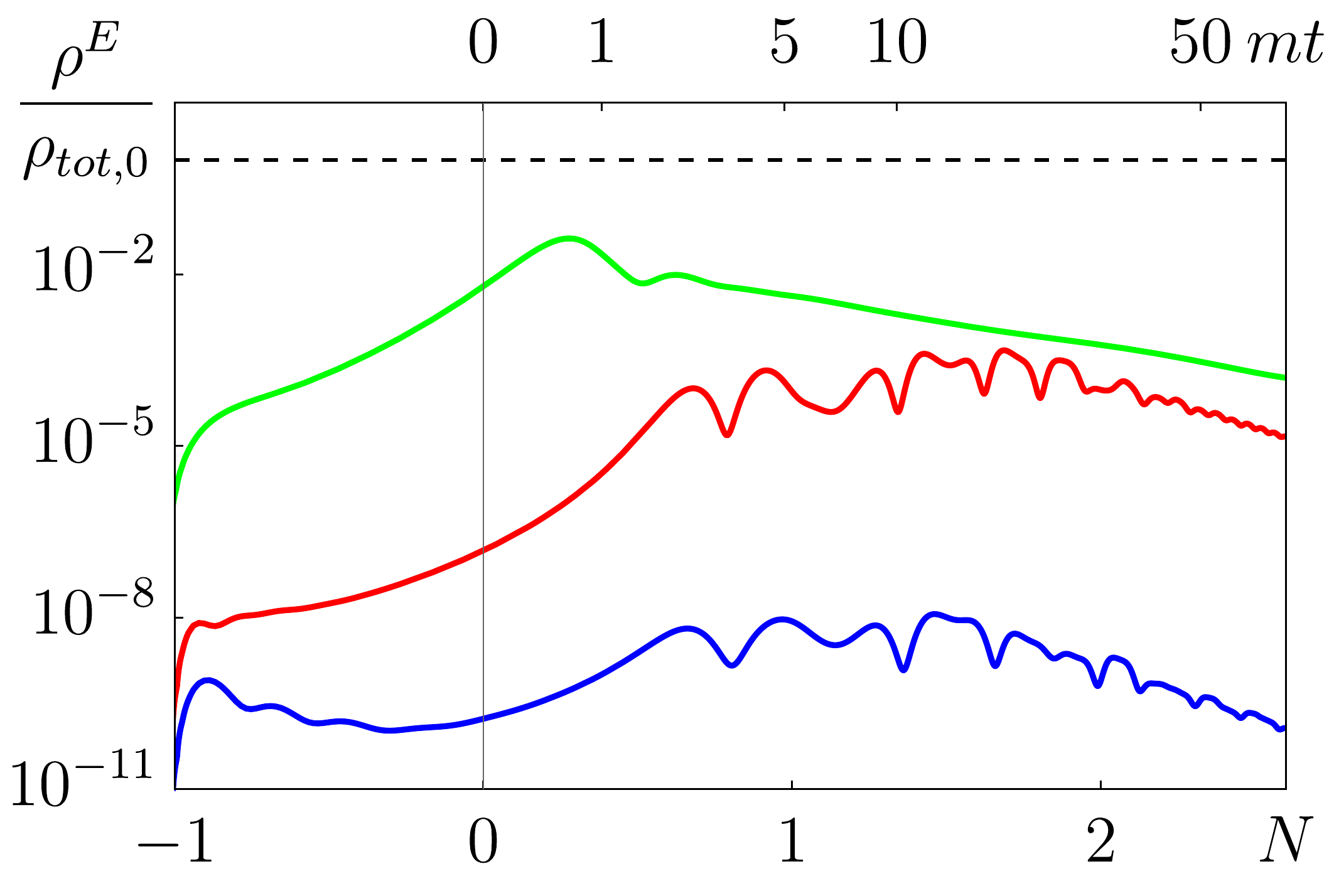}
		\caption{Electric energy} \label{fig:Dynamics_Mag_3couplings}
	\end{subfigure}	\hspace*{\fill}
	\begin{subfigure}{0.32\textwidth}
		\includegraphics[width=\linewidth]{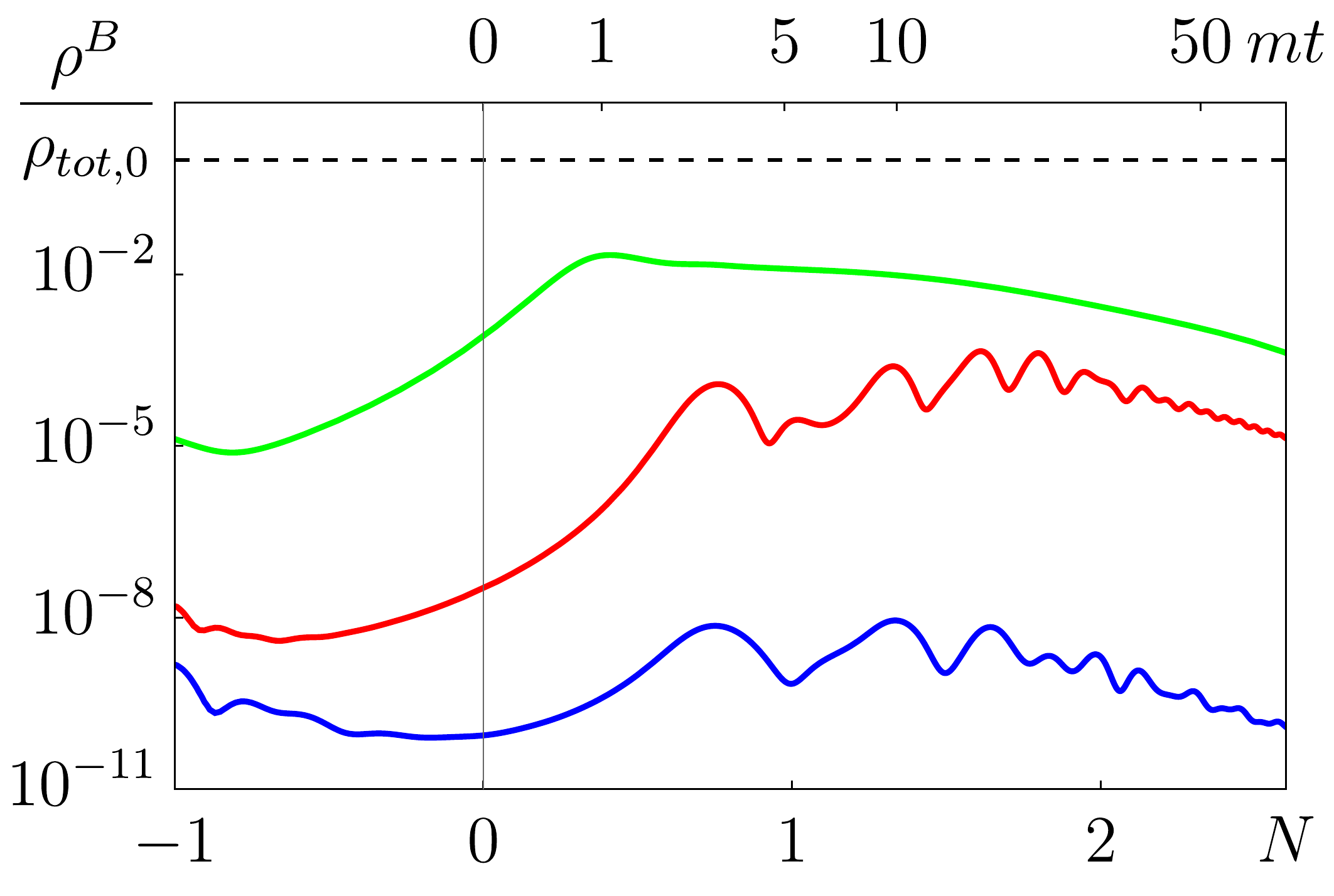}
		\caption{Magnetic energy} \label{fig:Dynamics_Grad_3couplings}
	\end{subfigure}\hspace*{\fill}
	\begin{subfigure}{0.32\textwidth}
		\includegraphics[width=\linewidth]{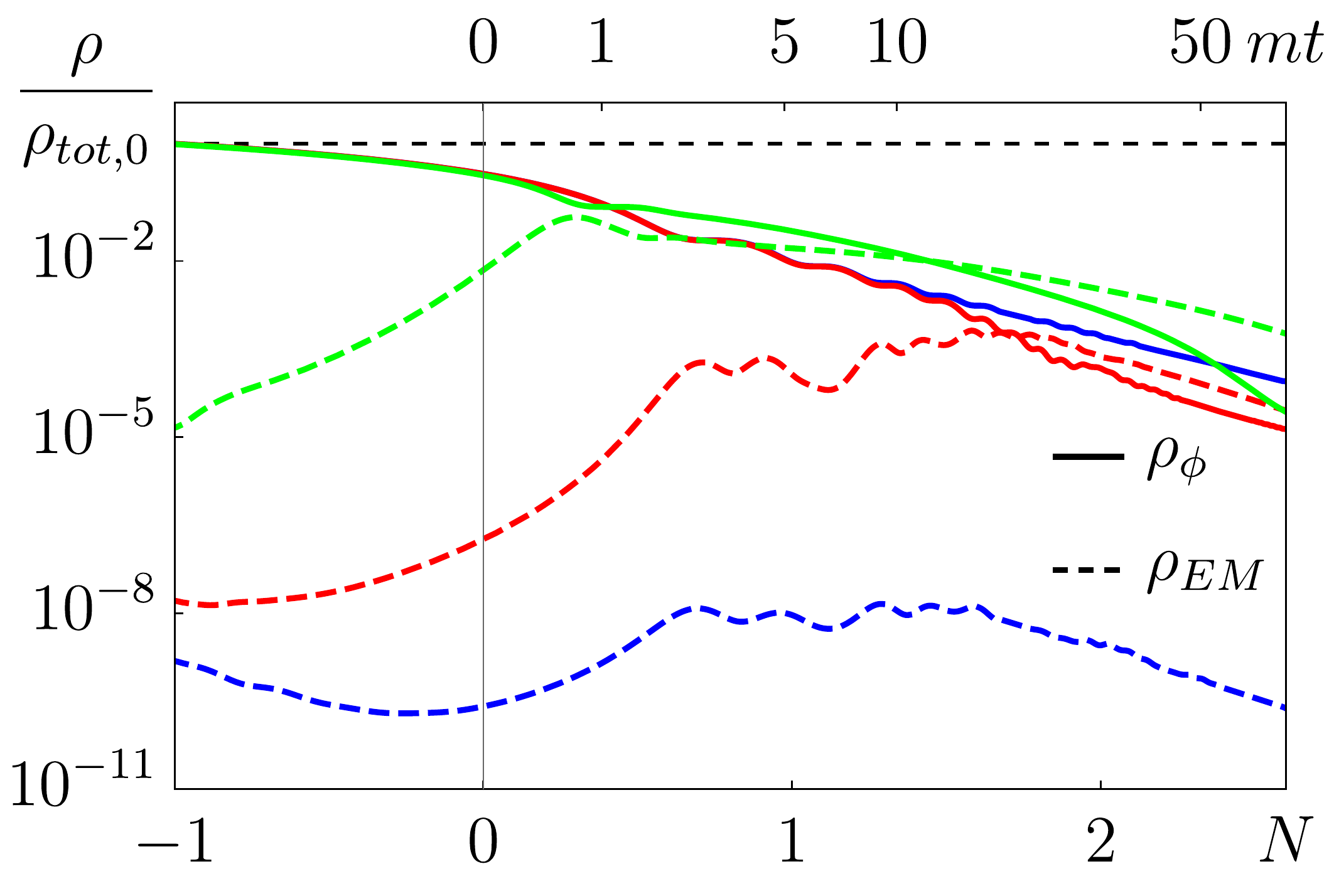}
		\caption{Axionic and EM energies} \label{fig:Dynamics_TotalEnergies_3couplings}
	\end{subfigure}	
	\caption{The different mean energy densities components, divided by the initial one, for three couplings: $1/\Lambda = 6~m_{pl}^{-1}$ in blue, $1/\Lambda = 9.5~m_{pl}^{-1}$ in red and $1/\Lambda = 15~m_{pl}^{-1}$ in green.}  \label{fig:Dynamics_energycomponents}
\end{figure}

The energy densities for $1/\Lambda = 9.5~m_{pl}^{-1}$ are shown in Fig.~\ref{Total_energies_1_95}. This case  corresponds to the critical coupling value, as the gauge field is well excited and eventually, from $mt \sim 20$ ($N\sim 2$ efolds after inflation), its mean energy density becomes comparable to the axion energy density. Furthermore, we notice an enhancement of the gradient mean energy density, which reflects a larger degree of inhomogeneity of the axion field. This is due to the backreaction of the gauge field into the scalar sector, which is not completely negligible anymore. Once the total energy density becomes almost equivalently split between the axion and the gauge sector, we notice a change in the scaling of the kinetic and potential mean energy densities, as expected. In particular, the observed scaling is neither characteristic of matter- or radiation-domination, as the Universe is neither axion or gauge field dominated, but rather dominated by an even mixture of the two. We note nonetheless that the axion field mean value still oscillates around the minimum of the potential, as the axion inhomogeneities still represent only a small perturbation over the homogeneous background configuration.

Finally in Fig.~\ref{Total_energies_1_15}, we present the energy densities for a super-critical coupling, $1/\Lambda = 15~m_{pl}^{-1}$. In this case the gauge field mean energy densities rapidly become dominant. We see that the oscillations of the mean kinetic and potential energy densities have disappeared and that the gradient energy also becomes rapidly comparable to the others. This shows how for a super-critical coupling the gauge field backreaction, and correspondingly the inhomogeneity of the axionic field, are very noticeable, clearly impacting on the evolution of the system dynamics. We see that for this large coupling, the wiggles that appeared during the evolution of electric and magnetic energy densities for sub-critical and critical couplings, have now disappeared. This can be explained by the fact that the gauge field are not being excited periodically by the axion since the latter cannot be considered anymore an oscillating homogeneous condensate. We also note that at late times the gauge sector reaches equipartition, with both the electric and magnetic energy densities correctly scaling as $\rho~\propto~a^{-4}$, while the scale factor scales as $a~\propto~t^{1/2}$. This is expected, as it corresponds to a radiation dominated Universe with expansion rate dictated by a dominating massless gauge field.

For further clarity, we plot in Fig.~\ref{fig:Dynamics_energycomponents} the different energy components of the system separately, comparing in each panel the evolution of a given component for the same three couplings $1/\Lambda$. The amount of gauge field excitation and corresponding impact in the system dynamics, comparing sub-critical, critical and super-critical couplings, is clearly appreciated in the different panels, in an component by component basis. 

\begin{figure} % "[t!]" placement specifier just for this example
	\begin{subfigure}{0.49\textwidth}
		\includegraphics[width=\linewidth]{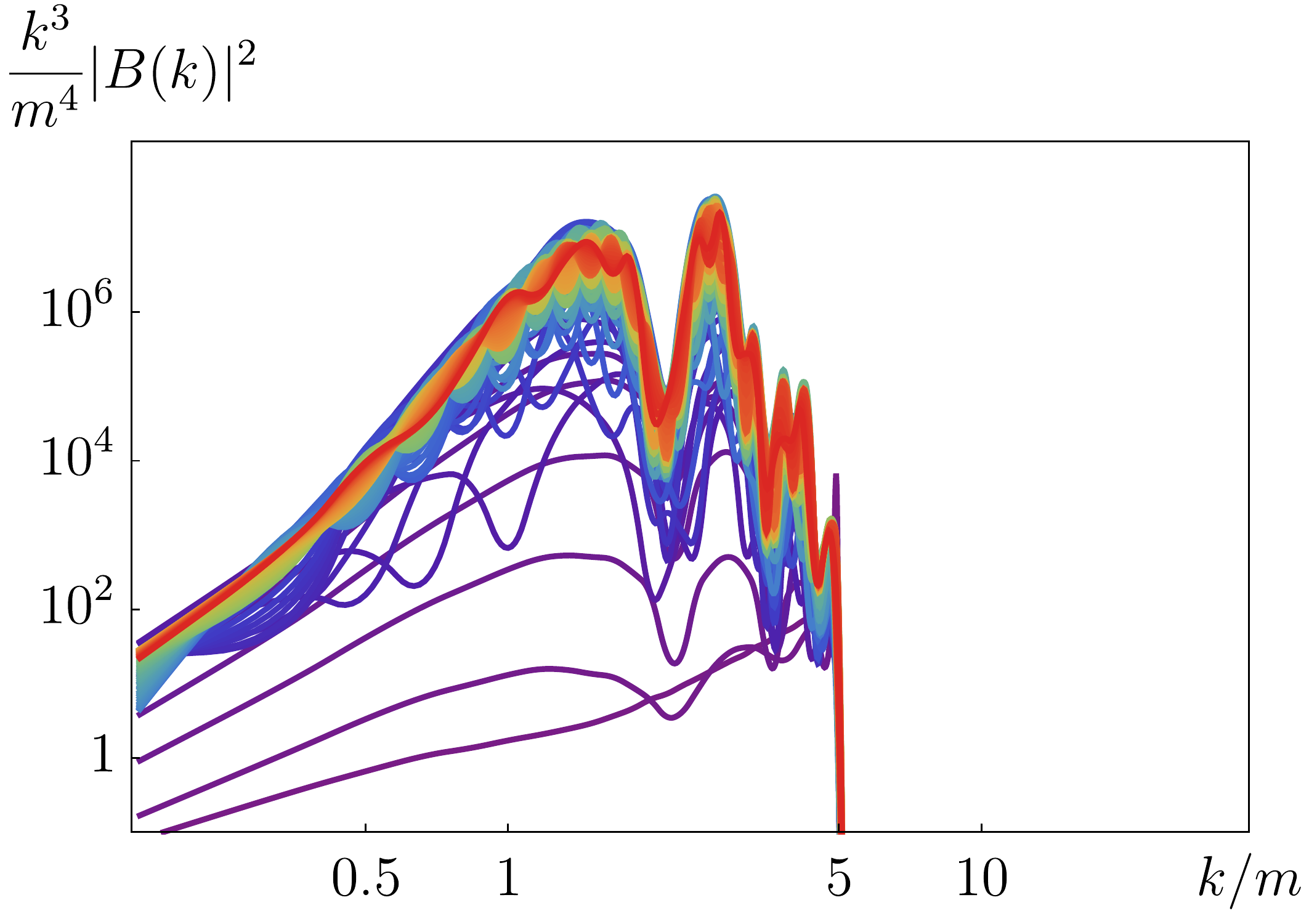}
		\caption{$1/\Lambda = 6~m_{pl}^{-1}$} \label{fig:B2PS_6}
	\end{subfigure}\hspace*{\fill}
	\begin{subfigure}{0.49\textwidth}
		\includegraphics[width=\linewidth]{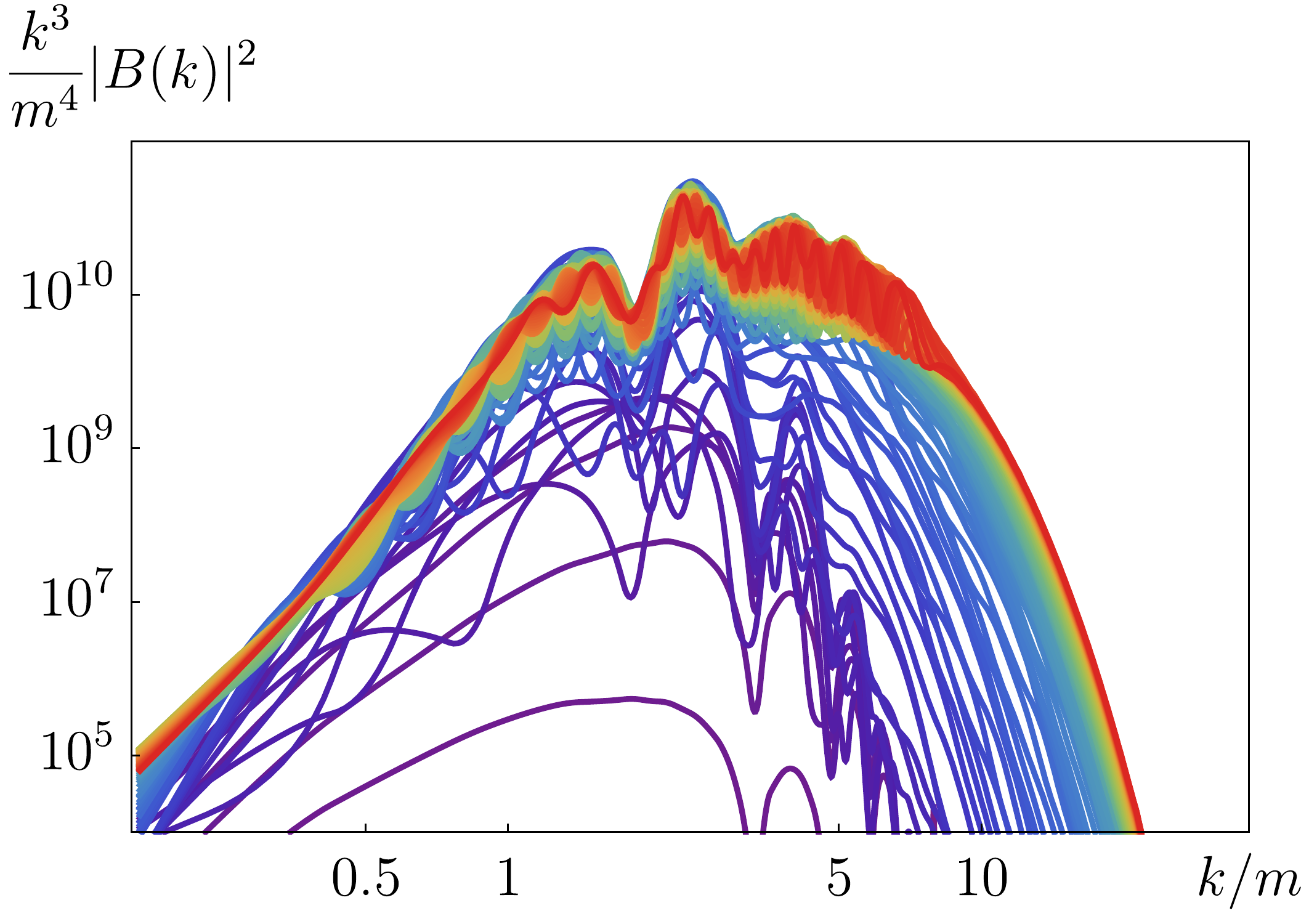}
		\caption{$1/\Lambda = 9.5~m_{pl}^{-1}$} \label{fig:B2PS_95}
	\end{subfigure}	
	\medskip
	\begin{subfigure}{0.49\textwidth}
		\includegraphics[width=\linewidth]{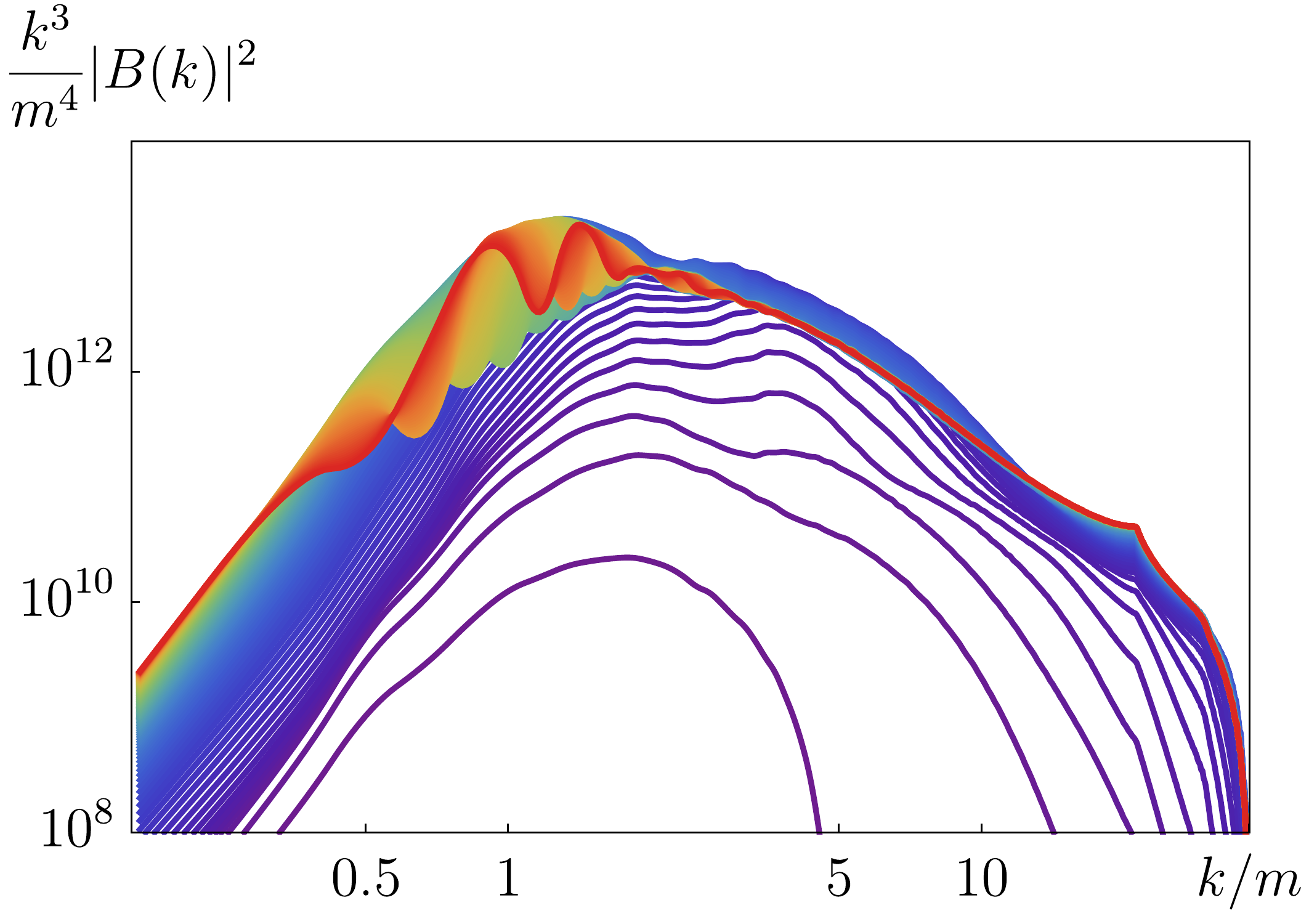}
		\caption{$1/\Lambda = 15~m_{pl}^{-1}$} \label{fig:B2PS_15}
	\end{subfigure}\hspace*{\fill}
	\begin{subfigure}{0.49\textwidth}
		\includegraphics[width=\linewidth]{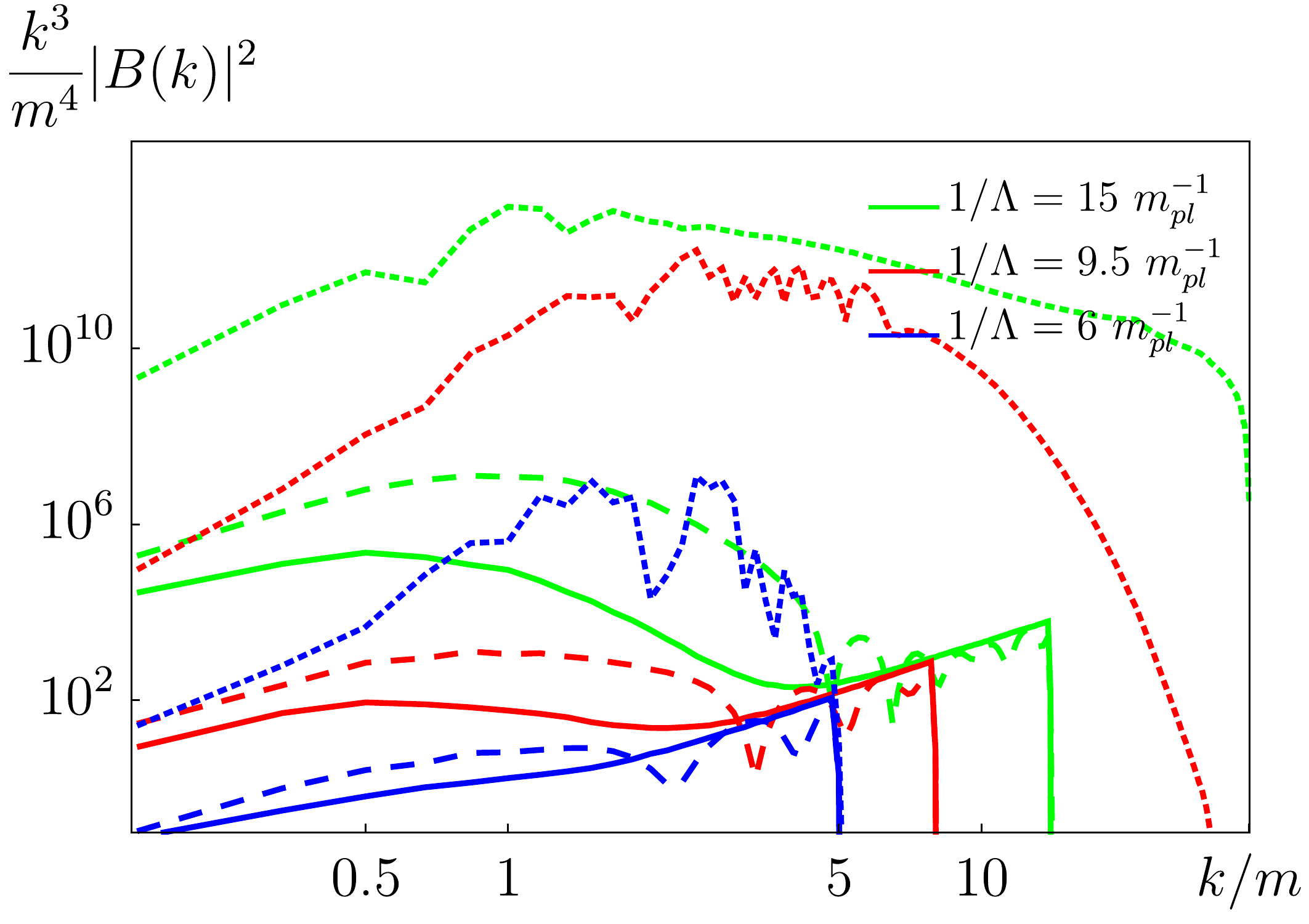}
		\caption{} \label{3_couplings_magnetic}
	\end{subfigure}
	
	\caption{The growth, from purple to red, of the magnetic energy power spectra of the three different
		couplings is shown in panels (a),(b) and (c). The time between each line
		is $m\Delta t = 1$. The panel (d) shows a comparison between the three at
		different times: at the beginning of the simulation (continuous line), at the end of inflation (dashed line) and at $mt = 100$ (dotted line).}  \label{fig:magnetic_PS}
\end{figure}

\begin{figure} % "[t!]" placement specifier just for this example
	\begin{subfigure}{0.49\textwidth}
		\includegraphics[width=\linewidth]{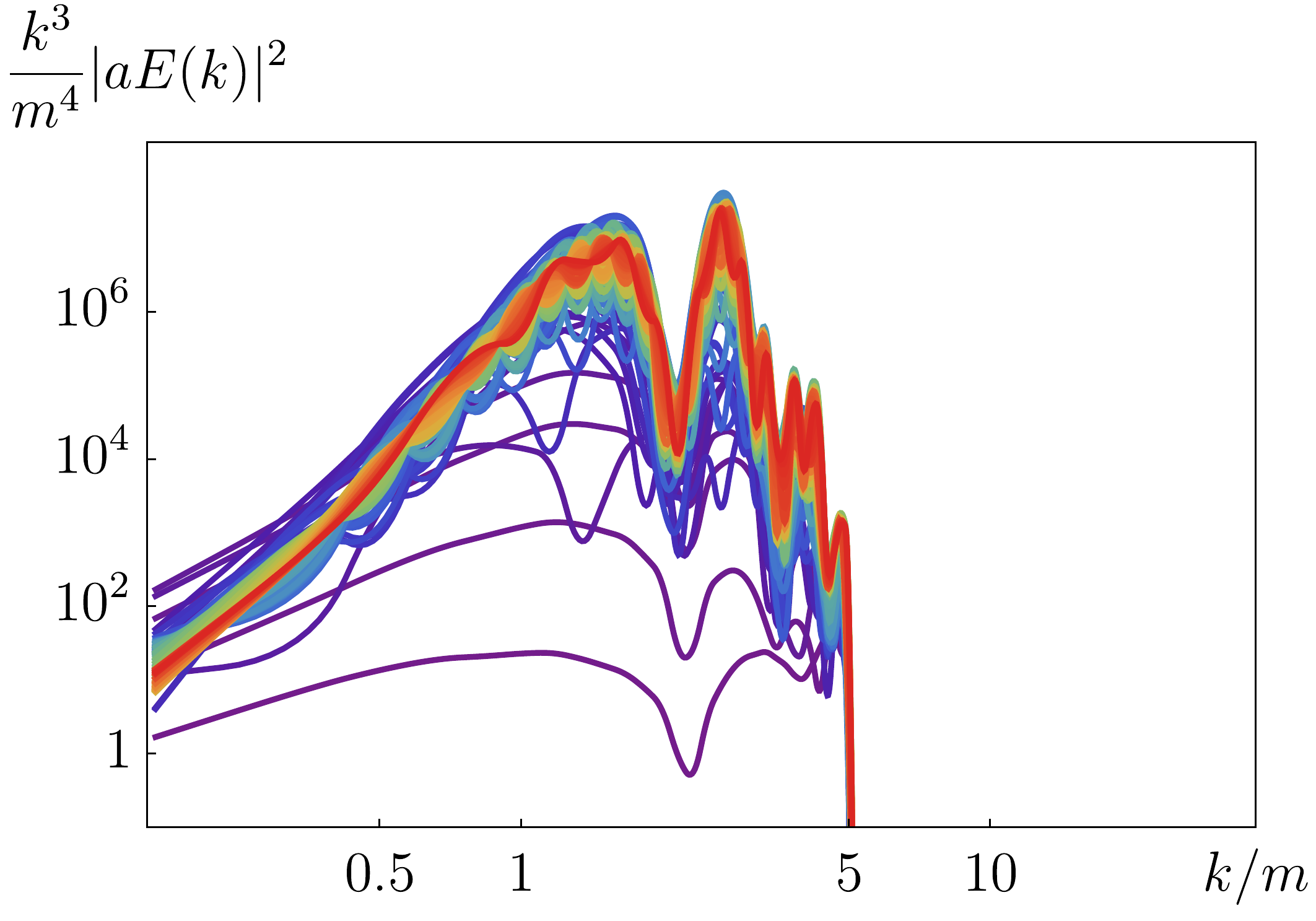}
		\caption{$1/\Lambda = 6~m_{pl}^{-1}$} \label{fig:E2PS_6}
	\end{subfigure}\hspace*{\fill}
	\begin{subfigure}{0.49\textwidth}
		\includegraphics[width=\linewidth]{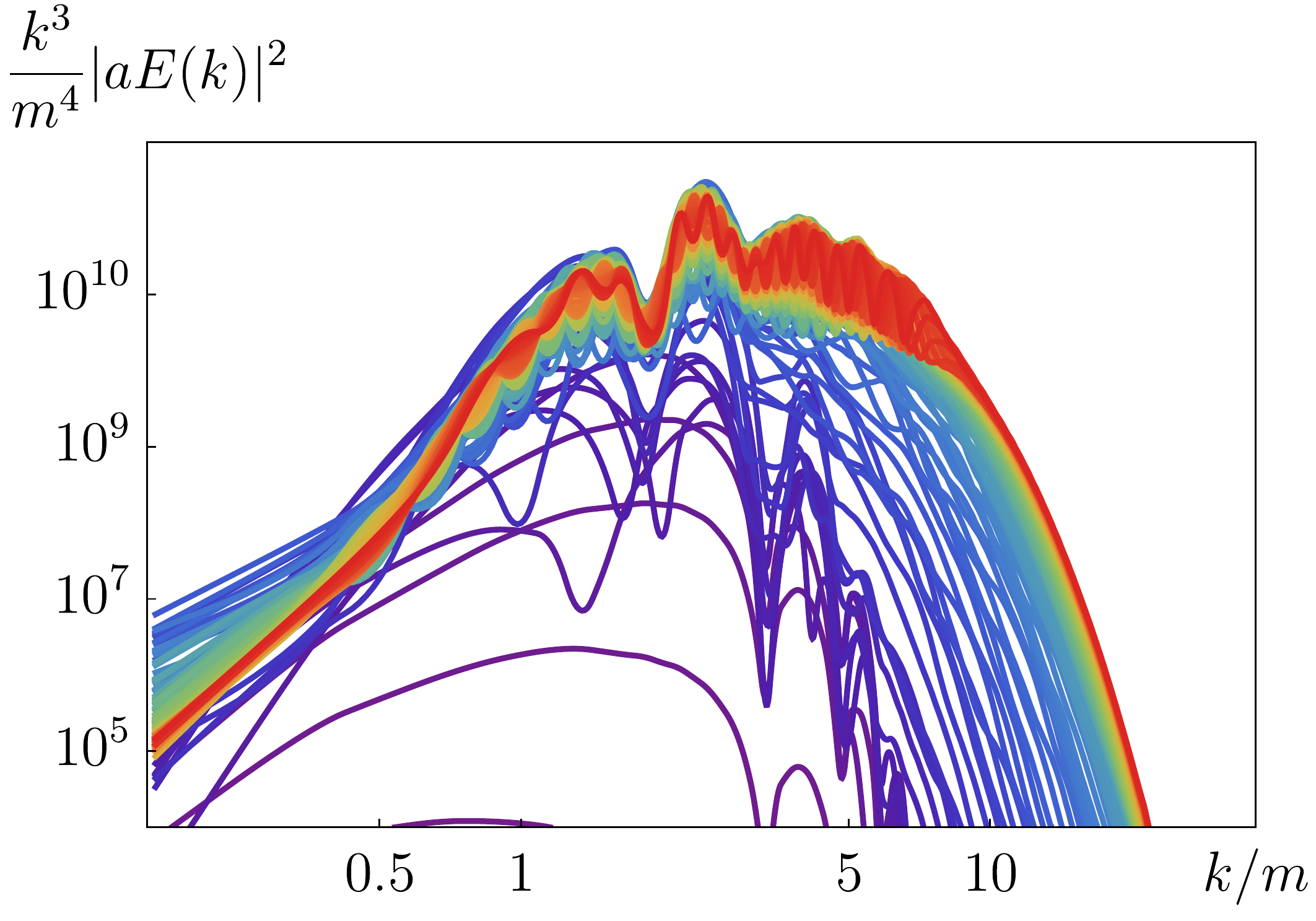}
		\caption{$1/\Lambda = 9.5~m_{pl}^{-1}$} \label{fig:E2PS_95}
	\end{subfigure}	
	\medskip
	\begin{subfigure}{0.49\textwidth}
		\includegraphics[width=\linewidth]{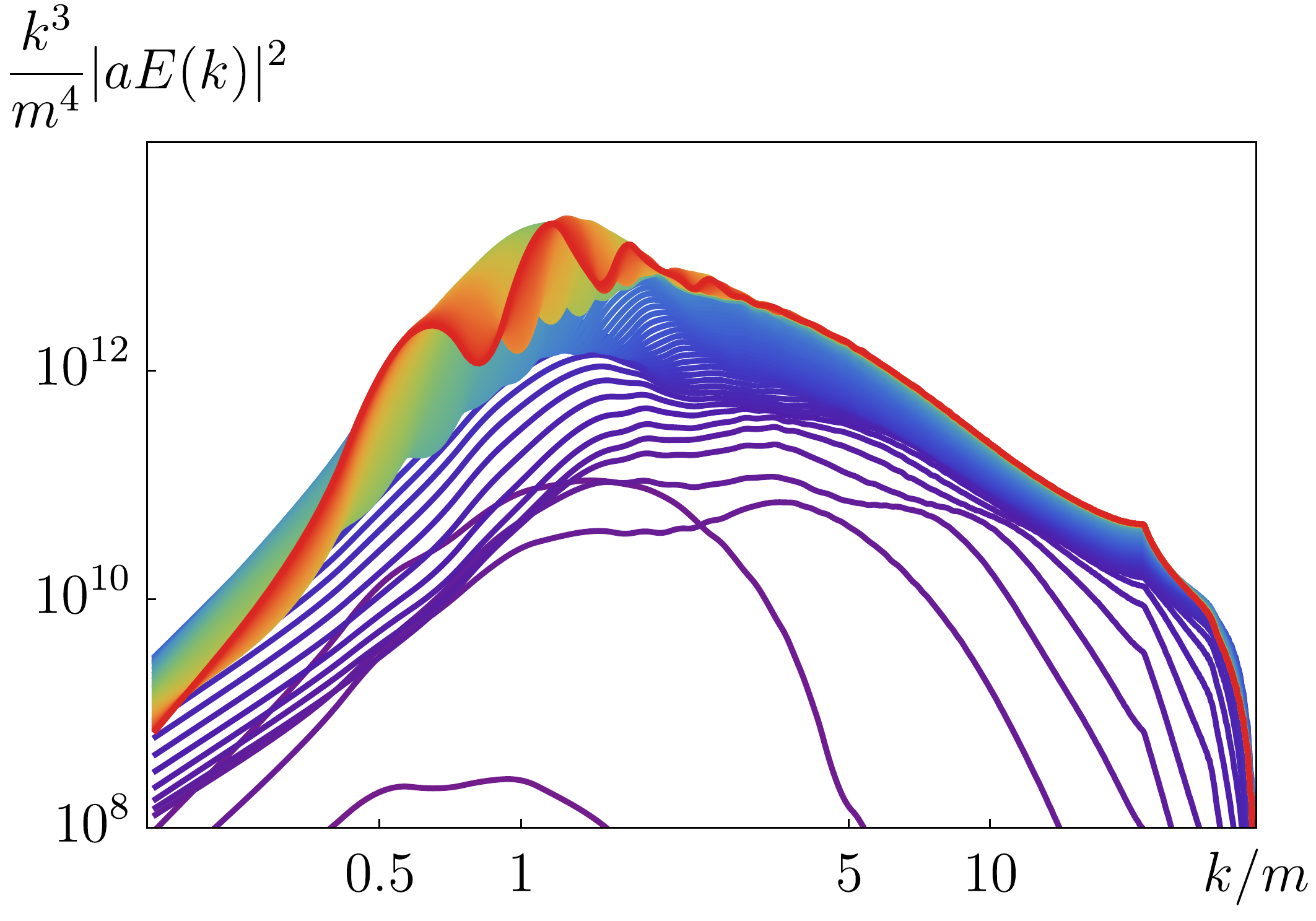}
		\caption{$1/\Lambda = 15~m_{pl}^{-1}$} \label{fig:E2PS_15}
	\end{subfigure}\hspace*{\fill}
	\begin{subfigure}{0.49\textwidth}
		\includegraphics[width=\linewidth]{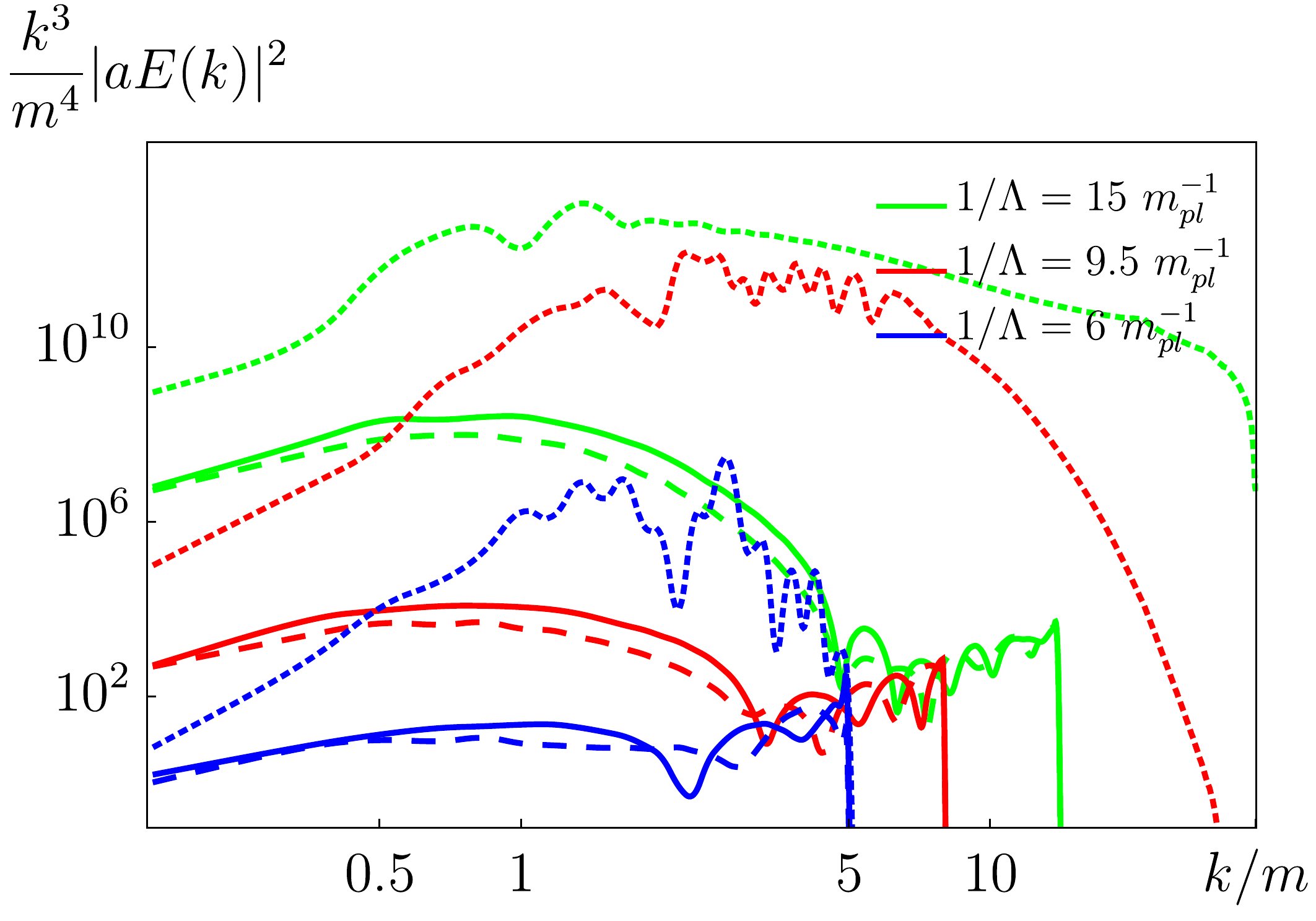}
		\caption{} \label{3_couplings_electric}
	\end{subfigure}
	\caption{The growth, from purple to red, of the electric energy power spectra of the three different
		couplings is shown in panels (a),(b) and (c). The time between each line
		is $m\Delta t = 1$. The panel (d) shows a comparison between the three at
		different times: at the beginning of the simulation (continuous line), at the end of inflation (dashed line) and at $mt = 100$ (dotted line).}  \label{fig:electric_PS}
\end{figure}

For a better understanding of the dynamics of preheating, we analyse now the time evolution of the power spectra of the gauge field. We will actually consider only the power spectra of the ``comoving'' electric and magnetic energy densities, removing the appropriate scale factor powers from the physical expressions, so that the dilution due to the expansion of the Universe is not visible, and we can focus only in the gauge field excitation due to its coupling to the axion. In Fig.~\ref{fig:magnetic_PS} the evolution of the magnetic energy power spectra for the previous three representative couplings, while the same analysis for the electric power spectrum is shown in Fig.~\ref{fig:electric_PS}. We will focus our discussion about power spectra only based on the magnetic energy spectrum, as the details on the electric counterpart are rather similar. 

In Figs.~\ref{fig:B2PS_6}-\ref{fig:B2PS_15} we see that the growth of the three coupling cases is, as expected, visibly distinct. This fact is highlighted by Fig.~\ref{3_couplings_magnetic}, where the line-styles represent different stages of the lattice evolution. We see that for 
the sub-critical case $1/\Lambda = 6~ m_{pl}^{-1}$ plot in Fig.~\ref{fig:B2PS_6}, only infrared modes $k < k_{*} = 5~m$ are excited, whereas higher modes are barely excited and don not play any role in the dynamics\footnote{Ultraviolet modes $k > k_{*} = 5~m$ in the sub-critical coupling example are interpreted as remaining in vacuum and hence should not be considered in a classical  simulation. Their energy density however is totally sub-dominant compared to the infrared modes, and their presence in our simulation is harmless.}.

The picture changes for the critical couplings $1/\Lambda = 9.5 m_{pl}^{-1}$ shown in Fig.~\ref{fig:B2PS_95}, where modes  above the initial cutoff are excited through the simulation, due to the non-linearity of the system. Let us recall that during inflation higher modes were successively excited as they approached the Hubble scale from initially sub-horizon scales, what corresponds to a linear regime of the system, where modes of different wavelength are not influenced by each other. However, after inflation, when the axion starts oscillating, the backreaction between the gauge field and the axion becomes more noticeable, so the dynamics become non-linear as both fields influence each other, leading to a spread of power into higher momenta. We also note that for both sub-critical and critical couplings,  the power spectra grow towards a stationary configuration, where the modes start to oscillate until they relax. These oscillations are due to the change of sign and amplitude of the axion momenta $\pi_{\phi}$, which is still oscillating around the bottom of its potential. 

Finally we can take a look at the super-critical case $1/\Lambda = 15~m_{pl}^{-1}$ shown in Fig.~\ref{fig:B2PS_15}, where the excitation of ultraviolet-modes above the initial cutoff takes place very rapidly, immediately after the end of inflation. The spread of power to higher modes occurs from the beginning of preheating (before the axion can even reach its minimum), as the growth of the mode amplitudes is actually significantly larger compared to the critical and sub-critical cases. This is precisely the spectral counterpart of what we observed in terms of the energy densities [c.f.~Figs.~\ref{Total_energies_1_15}, \ref{fig:Dynamics_TotalEnergies_3couplings}], as in supercritical cases the energy is transferred so efficiently to the gauge field that from the very end of inflation the systems becomes non-linear, resulting in an immediate excitation of ultraviolet modes well above the initial cutoff. In summary, the gauge field excitation encompasses now a broader range of momenta, and it is actually significantly larger in amplitude, as it can be immediately appreciated by comparing the range of scales in both $x$ and $y$ axis of Figs.~\ref{fig:B2PS_6}, \ref{fig:B2PS_95} and \ref{fig:B2PS_15}. Most of the excited modes in the dynamical range of supercritical coupling, have actually amplitudes within a range of $3$ orders of magnitude.  An interesting aspect appreciated in Fig.~\ref{fig:B2PS_15} is that the oscillations of the modes around the final stable configuration are not present, except for a small range around $k/m \approx 1$. %This is a complicate effect dur to the time evolution of the axion conjugate momenta $\pi_\phi = \dot \phi$, since the transfer of energy to the gauge field is so abrup that the axion velocity falls down rapidly (due to conservation of energy), and the tachyonic excitation is then reduced. We will comment further on this in section~\ref{subsec:Efficiency}, when we discuss the efficiency of energy transfer as a function of the coupling $1/\Lambda$.

In Fig.~\ref{3_couplings_magnetic} we present finally a direct comparison of the power spectra for the three chosen couplings, showing them at the beginning of the simulation $\sim 1$ efold before the end of inflation (continuous lines), at the end of inflation (dashed lines), and at the end of the simulation when the system has reached a stationary configuration (dotted lines). The difference in the amplitudes and in the range of excited modes are quite evident. However, we would like also to point out two aspects that the three cases have in common. The first one is that the peak of the power spectra is around $k/m \approx 3$, and the second is that the highest rate of excitation happens during the very first stages after the end of inflation. The transfer of energy with an axionic coupling is therefore extremely efficient, and will lead to very efficient preheating for the highest couplings.

\subsubsection{Preheating efficiency}
\label{subsec:Efficiency}
\begin{figure} % "[t!]" placement specifier just for this example
	\begin{subfigure}{0.49\textwidth}
		\includegraphics[width=\linewidth]{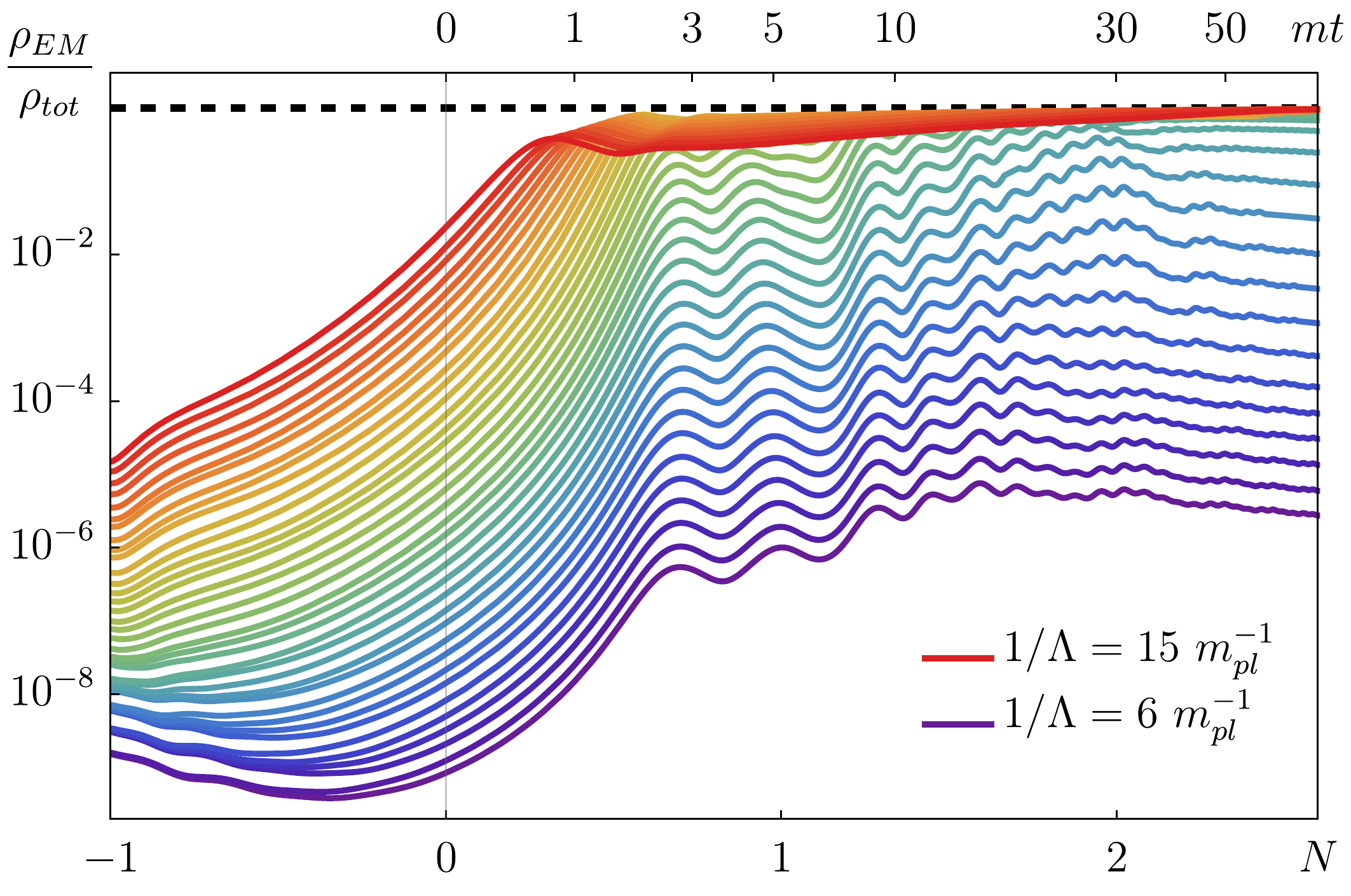}
		\caption{log scale} \label{fig:preheating_efficiency_log}
	\end{subfigure}\hspace*{\fill}
	\begin{subfigure}{0.49\textwidth}
		\includegraphics[width=\linewidth]{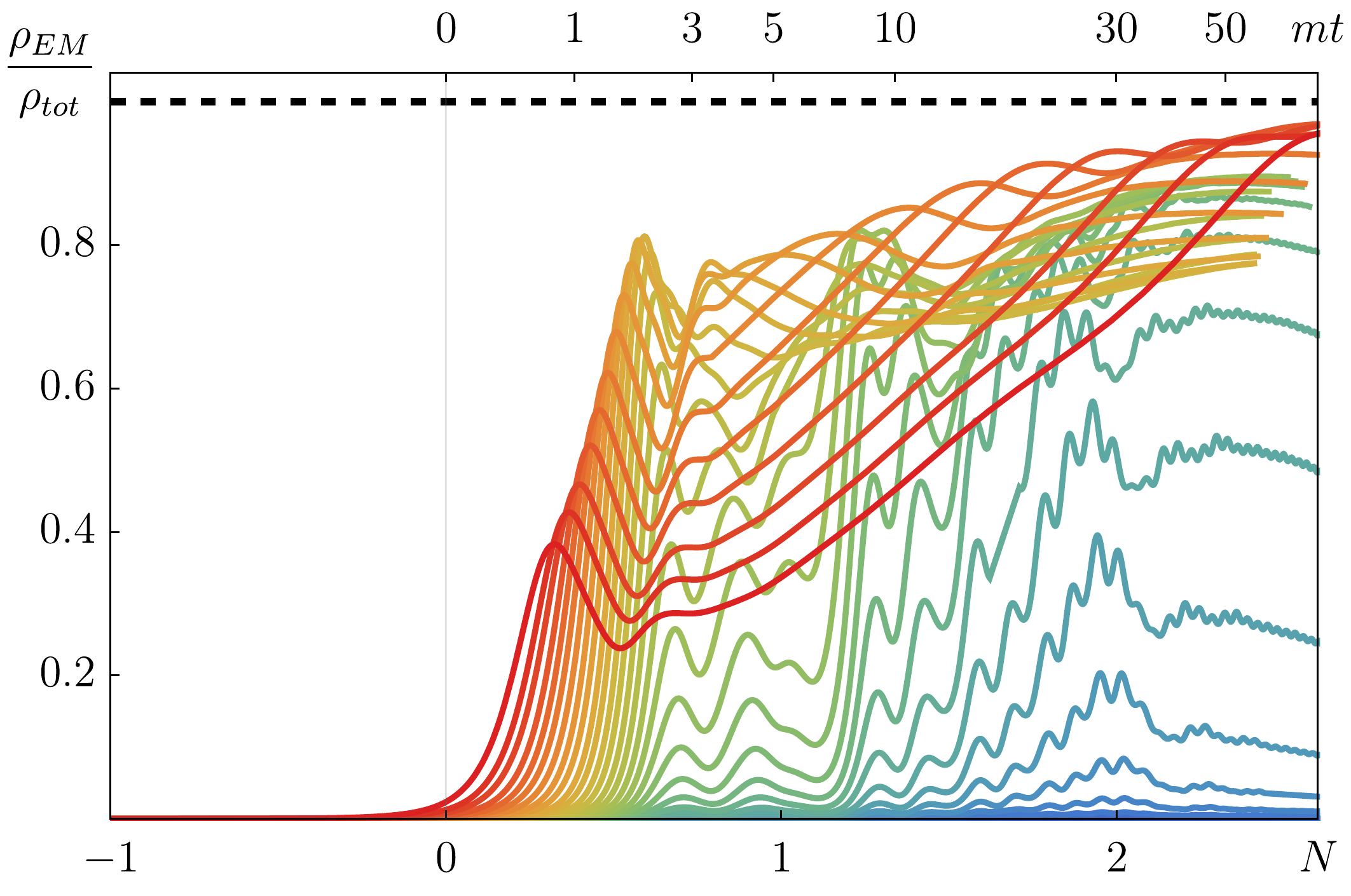}
		\caption{linear scale} \label{fig:preheating_efficiency_linear}
	\end{subfigure}
	\hspace*{\fill}
	\caption{The preheating efficiency evolution in semi-logarithmic (a) and linear (b) scales for couplings going from $6 ~m_{pl}^{-1}$, in purple, to $15~m_{pl}^{-1}$, in red.}  \label{fig:preheating_efficiency}
\end{figure}
We analyze now the efficiency of preheating as a function of the coupling $1/\Lambda$. In particular we monitor the evolution in time of the ratio between the electromagnetic and total energy densities $\rho_{EM}/\rho_{tot}$. This is what is shown in Fig.~\ref{fig:preheating_efficiency} for $1/\Lambda$ ranging from $6~m_{pl}^{-1}$  to $15~m_{pl}^{-1}$, sampled in $\Delta \Lambda^{-1} = 0.25 ~m_{pl}^{-1}$ intervals. We notice that, for small couplings $1/\Lambda < 9~m_{pl}^{-1}$, the preheating efficiency is typically smaller than $\sim 10~\%$, while for large couplings couplings $1/\Lambda > 10~m_{pl}^{-1}$, the energy stored in the electromagnetic field is above $\sim 80~\%$, reaching an almost perfectly reheated Universe as soon as the couplings reach $1/\Lambda \gtrsim 14~m_{pl}^{-1}$. The transition between inefficient ($\lesssim 50\%$) and efficient ($\gtrsim 50\%$) preheating occurs precisely around the critical value $1/\Lambda_c \approx 9.5~m_{pl}^{-1}$.

Looking in detail at Fig.~\ref{fig:preheating_efficiency_log} we can appreciate different patterns of the preheating efficiency. A first pattern concerns the sub-critical coupling regime $1/\Lambda \lesssim 9.5~m_{pl}^{-1}$, for which the gauge energy density growth is dictated in first place by the tachyonic instability, but once the scalar sector starts oscillating, the growth suddenly decreases drastically and wiggles appear. We can see that for such small couplings the growth continues at a very slow rate until $mt \approx 20-30$. From that moment onward the electromagnetic energy density starts decreasing in time, as the instability due to the axionic coupling cannot compensate the dilution due to the expansion of the Universe (which effectively behaves as matter domination after $mt \approx 20-30$). 

For supercritical couplings $1/\Lambda \gtrsim 9.5~m_{pl}^{-1}$, the transfer of energy into the gauge field is very efficient immediately after the end of inflation, the more efficient the larger the coupling $\Lambda^{-1}$. However, for the largest couplings $1/\Lambda \gtrsim 12 ~m_{pl}^{-1}$, the transfer is so efficient that the growth of the gauge energy reaches rapidly a local maximum around $mt \sim  1-2$, and the growth then ceases. This pattern can be clearly appreciated in Fig.~\ref{fig:preheating_efficiency_linear}. The larger the coupling $\Lambda^{-1}$ the sooner the maximum in $\rho_{EM}/\rho_{\rm tot}$ is reached, and the smaller its amplitude is. The reason for this is that for larger couplings in the supercritical regime, the earlier the kinetic energy density of the axion drops below the electromagnetic energy density. When the kinetic axion energy becomes very sub-dominant (typically $\sim 10\%$ of the electromagnetic energy), the tachyonic instability looses efficiency and the excitation of gauge field stops, until the kinetic energy is eventually restored, inducing again the tachyonic instability. In order words, after reaching the local maximum, the ratio $\rho_{EM}/\rho_{tot}$ falls for short while, but begins to grow again soon afterwards, as the tachyonic instability becomes again active. This pattern can be observed for instance in Fig.~\ref{Total_energies_1_15}, and we certify that we have always observed it  for sufficiently large couplings. It is not clear to us, however, whether the gradients of the axion field are also playing any role in this pathological behaviour, as the term $\propto\vec{\nabla}\phi$ is present in the equation of motion of the gauge field Eq.~(\ref{eq:gaugeEOM_vec_C_curved}).

\begin{figure}[t]
	\centering
	\includegraphics[width=0.7\linewidth]{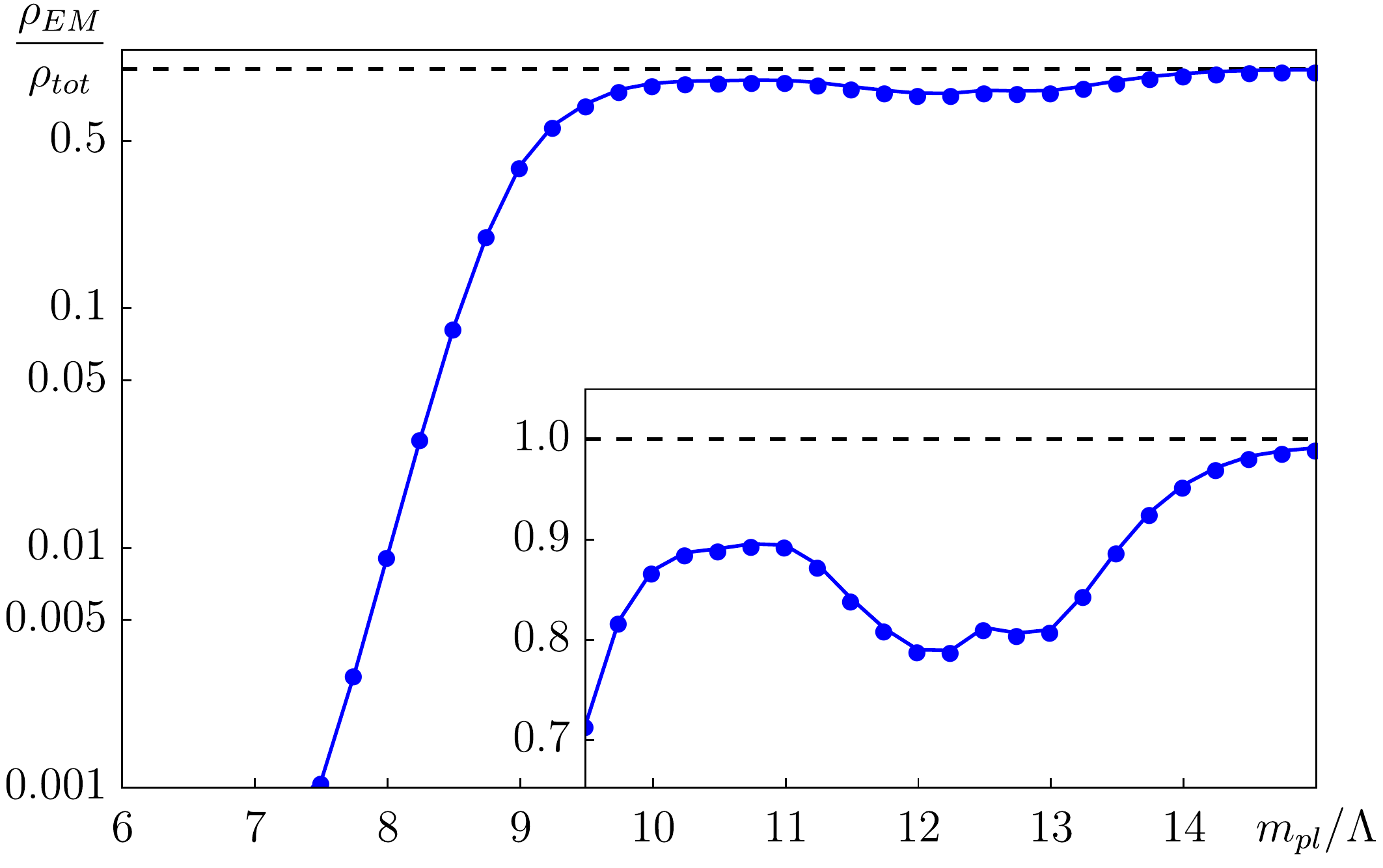}
	\caption{The maximal value of the efficiency ratio $\rho_{EM}/\rho_{tot}$ for the different couplings. In the left-lower part we present a magnified plot for the $1/\Lambda \geq 9.5~m_{pl}^{-1}$ couplings, for which preheating is efficient.}
	\label{fig:maxratios}
\end{figure}

Finally, let us note that the precise value of the critical coupling $\Lambda_c^{-1} \simeq 9.5 m_{pl}^{-1}$ can be actually inferred from Fig.~\ref{fig:maxratios}, where we show the maximum values that the ratio $\rho_{EM}/\rho_{tot}$ reaches for each simulation. Looking at Fig.~\ref{fig:maxratios} it becomes very clear why we speak in first place about sub-critical $\Lambda^{-1} \lesssim \Lambda_c^{-1}$ and supercritical $\Lambda^{-1} \gtrsim \Lambda_c^{-1}$ coupling values. For couplings below the critical value the preheating efficiency decreases actually exponentially fast. We also notice that only when the system transfers at least $\sim 80~\%$ of its energy into the gauge field, then it reaches eventually a stationary configuration where the energy fractions remain constant. Since in the supercritical case the Universe energy budget is dominated by the massless gauge field, the energy density of the gauge fields scales as radiation. The Universe is therefore fully reheated, entering into a radiation dominated epoch very soon after the end of inflation, for couplings $\Lambda^{-1} \gtrsim 14~m_{pl}^{-1}$ (see zoomed plot in Fig.~\ref{fig:maxratios}).

\subsection{Verification of dynamical constraints}
\label{subsec:Constraints}

\label{sec:constraints}

In order to test the correctness of our lattice approach, we are going to study the evolution of the Gauss and Hubble constraints of the system, which are given by Eqs.~(\ref{eq:GL_EOM_curved_L}), (\ref{eq:Hubble_constraint_L}). To be more precise, we will obtain the violation of these two constraints at each time step by computing the following quantity,
\begin{eqnarray}
\Delta_{G,H} = \frac{\sqrt{(lhs - rhs)^2}}{\sqrt{(lhs)^2 + (rhs)^2}} \text{  ,  }
\end{eqnarray}
where $lhs$ and $rhs$ refer to the left and right hand sides of Eqs.~(\ref{eq:GL_EOM_curved_L}), (\ref{eq:Hubble_constraint_L}) respectively. Then a total violation of the constraint corresponds to $\Delta_{G,H} = 1$, while the exact conservation would imply values of $\Delta_{G,H}$ close to machine precision. 

\begin{figure} % "[t!]" placement specifier just for this example
	
	\begin{subfigure}{0.49\textwidth}
		\includegraphics[width=\linewidth]{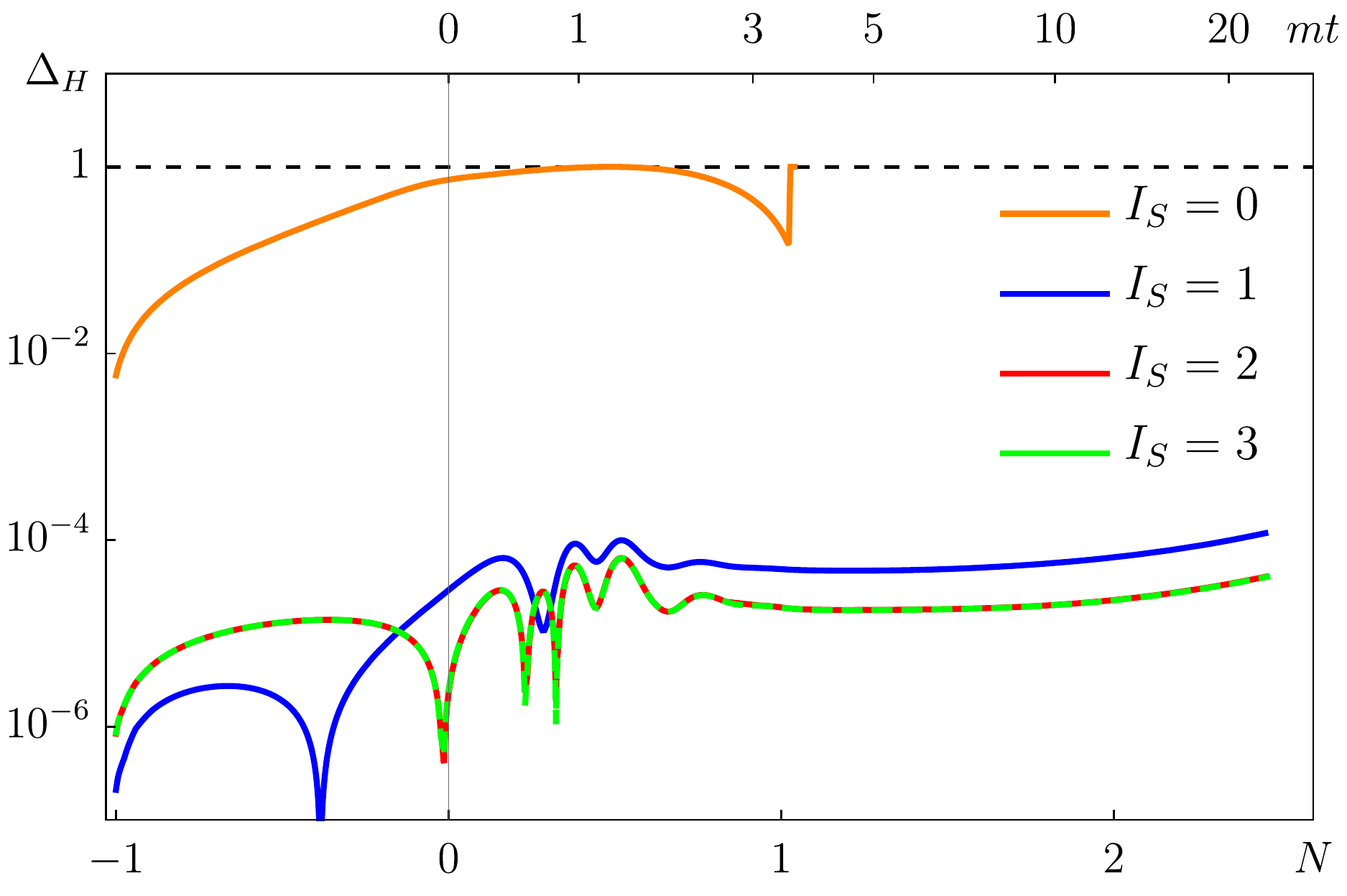}
		\caption{} \label{fig:HubbleConstraint_15}
	\end{subfigure}
	\hspace*{\fill}
	\begin{subfigure}{0.49\textwidth}
		\includegraphics[width=\linewidth]{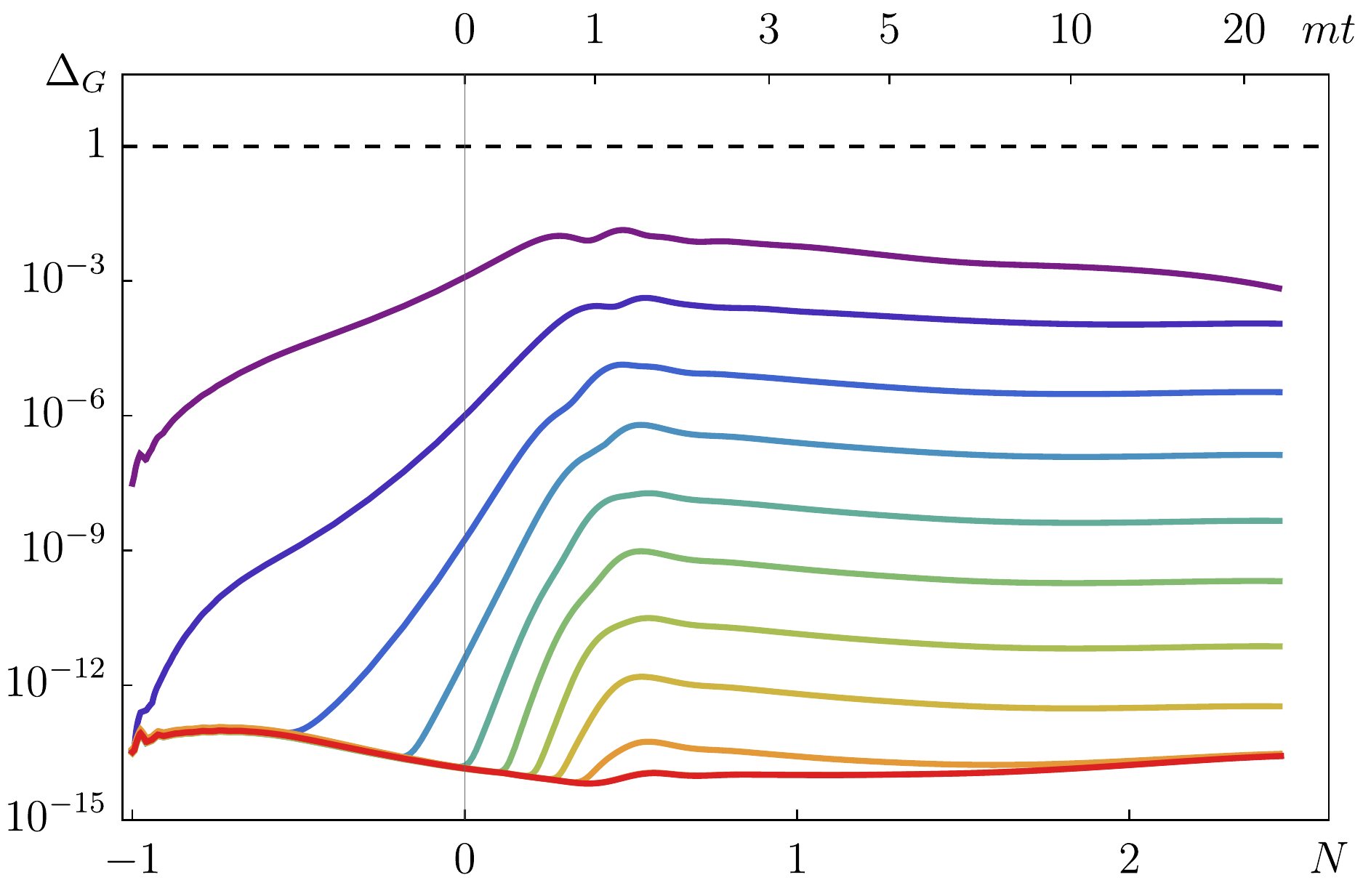}
		\caption{} \label{fig:GaussConstraint_15}
	\end{subfigure}
	
	\caption{Panel (a): Hubble constraint violation through the simulation for different scalar iterations $I_S$. $I_E=2$ is fixed. Panel (b): Gauss constraint violation through the simulation for different number of electric iterations, $I_E = 0~\text{(purple)},..., 11~\text{(red)}$. $I_S=2$ is fixed. }  \label{fig:Constraints}
\end{figure}

We start by studying the dependence of the Hubble constraint on the number of scale factor iterations, which will be denoted by $I_S$, while keeping fixed the number of electric iterations, $I_E$, at $2$. In Fig.~\ref{fig:HubbleConstraint_15} we can inspect its violation for a simulation with coupling $1/\Lambda = 15~m_{pl}^{-1}$. We notice that for $I_S = 0$ the system reaches the complete violation of the constraint, which corresponds to the development of an instability. Indeed in this case the dynamics of the fields are completely spoiled as the values quickly diverge towards infinities. The end of the orange line, $I_S = 0$, corresponds to the appearance of a \textit{non-numerical value} in the outputs of the simulation. This implies that the explicit approximation of the scale factor, Eq.~(\ref{eq:explicit_approx_a_L}), makes the system of equations unstable. Thus the results will be totally spoiled if the iterative method presented in Sect.~\ref{subsec:iterative_method_for_the_scale_factor} is not used.
Concerning the other three cases we notice that the conservation of the Hubble constraint has almost the same evolution. More precisely, the case $I_S = 1$ presents a small deviation compared to $I_S = 2,3$, which instead are indistinguishable. This behaviour repeats itself also for $I_E \neq 2$. From these considerations we conclude that $I_S = 2$ is the minimal number of iterations of the scale factor in order to have the best possible conservation of the Hubble constraint.

Having 
fixed the number of scale factor iterations, we can study the evolution of the Gauss constraint as a function of the number of electric iterations.  This dependence is shown in Fig.~\ref{fig:GaussConstraint_15}, for $I_E$ going one by one from $0$ (purple) to $11$ (red) and the coupling $1/\Lambda = 15~m_{pl}^{-1}$. We see that each extra iteration reduces the violation of the Gauss constraint by a 
factor $\sim 10$, until a lower limit is reached. In fact the lines given by $I_E = 10$ and $I_E = 11$ are superimposed and higher iterations give the same result.  This demonstrates  that, for a sufficiently high number of electric iterations, our lattice approach conserves the Gauss constraint up to machine precision.
The question now is to know if all these iterations and the consequently increased precision in the Gauss constraint are really necessary to capture the dynamics of the problem. 
This is important because, for each iteration, we are adding loops of computations over the whole lattice grid, hence increasing the computational cost. In fact we have found out that each iteration increases of $\sim 7~\%$ the total time of simulation, as shown in Appendix \ref{app:timedemand_study}. The conclusion is that the dynamics are almost insensitive to the number of electric iterations, as opposed to what happens with $I_S$. In fact we only noticed a small departure from the physical trajectory for $I_E = 0$, while for $I_E \geq 1$ substantially we have the same evolution through the whole simulation. We then decided to set both electric and scale iterations to $2$ during our simulations.

\section{Discussion}
\label{sec:Discussion}

In this paper we have introduced a lattice formulation of an axionic coupling ${\phi\over\Lambda} F\tilde F$, generalizing the original flat-space formulation from Ref.~\cite{Figueroa:2017qmv} to curved space with self-consistent expansion of the Universe, so that both (pseudo-)scalar and gauge field degrees of freedom contribute to the expansion rate. This formulation preserves (at the lattice level) all the properties of the continuum theory: gauge invariance, shift symmetry and the topological nature of $F_\mn\tilde F^\mn = \partial_\mu K^\mu$. We have proposed iterative methods to deal with the fact that we can no longer solve the lattice equations of motion by an explicit method. We have demonstrated that despite this technical complication, our numerical evolution is capable of preserving the Gauss constraint down to machine precision, and the Hubble expansion constraint down to the usual $\sim 10^{-4}$ accuracy in standard leap-frog evolution algorithms. Our lattice formulation represents therefore an optimal tool for studying the dynamics of axion-driven scenarios with a derivative coupling to an Abelian gauge field, as we approximate correctly the continuum theory  up to quadratic order corrections in the lattice spacing, while preserving all of its relevant properties on the lattice.

Using the proposed lattice formulation we have studied the last efolds of inflation and the onset of preheating in an axion-inflation model with quadratic potential. We have characterized in detail the energy transfer from the inflaton to the Abelian gauge field as a function of the strength of the coupling $1/\Lambda$. We have first observed that the analytical solution provided in the literature (for $\xi$ constant), fails significantly towards the end of inflation. Hence, in order to set the initial condition on the lattice, it is better to solve numerically in Fourier space the evolution of the gauge field deep inside inflation, when the backreaction is still negligible. Once the lattice code is initialized, full account of the backreaction of the gauge field is automatically considered in our simulations, including the influence of the gauge field on the axion dynamics, on the expansion rate, and even on the possible development of axion inhomogeneities.

We have quantified in detail  for quadratic axion-inflation the efficiency of preheating, i.e.~the ability of the system to transfer a given percentage of energy from the axion into the gauge field, as a function of the strength of the coupling $1/\Lambda$. Two coupling regimes are clearly identified, sub-critical $\Lambda^{-1} \lesssim  \Lambda_c^{-1}$ and super-critical $\Lambda^{-1} \gtrsim \Lambda_c^{-1}$, depending on whether the final energy fraction stored in the gauge field is below or above $\sim 50\%$. The approximated value of the critical coupling $\Lambda_c^{-1} \simeq 9.5~m_{pl}^{-1}$ can be extracted from Fig.~\ref{fig:maxratios}, which summarizes well the preheating efficiency in this scenario. For sub-critical couplings the efficiency of preheating drops exponentially when reducing the coupling, as the tachyonic growth due to the axionic coupling can never compete with the dilution due to the expansion of the Universe. For super-critical couplings the efficiency grows very fast when increasing the value of the coupling, reheating the Universe very efficiently (with more than $\sim 95\%$ of the total energy in the gauge field) for couplings $\Lambda^{-1} \gtrsim 14\,m_{pl}^{-1}$.

Despite of all the care put in the construction of a lattice formulation that respects all the necessary symmetries of the continuum theory, our numerical results on preheating are however very similar to those presented already in Ref.~\cite{Adshead:2015pva}, where lattice simulations of the same system were also explored with a less refined lattice formulation. Actually, a flat-space version of our lattice formulation has been used before in the context of the anomalous non-conservation of fermion number in Abelian gauge theories at finite temperature~\cite{Figueroa:2017hun}, where the statistical study of the Chern-Simons number $Q(t) \propto \int_{0}^t dt' \int d^3x F\tilde F = N_{\rm CS}(t) - N_{\rm CS}(0)$, with $N_{\rm CS}(t) \propto \int d^3x \vec A \vec B$, proved to be very sensitivity to the lattice formulation used. It seems therefore that depending on the observable, and perhaps on the context, the introduction of a lattice representation of $F\tilde F$ that can be expressed as a total derivative (hence preserving exactly the shift symmetry on the lattice), matters or not. Clearly, in the context where we have applied our formalism here -- preheating after axion inflation --, we are forced to conclude that the precise conservation on the lattice of gauge invariance and shift symmetry, seems not to be particularly relevant. We consider our analysis however, the demonstration of this fact; something that, admittedly, we were not expecting.

Our lattice formulation represents in any case a powerful tool for studying the non-linear dynamics of any system composed by an axion-like field (with arbitrary potential) and an Abelian gauge field coupled to the latter via a Chern-Simons interaction, considering self-consistent expansion of the Universe, and preserving exactly gauge and shift symmetries on the lattice. We expect therefore that our present paper will be the first one of a series, where we plan to apply our lattice formulation to review and investigate further phenomenological aspects of axion-driven inflation scenarios with a $\phi F\tilde F$ coupling, particularly in the non-linear regime where analytical studies cannot be precise.

\acknowledgments
We are very grateful to M. Shaposhnikov and  V. Domcke for discussions on the project. The work of DGF was supported partially by the ERC-AdG-2015
grant 694896 and partially by the Swiss National Science Foundation (SNSF).

\appendix
\section{Appendix}
\subsection{Dynamical range study}
\begin{figure} % "[t!]" placement specifier just for this example
	\begin{subfigure}{0.32\textwidth}
		\includegraphics[width=\linewidth]{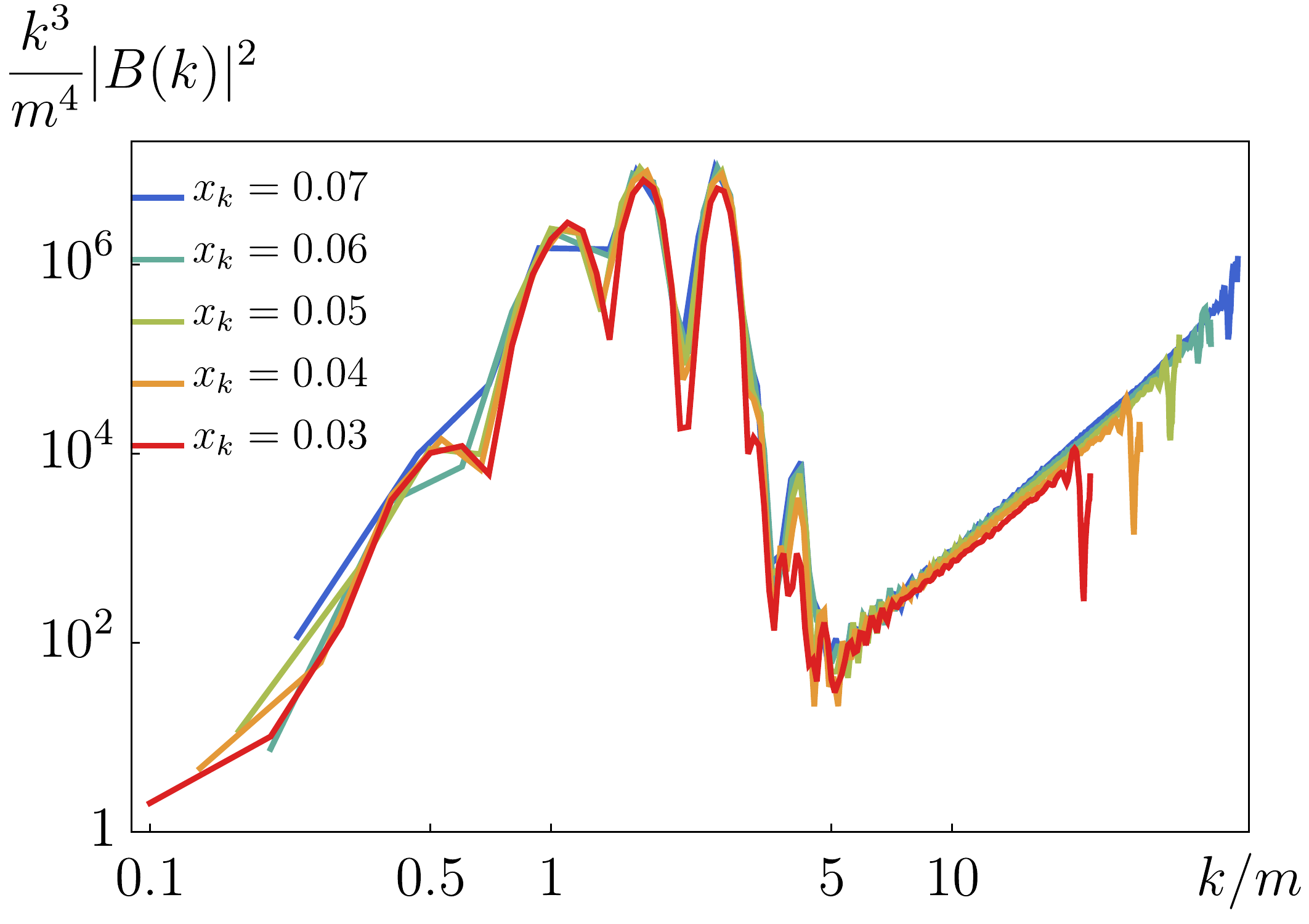}
		\caption{$1/\Lambda = 6~m_{pl}^{-1}$} \label{fig:DynamicalRange_Study_B_6}
	\end{subfigure}\hspace*{\fill}
	\begin{subfigure}{0.32\textwidth}
		\includegraphics[width=\linewidth]{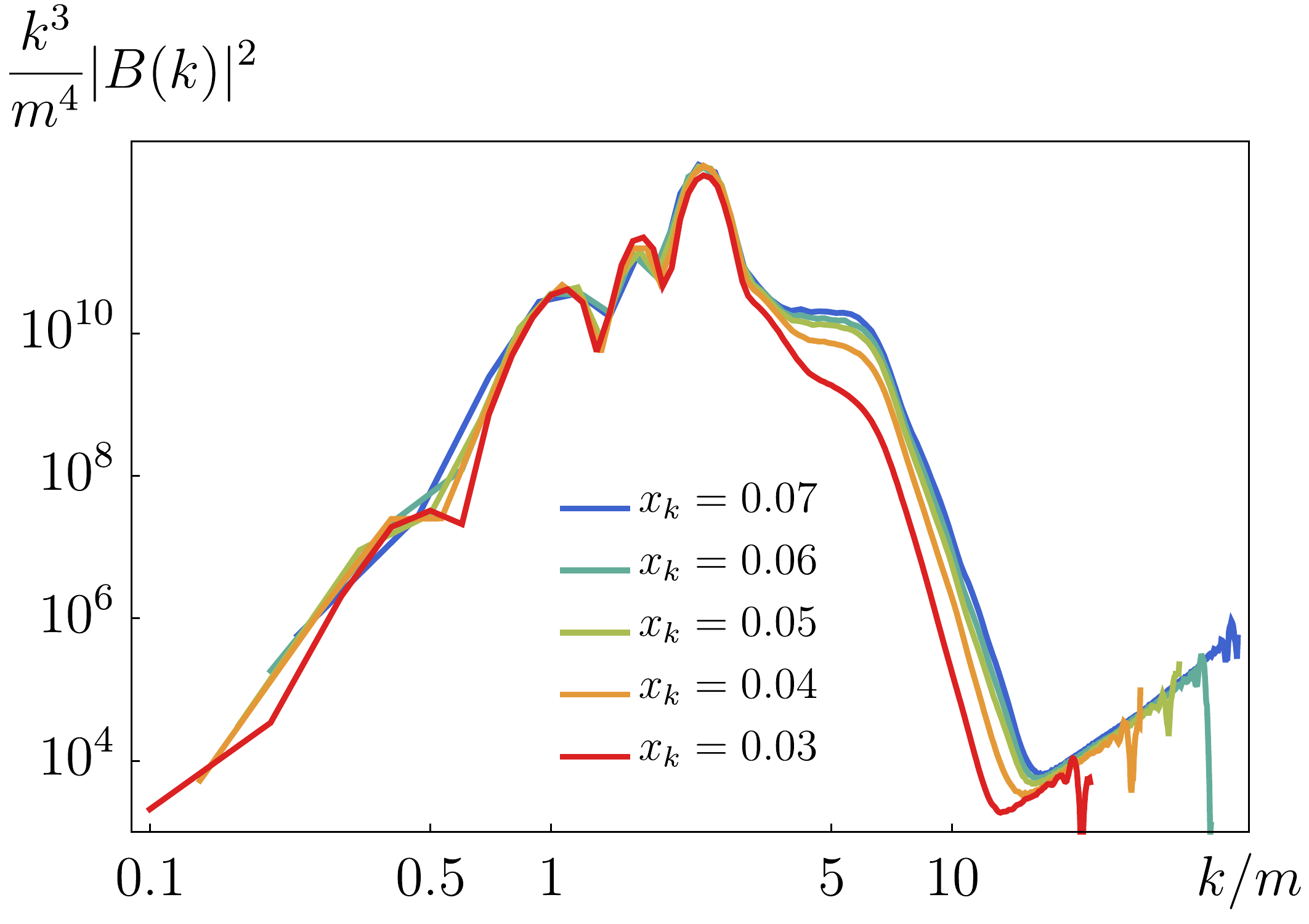}
		\caption{$1/\Lambda = 9.5~m_{pl}^{-1}$} \label{fig:DynamicalRange_Study_B_95}
	\end{subfigure}\hspace*{\fill}
	\begin{subfigure}{0.32\textwidth}
		\includegraphics[width=\linewidth]{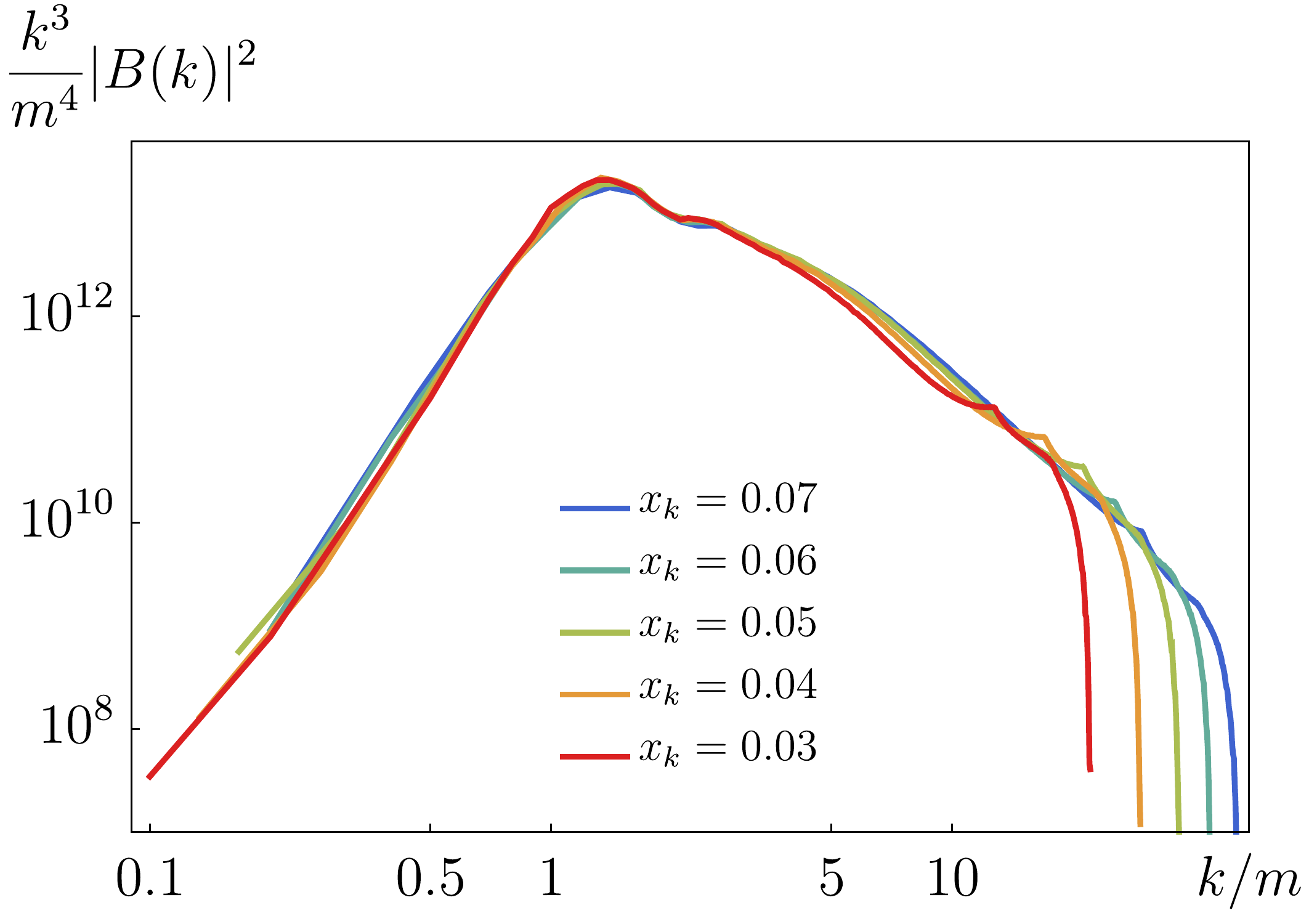}
		\caption{$1/\Lambda = 15~m_{pl}^{-1}$} \label{fig:DynamicalRange_Study_B_15}
	\end{subfigure}
	\caption{Power spectra comparison for different lattice spacings $dx$ at $t = 50~m^{-1}$. The study has been done with $1/\Lambda = 8.25~m_{pl}^{-1}, 11~m_{pl}^{-1}, 15~m_{pl}^{-1}$.}  \label{fig:DynamicalRange_analysis}
\end{figure}

An important aspect which can be analysed from the power spectra of the fields is how well we are covering the dynamical range of momentum of our system. In order to do this we run simulations with slightly different infrared lattice cutoff which we define by $k_{IR} = x_k aH$, where $x_k < 1$ is the arbitrary parameter and $aH = a(t_i)H(t_i)$ is defined by the initial conditions. We know that once $k_{IR}$ is defined, also the UV cutoff $k_{UV}$ is fixed by the relation $k_{UV} = \frac{\sqrt{3}N}{2}k_{IR}$. Then, by comparing the resulting power spectra we will be able to know how different values of $x_k$ affect the dynamics. If no evident distortion of the spectra appear, it means that our momentum coverage is sufficient to capture the dynamics of the system. The results of this study are presented in Fig.~\ref{fig:DynamicalRange_analysis}, where we show the magnetic energy density power spectra for three representative couplings at $mt = 50$. We see that the range of momentum is shifted for each different $x_k$, accordingly to our definition of $k_{IR}$. Nevertheless, the picture is broadly the same and no particular difference can be spotted, even though $x_k = 0.3$ is borderline in the UV. We then conclude that with our choice of $x_k = 0.5$ we are well capturing the dynamical range of the system.  

\begin{figure}
	\begin{subfigure}{0.49\textwidth}
		\includegraphics[width=\linewidth]{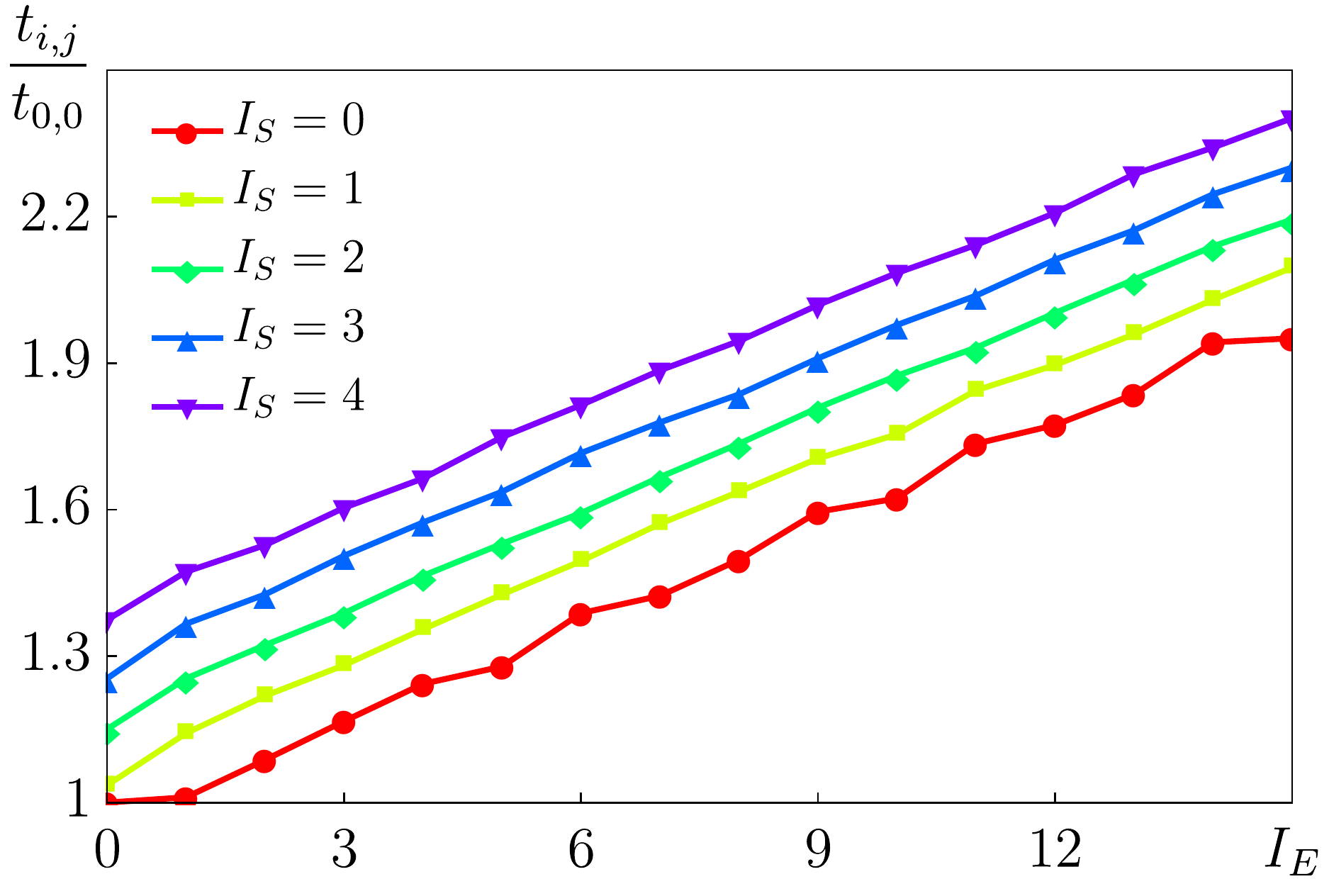}
		\caption{} \label{fig:timeconsume}
	\end{subfigure}\hspace*{\fill}
	\begin{subfigure}{0.49\textwidth}
		\includegraphics[width=\linewidth]{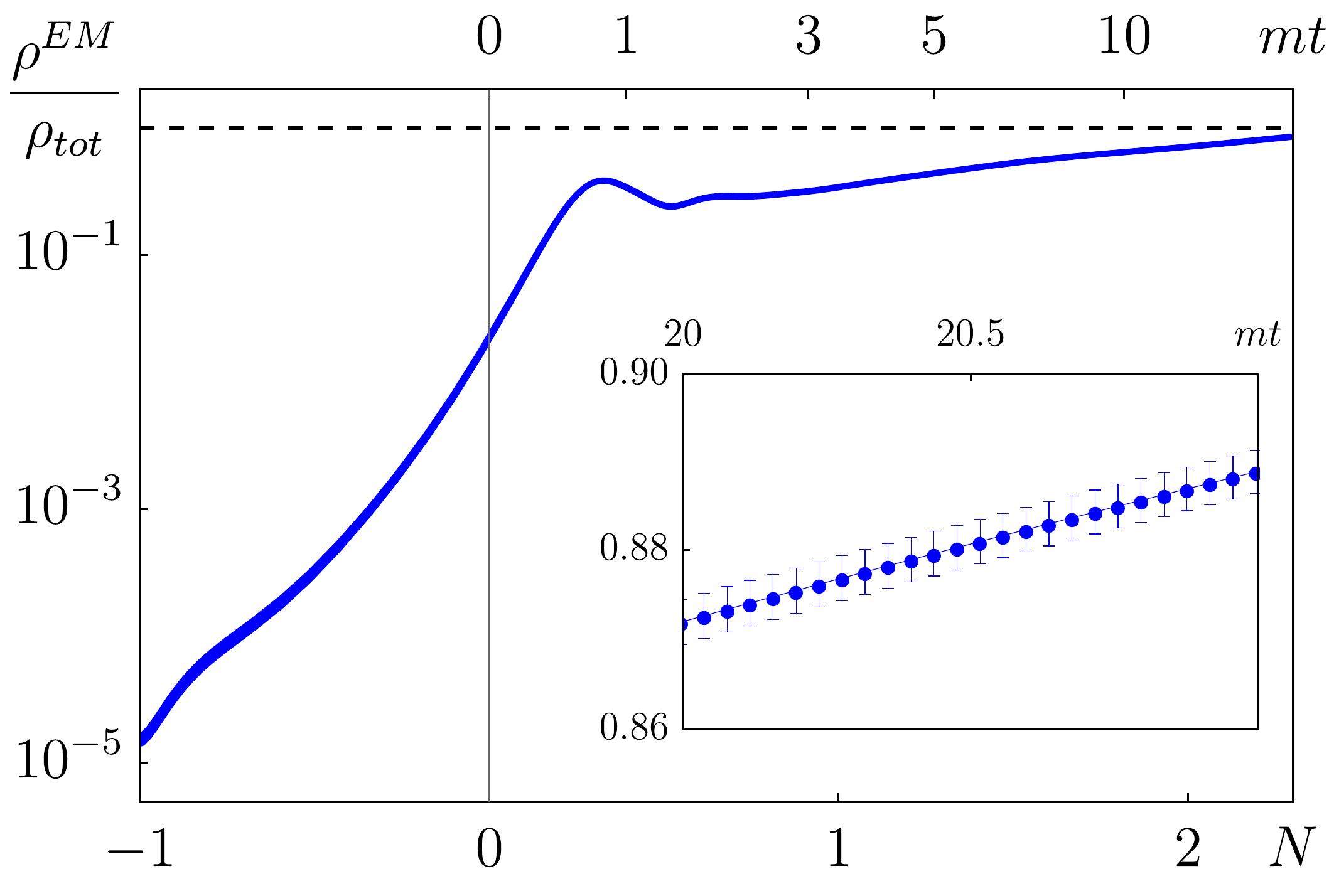}
		\caption{} \label{fig:statstudy}
	\end{subfigure}\hspace*{\fill}
	
	\caption{Panel (a) : Time-demand increase as a function of the number of electric and scale factor iterations, respectively $I_E$ and $I_S$. Panel (b) : Statistical variation in evolution due to random realizations of the initial conditions for $1/\Lambda = 15~m_{pl}^{-1}$. The statistics have been computed over 112 runs.}
	\label{fig:appendixbc}
\end{figure}

\subsection{Time-demand of iterative methods}\label{app:timedemand_study}

In this appendix we present a study on the time-demand of iterative methods. More precisely we would like to quantify how much costs, from the computational point of view, each extra loop imposed by these methods. In order to do this we will compute, for each combination of iterations, the ratio $\frac{t_{i,j}}{t_{0,0}}$,
where $t_{i,j}$ is the time needed for a code implementing $I_E = i$, $I_S = j$ iterations to run the simulation. 

In figure \ref{fig:timeconsume} we show the results for $I_E = 0,1,\dots,15$ and $I_S = 0,1,2,3,4$. From it we see that the increase on time is proportional to both number of electric and scale factor iterations. This is not surprising since each iteration adds always the same amount of computations to the simulation. From fitting the data, we find that the mean incremental percentage for each extra electric iteration is
\begin{eqnarray}
I_S = 0 &\rightarrow& 5.7~\% \nonumber\\
I_S = 1 &\rightarrow& 7.8~\% \nonumber\\
I_S = 2 &\rightarrow& 7.6~\% \nonumber\\
I_S = 3 &\rightarrow& 7.7~\% \nonumber\\
I_S = 4 &\rightarrow& 7.3~\%\text{  .  }  \nonumber
\end{eqnarray}
Moreover, we figured out that for each extra scale factor iteration the computational time increases by $\sim 8 \%$.

\subsection{Statistical fluctuations}
Here we present a statistical study to understand how much the fluctuations due to randomness in the initial conditions affect the dynamics of the system. In fact, each initialization of gauge field follows a random realization of a Rayleigh distribution, which \textit{rms} value coincides with the input value of the initial spectra computed following the prescription of Sect.~\ref{subsec:BackreactionLess}. In Fig.~\ref{fig:statstudy} we see that basically the dynamical evolution in insensitive to the statistical realization, as the standard deviation at each time step is essentially negligible.

\bibliographystyle{h-physrev4}
\bibliography{AxionPreheatBiblio} 

\end{document}